\documentclass[usenatbib]{mn2e}
\input{psfig}
\usepackage{natbib}
\usepackage{varioref}

\newcommand{\galics}{GalICS}
\begin{document}
\def\lsim{\mathrel{\hbox{\rlap{\hbox{\lower4pt\hbox{$\sim$}}}\hbox{$<$}}}}
\def\gsim{\mathrel{\hbox{\rlap{\hbox{\lower4pt\hbox{$\sim$}}}\hbox{$>$}}}}
\def\simlt{\mathrel{\rlap{\lower 3pt\hbox{$\sim$}}\raise 2.0pt\hbox{$<$}}}
\def\simgt{\mathrel{\rlap{\lower 3pt\hbox{$\sim$}}\raise 2.0pt\hbox{$>$}}}

\title[Accretion, feedback and galaxy bimodality]
{Accretion, feedback and galaxy bimodality:  a
comparison  of the GalICS semi-analytic model and cosmological SPH simulations}
\author[Cattaneo et al.]
 {\parbox[t]{\textwidth}{
Andrea~Cattaneo $^{1,2}$, J\'er\'emy~Blaizot $^3$, David~H.~Weinberg$^{2,4}$,
Dusan~Kere{\v s}$^5$
St\'ephane~Colombi$^2$, Romeel~Dav\'e$^6$, Julien~Devriendt$^7$,
Bruno~Guiderdoni$^7$, Neal~Katz$^5$, }
\vspace*{6pt}\\
$^1$Astrophysikalisches Institut Potsdam, an der Sternwarte 16, 14482 Potsdam, Germany\\
$^2$Institut d'Astrophysique de Paris, 98bis Boulevard Arago, 75014 Paris, France\\
$^3$Max-Planck-Institut f\"ur Astrophysik, Karl-Schwarzschild-Str.1, 85740 Garching, Germany\\
$^4$Ohio State University, Department of Astronomy, Columbus, OH 43210, USA\\
$^5$Astronomy Department, University of Massachusetts at Amherst, MA 01003, USA\\
$^6$University of Arizona, Steward Observatory, Tucson, AZ 85721, USA\\
$^7$Centre de Recherche Astronomique de Lyon, 9 Avenue Charles Andr\'e, 69561, St-Genis-Laval Cedex, France\\}
\maketitle
\begin{abstract}
We compare the galaxy population of a smoothed particle hydrodynamics (SPH)
simulation to those predicted by the \galics\ N-body + semi-analytic
model and a stripped down version of \galics\ that omits supernova and
AGN feedback. The SPH simulation and the no-feedback \galics\ model make similar
predictions for the baryonic mass functions of galaxies and for the dependence
of these mass functions on environment and redshift. The two methods also
make similar predictions for the galaxy content of dark matter 
%dw halo 
haloes
as a function of halo mass and
for the gas accretion history of galaxies. There is a fairly good correspondence
between the ``cold" and ``hot'' accretion modes of the SPH simulation and the
rapid and slow cooling regimes of the \galics\ calculation. Both the
SPH and no-feedback \galics\ models predict a bimodal galaxy population at
$z=0$. The ``red'' sequence of gas poor, old galaxies is populated mainly by
satellite systems, which are starved of fresh gas after they begin orbiting
in larger haloes, while, contrary to observations, the central galaxies of massive
haloes lie on the ``blue'' star-forming sequence as a result of continuing
hot gas accretion at late times. Furthermore, both models overpredict the
observed baryonic mass function, especially at the high mass end. In the full
\galics\ model, supernova-driven outflows reduce the masses of low and
intermediate mass galaxies by about a factor of two. AGN feedback suppresses
gas cooling in large haloes, producing a sharp cut-off in the baryonic mass
function and moving the central galaxies of these massive haloes to the red
sequence. Our results imply that the observational failings of the SPH
simulation and the no-feedback \galics\ model are a consequence of missing
input physics rather than computational inaccuracies, that truncating gas
accretion by satellite galaxies automatically produces a bimodal galaxy
distribution with a red sequence, but that explaining the red colours
of the most massive galaxies requires a mechanism like AGN feedback that
suppresses the accretion onto central galaxies in large haloes.
\end{abstract}


\section{Introduction}

The classification of galaxies into early-type and late-type goes back to the earliest studies \citep{hubble26,humason36},
but the clear demonstration that there is no smooth continuity between the two types
was only possible with the advent of large statistical surveys
(\citealp{strateva_etal01,kauffmann_etal03b,
baldry_etal04,balogh_etal04,hogg_etal04};
also see Dekel \& Birnboim 2006
for a comprehensive introduction to the bimodality problem).
Galaxy colour-magnitude diagrams from the
Sloan Digital Sky Survey (SDSS) show a clear distinction between a tight
red sequence with no recent star formation and a broader blue sequence, where star formation is still going on.
The blue sequence is truncated near the characteristic luminosity $L_*$
of the \citet{schechter76} luminosity function, while the red sequence
extends to luminosities several times higher.
Elliptical galaxies concentrate in the bright part of the red sequence ($L\gsim L_*$).
Galaxies in crowded environments are normally  red,
while blue galaxies reside mainly in the field
(\citealp{kauffmann_etal03b,blanton_etal05}; and numerous references
therein).
Deep surveys like COMBO-17 and DEEP-2 show that bimodality is
already well established at $z\sim 1$
\citep{bell_etal04,faber_etal06}, and even at $z \ga 3$ there
appears to be a distinction between actively star-forming galaxies
and passive systems dominated by redder stellar populations
(e.g., \citealp{vandokkum_etal06}).
Here we investigate the physical origin of this bimodality
in the context of modern galaxy formation theory, comparing
the results of two different modelling methods.

Cosmological hydrodynamic simulations (see
\citealp{frenk_etal99} for a review),
which integrate the equations of motions
for the baryons alongside those for the dark matter,
and hybrid simulations, which combine an N-body treatment of the dark matter
with semi-analytic modelling of the baryons
(e.g. \citealp{yoshida_etal02,hatton_etal03,helly_etal03,DeLuciaEtal04,croton_etal05,cattaneo_etal06}),
are the main methods for exploring the physics of
galaxy formation in a cosmological scenario while
retaining the spatial information on the galaxy distribution.
The fully numerical approach treats the dynamics and
the gas physics more self-consistently
because accretion and mergers come out naturally from the
physics of the simulations.
In the hybrid approach, the same phenomena are modelled through assumptions,
such as the hypothesis that the gas starts cooling from a spherical
distribution at the virial temperature of the halo or the dynamical friction formula to compute merging rates.
However, hydrodynamic simulations are expensive in CPU time.
That limits the resolution mass and the size of the computational box, the latter resulting in poor sampling of rare objects.
It also means that, for the same CPU time, hybrid methods allow one
to explore a  wider set of physical assumptions, such as different feedback scenarios.
In this paper, we compare results from a smoothed particle hydrodynamics (SPH)
simulation \citep{katz_etal96} to those of the
GalICS hybrid model (Galaxies In Cosmological Simulations; \citealp{hatton_etal03}).

This comparison tests the consistency between the gas dynamical treatment
of the SPH simulation and the idealised model incorporated in \galics,
since we run GalICS with merger trees extracted from the SPH simulation itself.
Two groups, centred
in Durham \citep{benson_etal01,helly_etal03} and Munich \citep{yoshida_etal02},
have already performed similar checks, verifying that the two
semi-analytic models investigated \citep{cole_etal00,kauffmann_etal99}
predict galaxy masses and luminosity functions in reasonable agreement
with cosmological SPH simulations run with the HYDRA \citep{pearce_etal01} and GADGET
\citep{springel_etal01} codes, when the physical assumptions are similar.
One motivation for our study was simply to conduct the same test for \galics.
However, we also focus on issues related to the galaxy bimodality, such as
the distinction between central and satellite galaxies and the possible
connection to the distinction between ``cold'' and ``hot'' gas accretion
\citep{katz_etal03,birnboim_dekel03,keres_etal05,dekel_birnboim06}.
To illuminate these issues, we consider two versions of the \galics\ model:
a ``stripped-down'' version with no feedback,
designed to approximately match the
physics of the SPH simulation, and the full version described by
\citet{hatton_etal03} and used in
\citet{blaizot_etal04}, \citet{lanzoni_etal05} and \citet{cattaneo_etal05},
which contains supernova and AGN feedback prescriptions designed to reproduce
the blue luminosity function of galaxies.\footnote{As we discuss later,
a newer version of \galics,
with an improved treatment of AGN and supernova feedback, is now available
\citep{cattaneo_etal06}.
The new version can fit the joint distribution of galaxy magnitudes and colours
at $z\sim 0$ and the luminosity function of Lyman-break galaxies
at $z\sim 3$ to
unprecedented accuracy. However, we carried out all of our analyses prior
to the development of this new version, which was partly motivated by
ideas discussed here.  The old version is adequate
for the purposes of this paper. We also wanted to provide tests of the
\galics\ version used for most of its results that have been published
to date.}
We show that the SPH simulation and the no-feedback \galics\ model produce
a red sequence of satellite galaxies in massive haloes,
but predict, contrary to observations,
that the central galaxies of massive haloes lie on the blue sequence.
The strong supernova feedback
incorporated in the full \galics\ model reduces galaxy baryonic
masses to a level consistent with observations. The combination
of supernova and (more importantly) AGN feedback shuts off gas
accretion in massive central galaxies, turning them red.

Section~2 describes
the astrophysical and computational assumptions used in
the SPH simulation together with the algorithms used
to identify galaxies and haloes, to attribute galaxies to haloes,
and to construct merger trees by linking haloes identified at different time-steps.
Section~3 summarises the \galics\ implementation used here, in particular
its assumptions about gas accretion, star formation and feedback.
In Section~4
we begin our comparison of the two methods
with a global view of the model galaxy populations. We investigate
the mass function, its dependence on redshift and environment, and
the relation between galaxies and dark matter haloes.
We then move into more specific aspects such as the growth mechanism
(cold flows versus hot flows; Section~5), the gas content, and the presence (or absence)
of a bimodality in the distributions of gas content and
star formation time-scales (Section~6).
In Section~7, we discuss how this work contributes to our understanding of
galaxy formation and of the processes that are at the basis of the observed galaxy bimodality.
Section~8 summarises our results.

\section{The SPH simulation}

\subsection{Cosmological model and simulation parameters}

The cosmology used for this study is a flat $\Lambda$CDM universe
(inflationary cold dark matter with a cosmological constant) with
$h\equiv H_0/(100{\rm\,km\,s^{-1}\,Mpc^{-1}})=0.65$,
$\Omega_m=0.4$, $\Omega_{\Lambda}=0.6$,
$\Omega_b=0.02h^{-2}=0.0473$, inflationary spectral index
$n=0.93$, and power spectrum normalisation $\sigma_8=0.8$.
We have used this combination of parameters in a series
of simulations over the course of a number of years
(beginning with \citealp{dave_etal99}), systematically varying
numerical resolution and simulation volume.
The values of $h$, $\Omega_b$, $n$, and $\sigma_8$ are similar
to those inferred from joint analyses of cosmic microwave
background (CMB) measurements from the WMAP satellite and
galaxy power spectrum measurements from the 2dF Galaxy Redshift
Survey and the SDSS (e.g., \citealp{spergel_etal03,tegmark_etal04,sanchez_etal06}).
These analyses generally favour a lower $\Omega_m$ value, in the
range $\sim 0.23-0.3$, but the qualitative features of galaxy
formation are not sensitive to the precise value of $\Omega_m$
(a point we have checked by running a comparable simulation with
$\Omega_m=0.3$).  Since we adopt identical parameters for the SPH and
\galics\ modelling, this simulation should be entirely adequate
for our purposes in this paper, despite its slightly outdated
parameter values.

The simulation volume is a comoving periodic cube
$22.22h^{-1} = 34.19$ Mpc on a side, modelled
using $128^3$ dark matter particles and $128^3$ gas particles.
 Gravitational forces are softened with a cubic spline kernel of
comoving radius $7.7\,$kpc, approximately equivalent to a Plummer
force softening of $\epsilon_{\rm grav} = 5.4\,$kpc,
while hydrodynamic pressure forces are calculated over 32 neighbours.
The baryonic mass threshold for resolved galaxies
 is $6.8 \times 10^9 M_\odot$, the mass
of 64 gas particles, and there are 1120 galaxies in the box
above this threshold at $z=0$.
Further details of the simulation can be found in %\citet{murali_etal02} and
\citet{keres_etal05}.

\subsection{The SPH code}

The simulation was performed with the parallel version of TreeSPH
(\citealt{hernquist_katz89}; \citealt{katz_etal96}; \citealt{dave_etal97}).
This code combines smoothed
particle hydrodynamics (SPH; \citealt{lucy77,gingold_mohaghan77})
with a hierarchical tree algorithm for computing gravitational
forces \citep{barnes_hut86,hernquist87}.
TreeSPH is a completely Lagrangian code, adaptive both in space and in time.
It contains three kinds of particles, representing
dark matter, gas and stars.
The dark matter and the stars are only subject to gravity, while the gas is also subject to pressure gradients and shocks.
The gas experiences adiabatic heating and cooling, shock heating,
inverse Compton cooling off the microwave background and radiative
cooling via free-free emission, collisional ionisation, collisional recombination,
and collisionally excited line cooling.
The cooling rate is calculated for primordial chemical abundances and, since there is
only atomic cooling, the gas cannot cool radiatively below $T \sim 10^4\,$K.
A uniform photoionising UV background
heats low temperature gas and suppresses cooling processes
involving neutral atoms at low gas densities \citep{hardt_madau96}.

The simulation also includes a simple treatment of star formation and
its associated supernova feedback.
Gas above a threshold hydrogen number density
$n_H=0.1{\rm cm}^{-3}$ turns
into stars on a time-scale set by the dynamical time or the cooling
time, whichever is longer.  Additional star formation conditions are that the gas
is Jeans unstable, part of a converging flow
and above the virial overdensity, but gas that
satisfies the density criterion usually also satisfies the other three.
Gas reaches this high density only
after cooling to $T \sim 10^4\,$K. The
subsequent molecular and metal-line cooling to lower temperatures
is implicitly included as part of the star formation process.  This star
formation prescription leads to a relation with the gas surface density
similar to a Schmidt law (\citealp{schmidt59,kennicutt98}; see
discussions by \citealp{katz_etal96} and \citealp{stinson_etal06}).

Stars more massive than $8M_\odot$ explode as supernovae.
For a \citet{miller_scalo79} initial mass function that means
$7.35\times 10^{-3}$ supernovae per solar mass of formed stars.
Each supernova releases $10^{51}\,$erg,
gradually distributed as heat to the gas particles nearby
with an exponential time decay of $2\times 10^7\,$yr.
The surrounding medium is usually dense, so the deposited energy is
typically radiated away before it can drive a galactic scale wind.
For this reason, we consider this a ``minimal'' feedback
algorithm, even though the full amount of expected supernova
energy is incorporated.

%JB
%\subsection{Identification of galaxies and dark matter haloes}
\subsection{Identification of galaxies}
%end JB
Cosmological simulations that incorporate cooling and star formation produce
dense groups of baryonic particles with the sizes and masses of
observed galaxies \citep{katz92,evrard_etal94}.
We identify these aggregations with the group finding algorithm
`Spline Kernel Interpolative DENMAX' (SKID)\footnote{We use the
implementation of J.~Stadel and T.~Quinn, available at
\tt http://www-hpcc.astro.washington.edu/tools/skid.html}
\citep{gelb_bertschinger94,katz_etal96}.
This algorithm involves five basic steps: (1)
determining the smoothed baryonic density field; (2) moving baryonic particles
towards higher density along the initial gradient of the baryonic
density field; (3) defining the initial group as a set of particles that
aggregate at a particular density peak; (4) linking together initial groups
that are very close together; (5) removing group particles with positive binding energy
relative to the group's centre of mass.
We apply SKID to the population of all star particles and to the gas
particles that have temperatures $T < 3 \times 10^4\,$K
and overdensities $\rho_{\rm gas}/{\bar\rho}_{\rm gas} > 10^3$.
Henceforth, we refer to the aggregations of stars and cold gas that
SKID identifies simply as ``galaxies.''
Tests on simulations with varying mass resolution show that
the simulated galaxy population becomes substantially incomplete
below a baryonic mass corresponding to $\sim 64 m_{\rm SPH}$
but is fairly robust above this limit (see, e.g., \citealt{murali_etal02}).
We therefore adopt $64 m_{\rm SPH}$ ($6.8 \times 10^9 M_\odot$) as our resolution threshold and ignore
lower mass galaxies in our analysis.
Because of our high overdensity threshold for star formation,
essentially all star formation in the simulation takes place
in galaxies, though some of these are below the resolution limit,
and some stars are tidally stripped from galaxies
during dynamical interactions.

% begin JB
% i moved the halo identification discussion to section 3.1.
%end JB

The minimum galaxy mass of $64 m_{\rm SPH}$ corresponds
to a minimum host halo mass of $\sim (\Omega_m/\Omega_b)64m_{\rm SPH}$,
since the fraction of cold gas in a halo never substantially exceeds the universal
baryon fraction.
The corresponding minimum virial temperature for our simulation is
$ \sim 80,000 (1+z)\,$K for the simulation considered in this article,
where the $(1+z)$ factor
arises from the increasing physical density at higher $z$ for
fixed virial overdensity.
The redshift dependence of the virial overdensity makes
the `$80,000\,$K' factor
slightly lower at high redshift ($\sim 75,000\,$K)
and slightly higher at low redshift ($88,000\,$K at $z=0$).
Any resolved galaxy in the simulation
resides in a dark matter halo with virial temperature higher than this
minimum temperature.

\section{The GalICS hybrid model}
%begin JB
The semi-analytic model that we use in this paper is exactly the one
described in \citet{hatton_etal03}. In this section, we therefore only
briefly summarise those of its features which are relevant to the present
study.

%end JB

\subsection{Dark matter haloes and their merger trees}
% begin JB : the next paragraph is re-written from what was before in Sec. 2.3.

We identify and characterise dark matter haloes the same way as in
\citet{hatton_etal03}. We use a friends-of-friends (FOF) algorithm
\citep{davis_etal85} to detect groups of particles with overdensity
$\sim 200$ times the mean density (using a fixed linking-length
parameter $b=0.2$). For each group with more than 20 particles, we
compute the total kinetic and potential energies, and only keep in the
halo catalogue the groups that are bound. The halo mass we refer to in
the rest of the paper is simply $N_p\times m_p$, where $N_p$ is the
number of particles in the group, and $m_p$ the mass of a dark matter
particle. The minimum halo mass is thus $\sim 1.8 \times 10^{10}
M_{\odot}$. We set the centres of groups to be at the positions of
their centres of mass.

The information about an individual halo measured from the SPH outputs
and passed to the semi-analytic model is contained in three
parameters: the virial mass, the spin parameter, and the virial
radius. The virial mass and the virial radius are measured in two steps. First,
we compute the inertia tensor of each halo. From this we determine the
three main axes of an ellipsoid that fits the mass distribution and is
centred on the centre of mass. Then we shrink this ellipsoid until
the halo particles that it contains satisfy the virial theorem.
The mass of the particles in the ellipsoid at this point is the virial mass.
The virial radius is the radius of a sphere with the same volume as the virial ellipsoid.

We compute merger trees by linking haloes identified in each SPH
snapshot with their progenitors in the previous one. All predecessors
from which a halo has inherited one or more particles are counted as
progenitors.

% end JB

\subsection{The cooling scheme}
%JB : modified significantly the content of this section ...

Newly identified haloes receive a gas mass determined by the universal
baryonic fraction $\Omega_b/\Omega_m$.
%{\bf Is my addition of ``above the resolution threshold'' correct? If not, specify threshold.}
%AC: correct
% JB : no ! All haloes receive gas in the SAM. That is, all groups of > 20 particles which are bound.
As in other semi-analytic models, all baryons start in a
zero-metallicity hot phase, shock-heated to the virial temperature.
We assume that the the hot gas density profile is described by a singular isothermal
sphere truncated outwards at the virial radius and inwards at a core
radius of 0.1 kpc.

The cooling time $t_{\rm cool}(r)$ is calculated from the hot gas
density distribution with the metal-dependent cooling function from
\citet{sutherland_dopita93}. There are two differences with the SPH
calculation here. First, the semi-analytic calculation does {\it not} include a
photoionising background, although photoionisation effects on galaxy
formation are small in the halo mass range resolved by the simulation
\citep{quinn_etal96,thoul_weinberg96}. Second, \galics{} does take
into account the enhancement of cooling by metals, contrary to the SPH
code, which assumes zero-metallicity cooling. However, in the
no-feedback \galics{} model, metals are never expelled from galaxies,
and the intra-cluster medium always remains of pristine
composition. Hence, the stripped down version of \galics{} reproduces a
cooling scheme that is in practice very similar and hence directly comparable
to that of the SPH simulation.

In small haloes, the cooling time is very short and all the gas gets
cold almost immediately.  Without thermal pressure support, the gas
collapses to the centre of the dark matter halo in free fall until the
centrifugal force determined by the conservation of angular momentum
balances the gravitational force and the gas settles into a disc.  In
this regime, the gas accreted by the disc in the time $\Delta t$
between two snapshots is the gas that is found within a radius $r_{\rm
infall}=\Delta t/v_{\rm c}$ from the halo centre (here $v_{\rm c}$ is
the halo's circular velocity).  In larger haloes the cooling time is
longer, and gas slowly flows to the centre as it cools. In this case
it is the cooling time that determines the rate at which the disc can
grow. The gas that can cool in the time $\Delta t$ between two
snapshots is the gas within a sphere of radius $r_{\rm cool}$ such
that $t_{\rm cool}(r_{\rm cool})=\Delta t$.  The $\Delta t$ interval
between redshift outputs is typically $\sim 0.3\,$Gyr ($\sim
0.1H^{-1}(z)$) at high redshift and $\sim 0.5\,$Gyr ($\sim 0.04
H^{-1}(z)$) at low redshift.

\citet{birnboim_dekel03}, \citet{katz_etal03}, and
\citet{keres_etal05} emphasise the distinction between ``hot'' gas
accretion, in which shocks heat gas to the halo virial temperature
near the virial radius, and ``cold'' accretion in which cold,
unshocked gas streams penetrate far inside the virial radius.  To
examine this distinction in \galics, we identify the cooling-limited
and infall-limited regimes with the ``hot'' and ``cold'' accretion
modes, respectively (see Section~5 below).  The boundary between these
regimes depends on $\Delta t$, and hence on our particular choice of the
redshift outputs used for constructing halo merger trees.  The physical
identification with the two accretion modes is therefore only
qualitative.  \citet{keres_etal05} and \citet{croton_etal05} discuss
the relation between the accretion modes and the cooling criteria of
semi-analytic models in greater detail.

\subsection{Star formation}

Star formation is activated when the gas surface density is
$\Sigma_{\rm gas}>20\,m_{\rm p}{\rm\,cm}^{-2}$, where $m_{\rm p}$ is the proton mass.
Disc gas forms stars at a rate $\dot{M}_*=M_{\rm cold}/t_*$, where
$M_{\rm cold}$ is the mass of the gas available to form stars
and $t_*=50\,t_{\rm dyn}$ (\citealp{guiderdoni_etal98};
the dynamical time $t_{\rm dyn}$ is the time in which
a star completes half a circular orbit at the disc half mass radius).
The disc is assumed to have an exponential profile and its radius is
determined from conservation of angular momentum.

Mergers and disc instabilities form bulges by transferring gas and stars from
the disc to the spheroidal component \citep{hatton_etal03}.
In this paper, we are not directly concerned with morphologies.
However, these morphological transformations can
temporarily accelerate the conversion of gas into stars,
and the bulge mass determines the effectiveness of AGN
feedback as described below.

\subsection{Feedback}

The formation of young stars is rapidly followed by supernova explosions.
The energy released by supernovae is used to reheat the cold gas and remove
it from the galaxy.
The version of the GalICS model used to produce this paper follows
\citet{hatton_etal03} and uses a feedback prescription in which supernova
outflows entrain more gas in massive galaxies owing to the larger porosity of
the interstellar medium \citep{silk01,silk03}.
This model leads to the very simple result that the outflow rate is comparable
to the star formation rate, independently of the depth of the potential well.

%{\bf Are there any other ingredients to the supernova feedback recipe,
%such as a threshold in SFR or SFR surface density?
%AC: See the addition at the beginning of the star formation subsection
%Are the SFR and outflow rate exactly equal?} %AC: no

Supernova feedback is not enough to compensate
the overcooling of gas in massive
haloes, which would produce too many massive galaxies.
Following ideas that have been around in the recent literature, we
postulate the existence of a second feedback mechanism, associated with the
growth of supermassive massive black holes in early-type galaxies.
The simplest way to model black hole feedback is to assume an energy
threshold.
When the total energy injected
by active galactic nuclei (AGN) in the intergalactic
medium (IGM) of a group or cluster of galaxies exceeds the threshold,
the hot gas is assumed to have
acquired so much entropy that it is no longer able to cool.
As black hole accretion releases an energy proportional
to the mass accretion rate,
and as the black hole mass is proportional to that of the host bulge,
this criterion translates into a requirement on the total mass of bulge stars
in the group or cluster. Following
\citet{hatton_etal03}, we assume that gas stops cooling
when $\Sigma M_{\rm bulge}>10^{11}M_\odot$, where the sum is
over all galaxies in the halo.

\begin{figure}
\noindent
\begin{minipage}{8.4cm}
  \centerline{\hbox{
      \psfig{figure=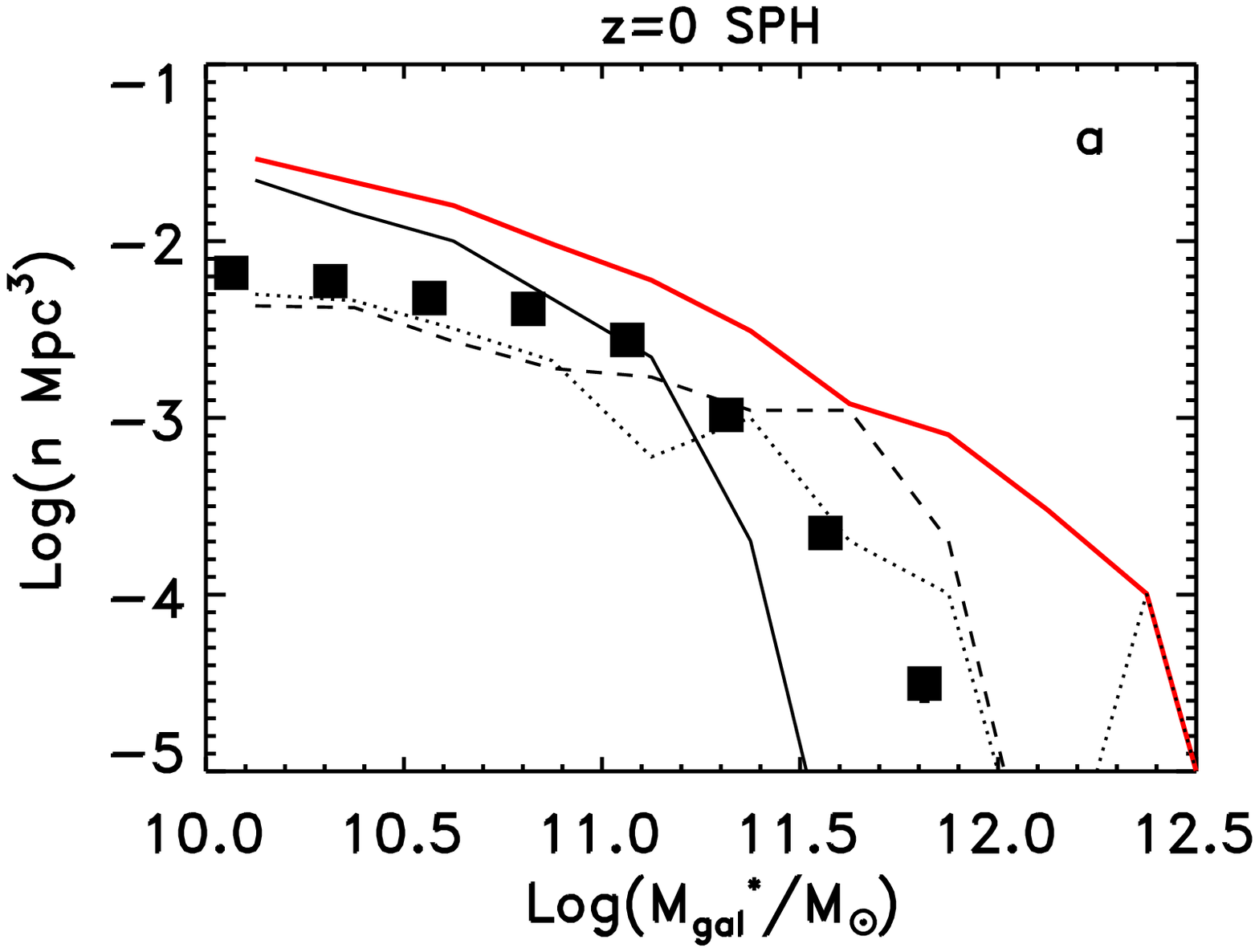,height=5.73cm,angle=0}
  }}
\end{minipage}\    \
%\hskip
\begin{minipage}{8.4cm}
 \centerline{\hbox{
  \psfig{figure=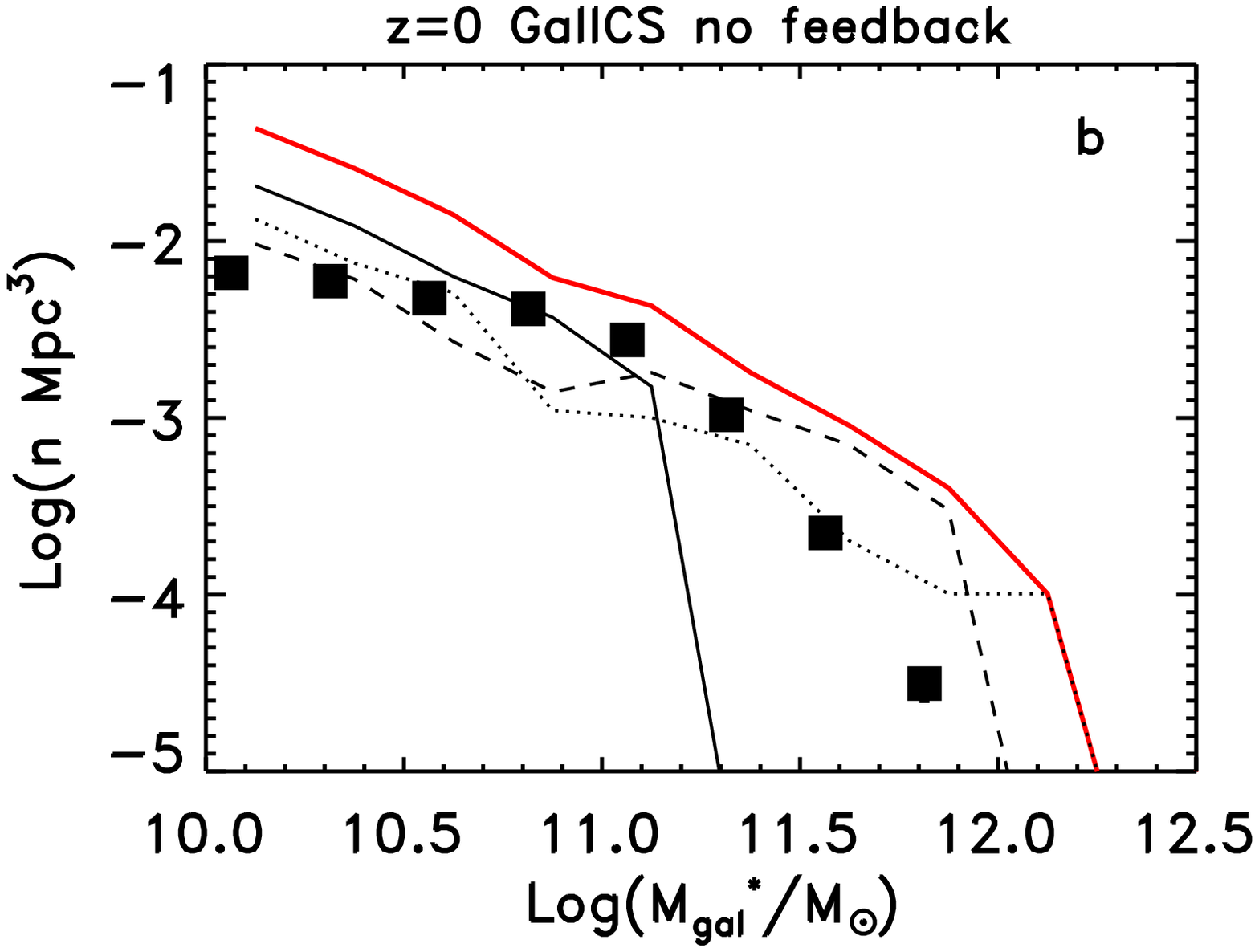,height=5.73cm,angle=0}
  }}
\end{minipage}\    \
%\hskip
\begin{minipage}{8.2cm}
\centerline{\hbox{
\psfig{figure=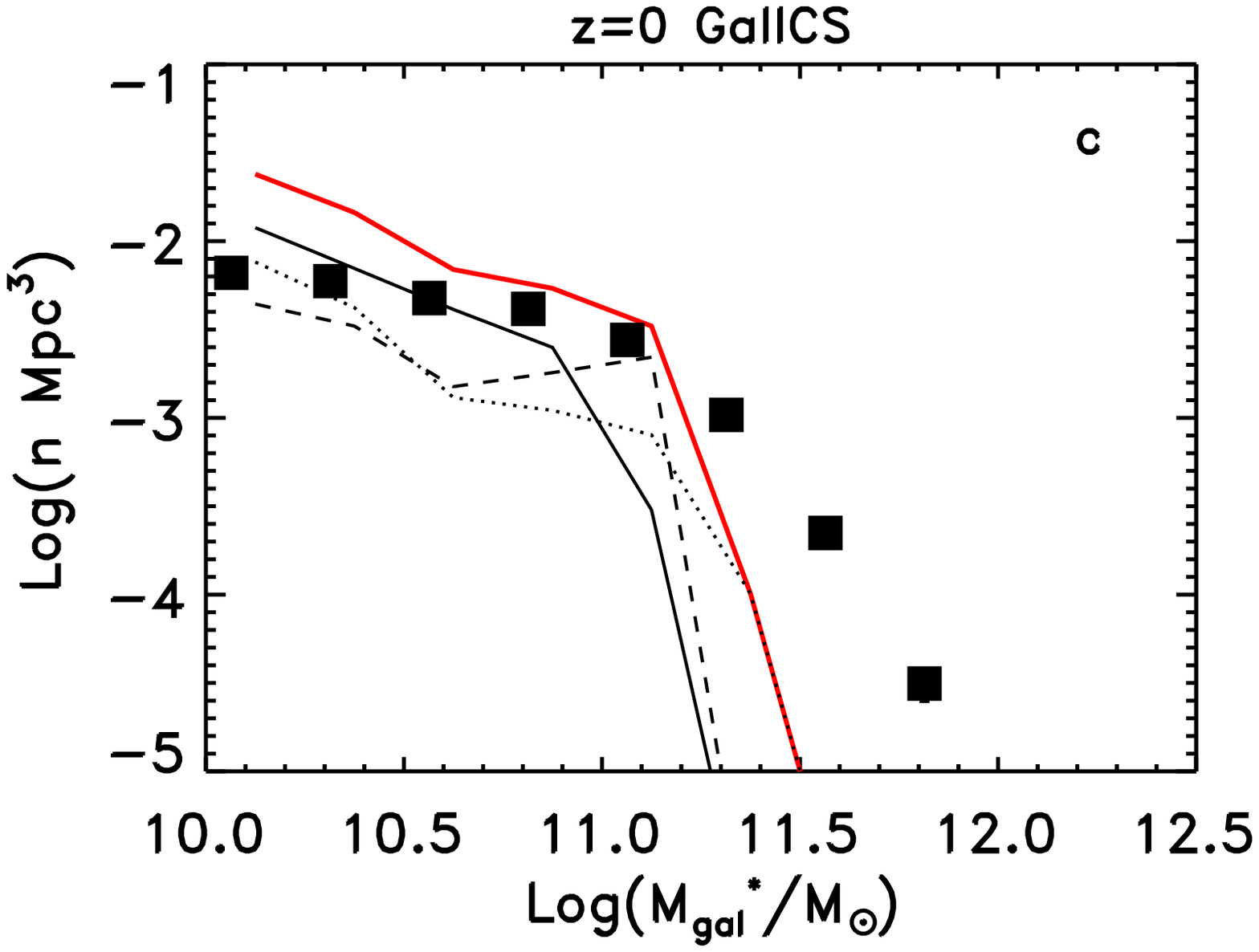,height=5.73cm,angle=0}
}}
\end{minipage}\    \
\caption{The galaxy mass functions computed with three different
  models (the SPH model, the no-feedback GalICS model,
  and the GalICS model with the feedback recipes of Section~3.4) compared with
  the observational estimate of \citet{bell_etal03}.
  In each of the
  three diagrams, the observational mass function is shown by the points
  and the theoretical one by the red line. The latter is
  decomposed into the mass functions of ``field'' galaxies (black solid line),
  ``group'' galaxies (dashed line) and ``cluster'' galaxies (dotted line),
  as described in the text.
}
\end{figure}

\section{Global view of the galaxy population}

We consider three models: the SPH model, the GalICS model without any type of
feedback, and the GalICS model with the feedback recipes described in
Section~3.4.
Because the feedback incorporated in the SPH simulation does not
generally drive galactic winds or suppress gas cooling in massive
haloes, it is the no-feedback \galics\ model that is most comparable
to the SPH simulation in its physical assumptions.

Figure~1 compares the galaxy baryonic
mass functions at $z=0$ in these three models.
(Here baryonic mass includes stars and the cold interstellar medium,
but it does not include hot gas in the galaxy or group halo.)
It also shows the separate contribution to the mass function of ``field''
galaxies
($M_{\rm halo}<3\times 10^{12}M_\odot$), ``group'' galaxies
($3\times 10^{12}M_\odot<M_{\rm halo}<10^{14}M_\odot$), and ``cluster''
galaxies ($M_{\rm halo}>10^{14}M_\odot$).
In fact, the $34.19\,{\rm Mpc}$ simulation cube contains only
a single ``cluster'' mass halo ($M=3\times 10^{14}M_\odot$), and the
most massive ``group'' haloes are $\sim 3\times 10^{13}M_\odot$.

The SPH and no-feedback \galics\ predictions agree remarkably well, both for
the total baryonic mass function and for the separate mass functions
in the three halo mass regimes.
Feedback in the full \galics\ model
reduces galaxy masses, dropping the low mass end of
the mass function and, most strikingly, producing sharp truncation
of the high ends of the mass functions in the group and cluster
haloes.

\begin{figure*}
\noindent
\begin{minipage}{8.6cm}
  \centerline{\hbox{
      \psfig{figure=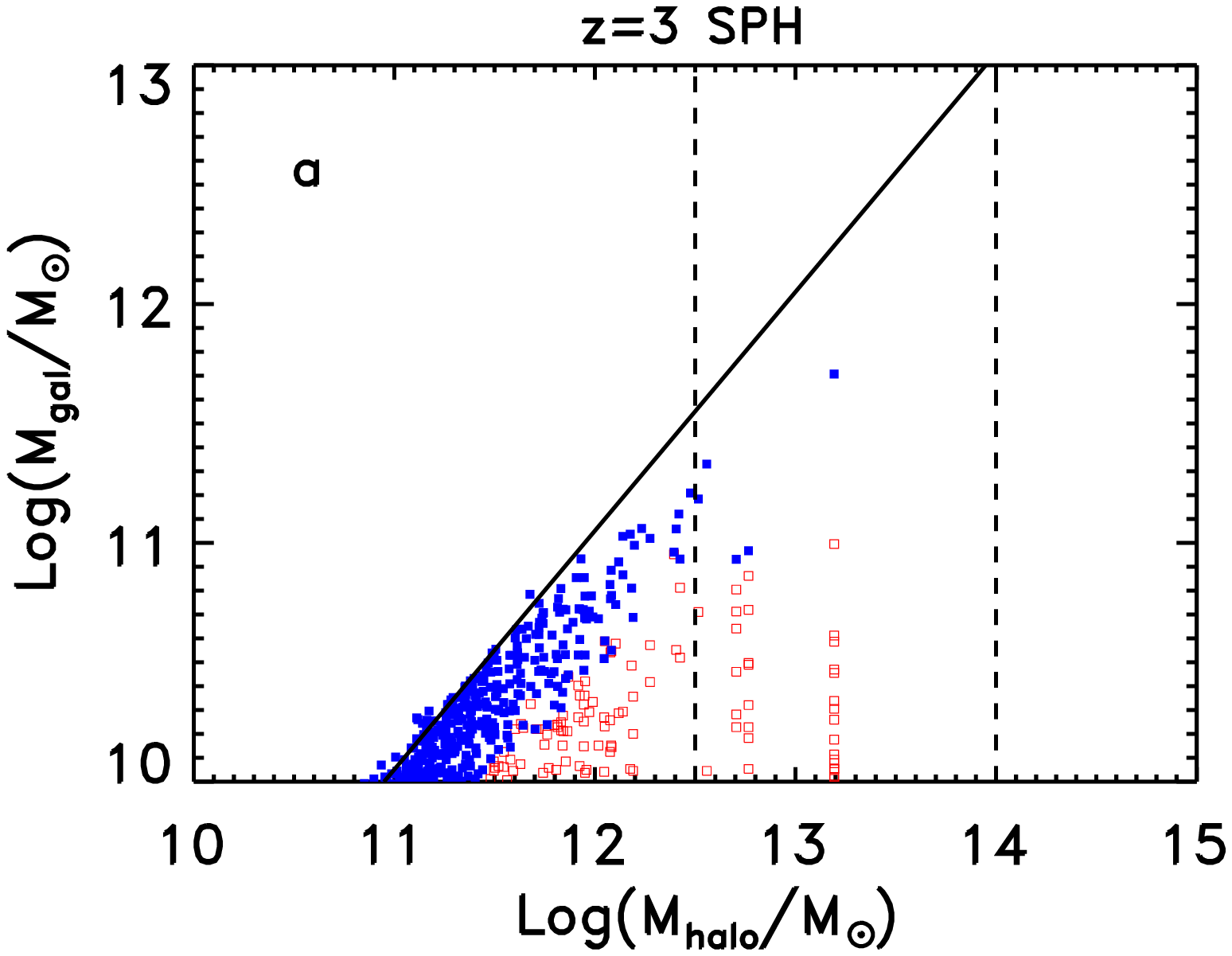,height=5.7cm,angle=0}
  }}
\end{minipage}\    \
%\hskip
\begin{minipage}{8.6cm}
  \centerline{\hbox{
      \psfig{figure=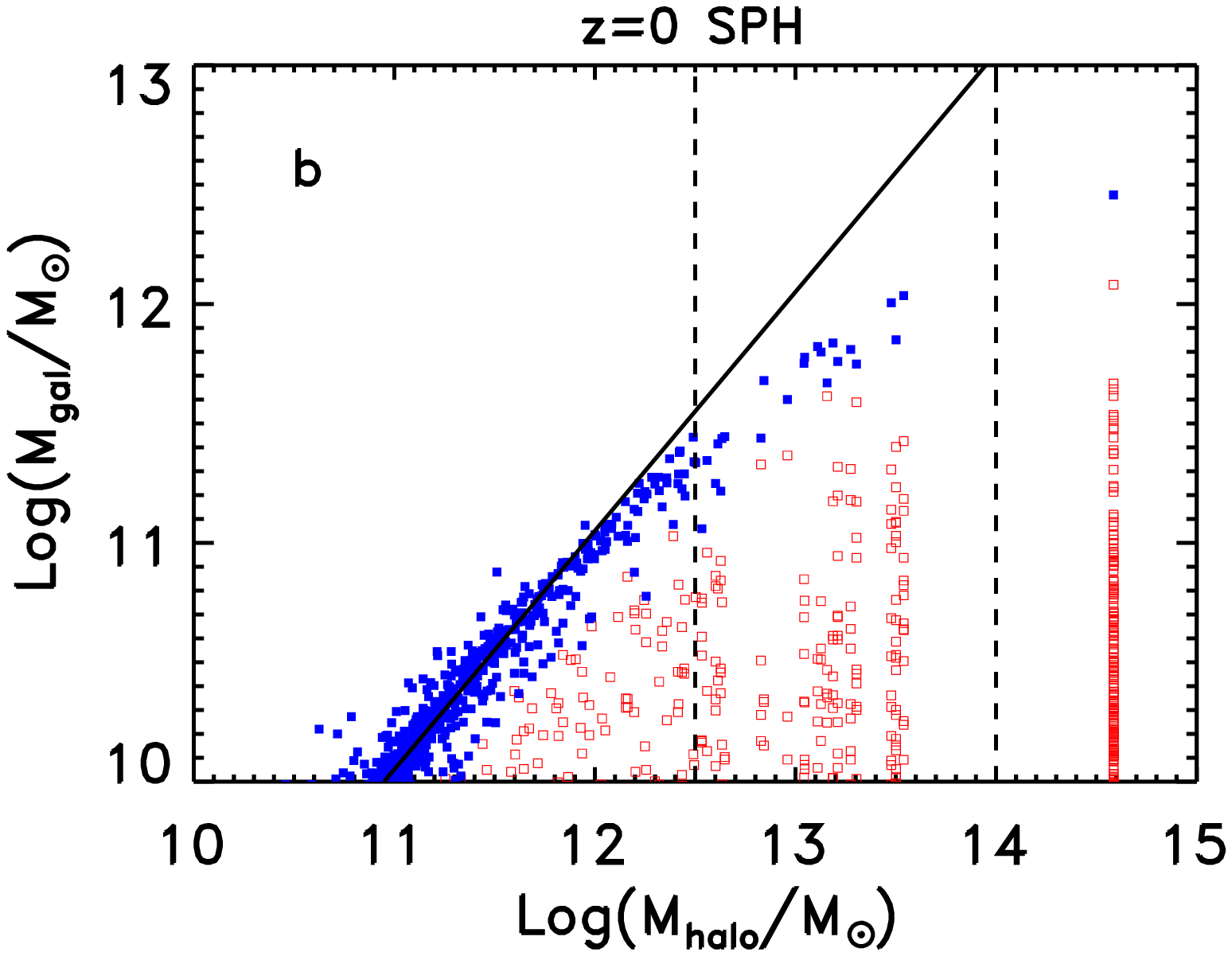,height=5.7cm,angle=0}
  }}
\end{minipage}\    \
%\hskip
\begin{minipage}{8.6cm}
  \centerline{\hbox{
      \psfig{figure=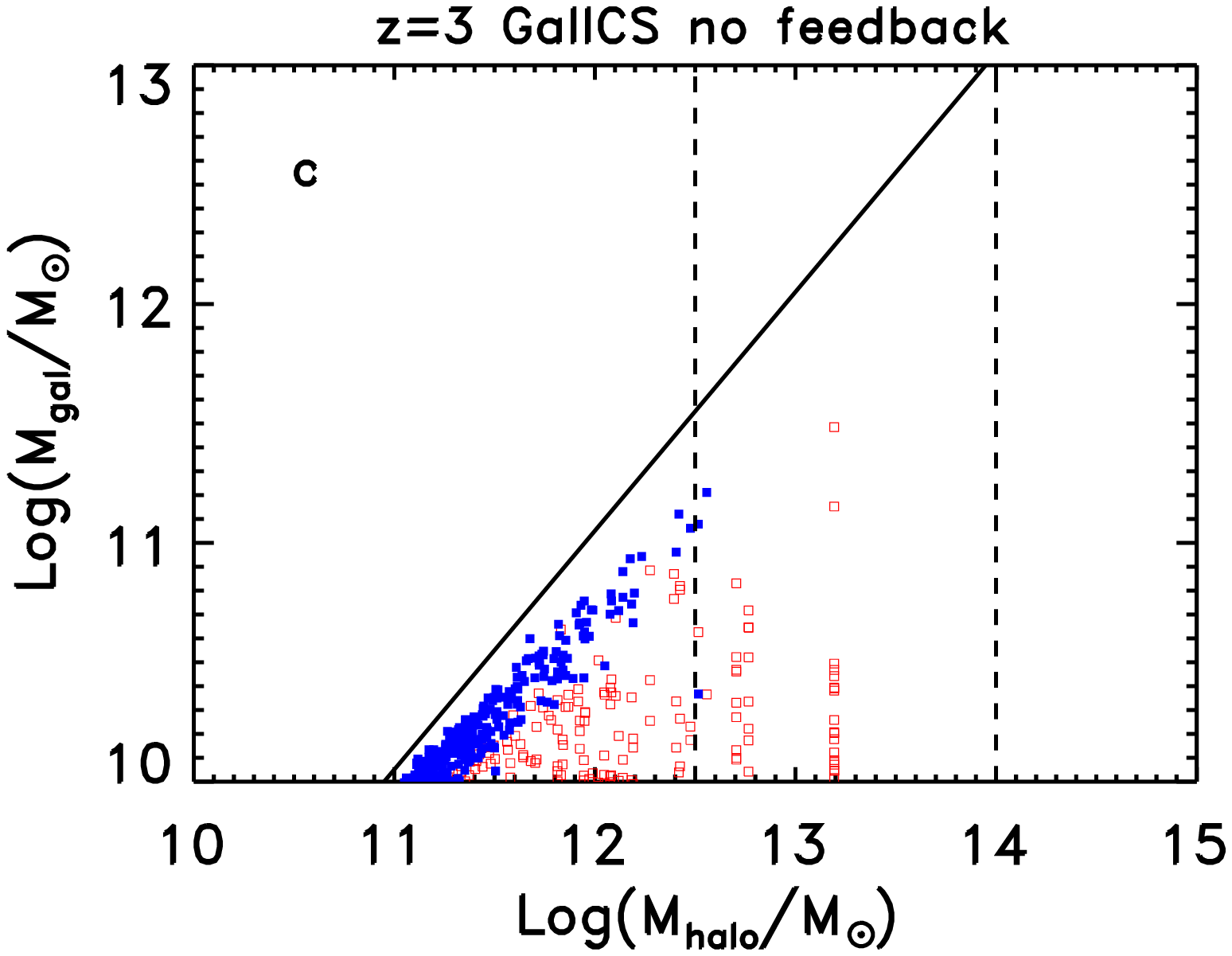,height=5.7cm,angle=0}
  }}
\end{minipage}\    \
\begin{minipage}{8.6cm}
  \centerline{\hbox{
      \psfig{figure=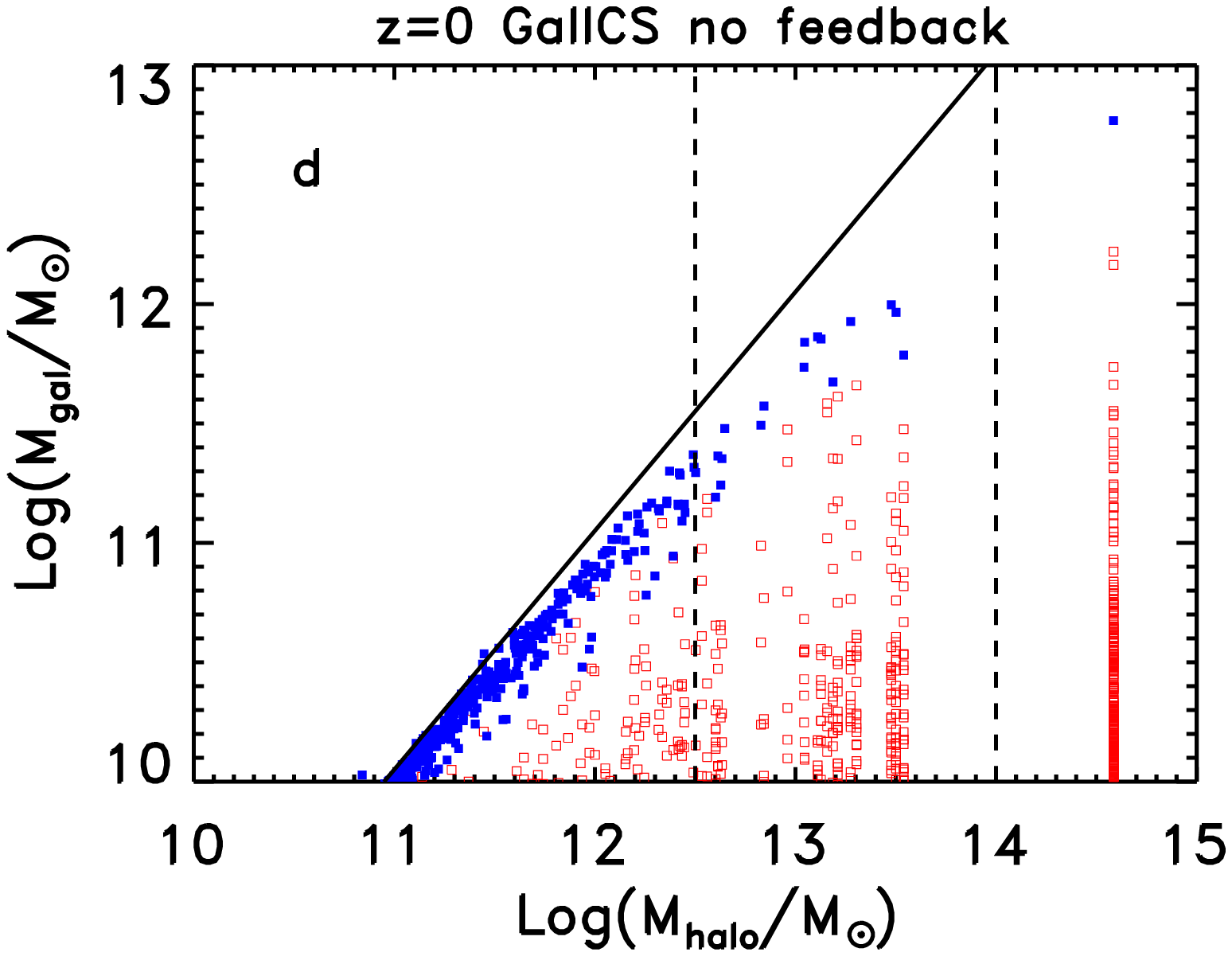,height=5.7cm,angle=0}
  }}
\end{minipage}\    \
%\hskip
\begin{minipage}{8.6cm}
  \centerline{\hbox{
      \psfig{figure=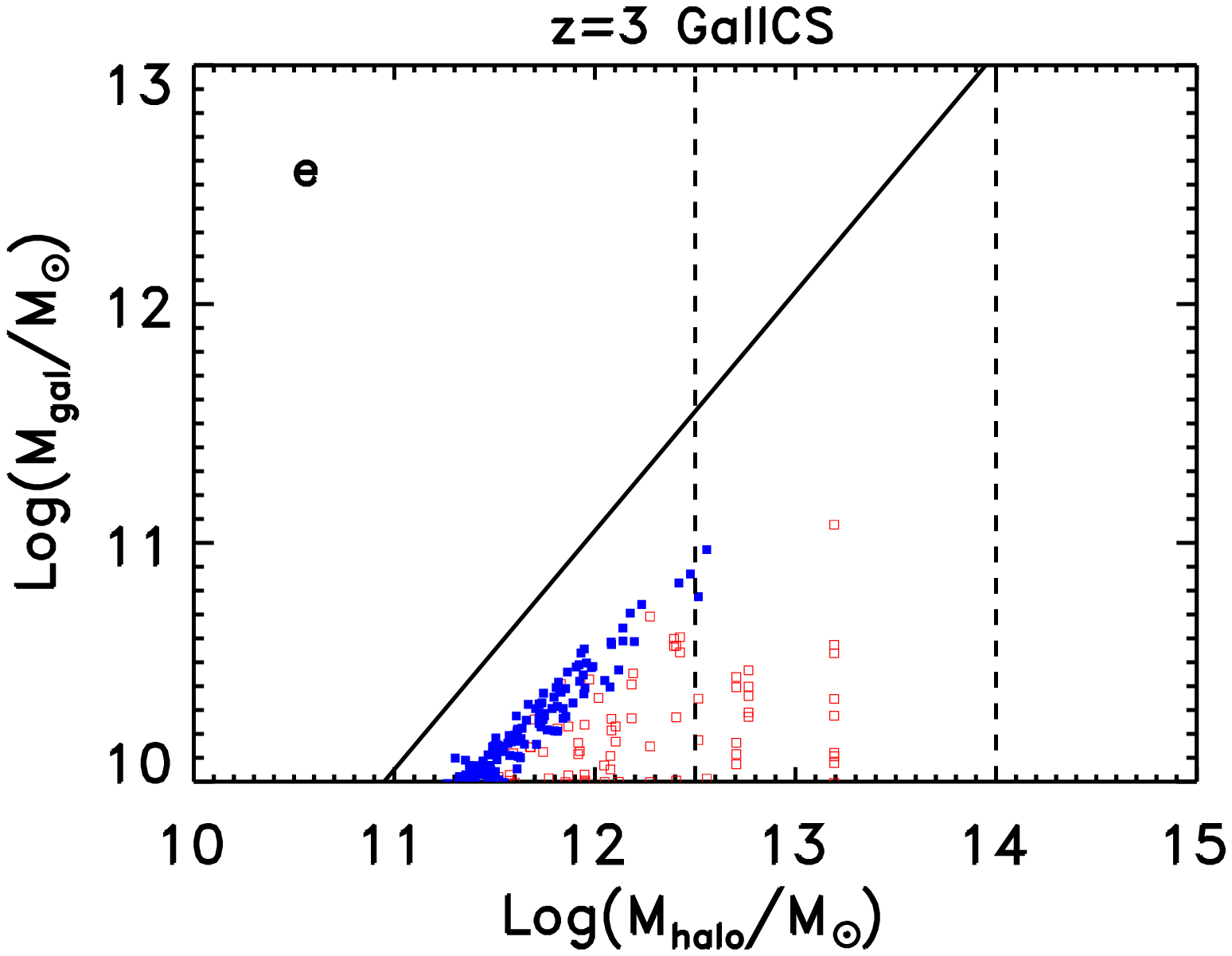,height=5.7cm,angle=0}
  }}
\end{minipage}\    \
%\hskip
\begin{minipage}{8.6cm}
  \centerline{\hbox{
      \psfig{figure=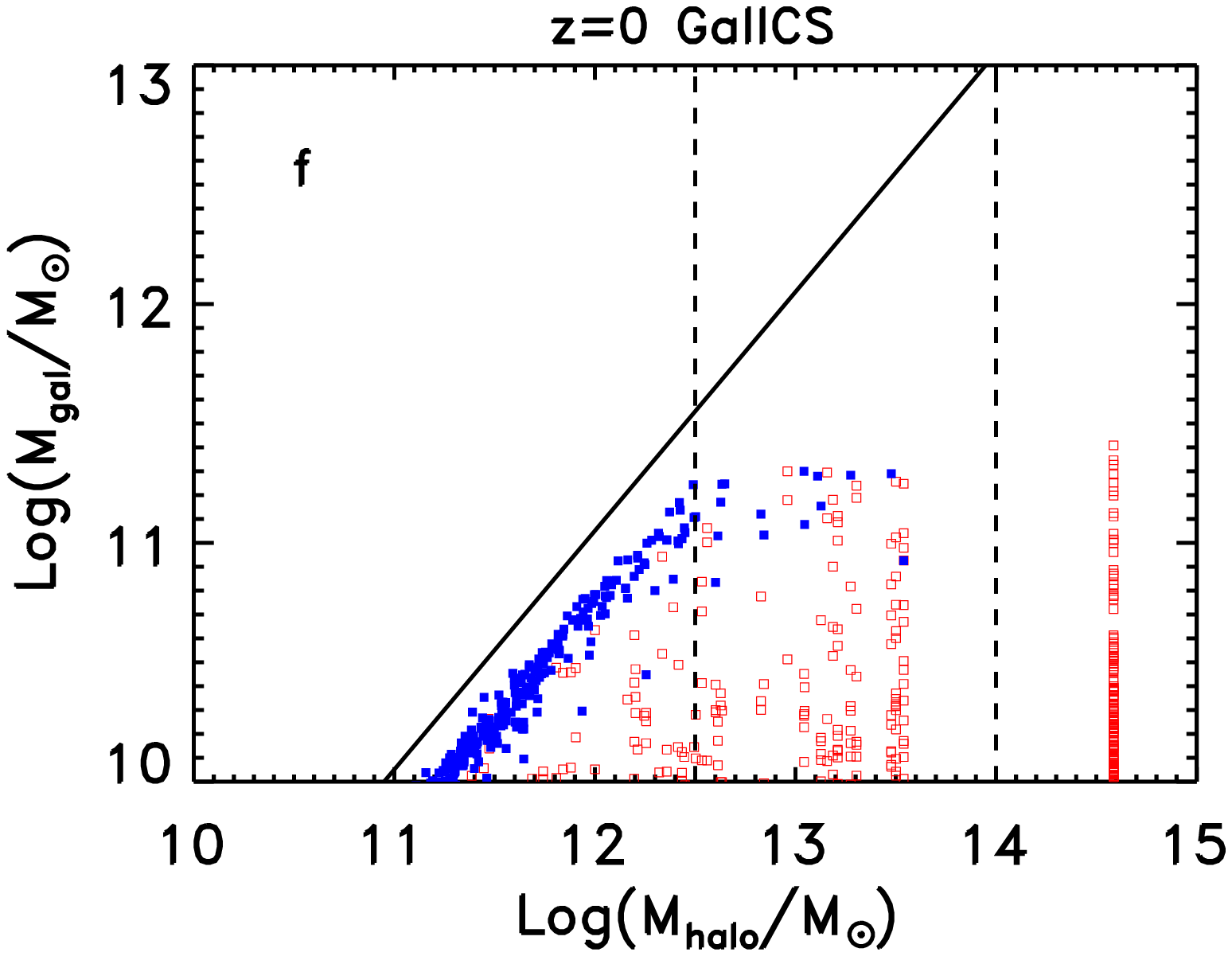,height=5.7cm,angle=0}
  }}
\end{minipage}\    \
\caption{The galaxy content of dark matter haloes in the SPH simulation and in
      GalICS. The blue points correspond to galaxies that are at the
      centre of their dark matter halo, while the red points are
      satellite galaxies. In the SPH simulation, we always define the
      most massive galaxy to be ``central.''
      In GalICS, a halo may have no central galaxy if it has
      had a recent merger, because dynamical friction takes time
      to make galaxies sink to the new centre of mass.}
\end{figure*}

Points show the observational estimate of the total baryonic mass
function from \citet{bell_etal03}, based on 2MASS and SDSS data.
Because of the limited simulation volume, we would not expect
perfect agreement with the data even for a model with exactly
correct input physics.  Nonetheless, the SPH and no-feedback
\galics\ predictions clearly exhibit the well known tendency
of models with efficient gas cooling and minimal feedback to
overproduce the baryonic mass function, a discrepancy that remains
with larger simulation volumes (see \citealp{murali_etal02}).
The full \galics\ model, by contrast, appears to underpredict
the high mass end of the mass function.  We believe that this is
a real discrepancy, arising because the \citet{hatton_etal03} feedback
prescriptions are tuned to match the $B$-band luminosity function,
{\it and} they produce massive galaxies that are too blue.
The newer version of \galics\ described by \citet{cattaneo_etal06}
shuts off gas accretion onto central galaxies of all high mass haloes
and thereby achieves a simultaneous match to the $u$- and $r$-band
luminosity functions.

\begin{figure*}
\noindent
\begin{minipage}{9.6cm}
  \centerline{\hbox{
      \psfig{figure=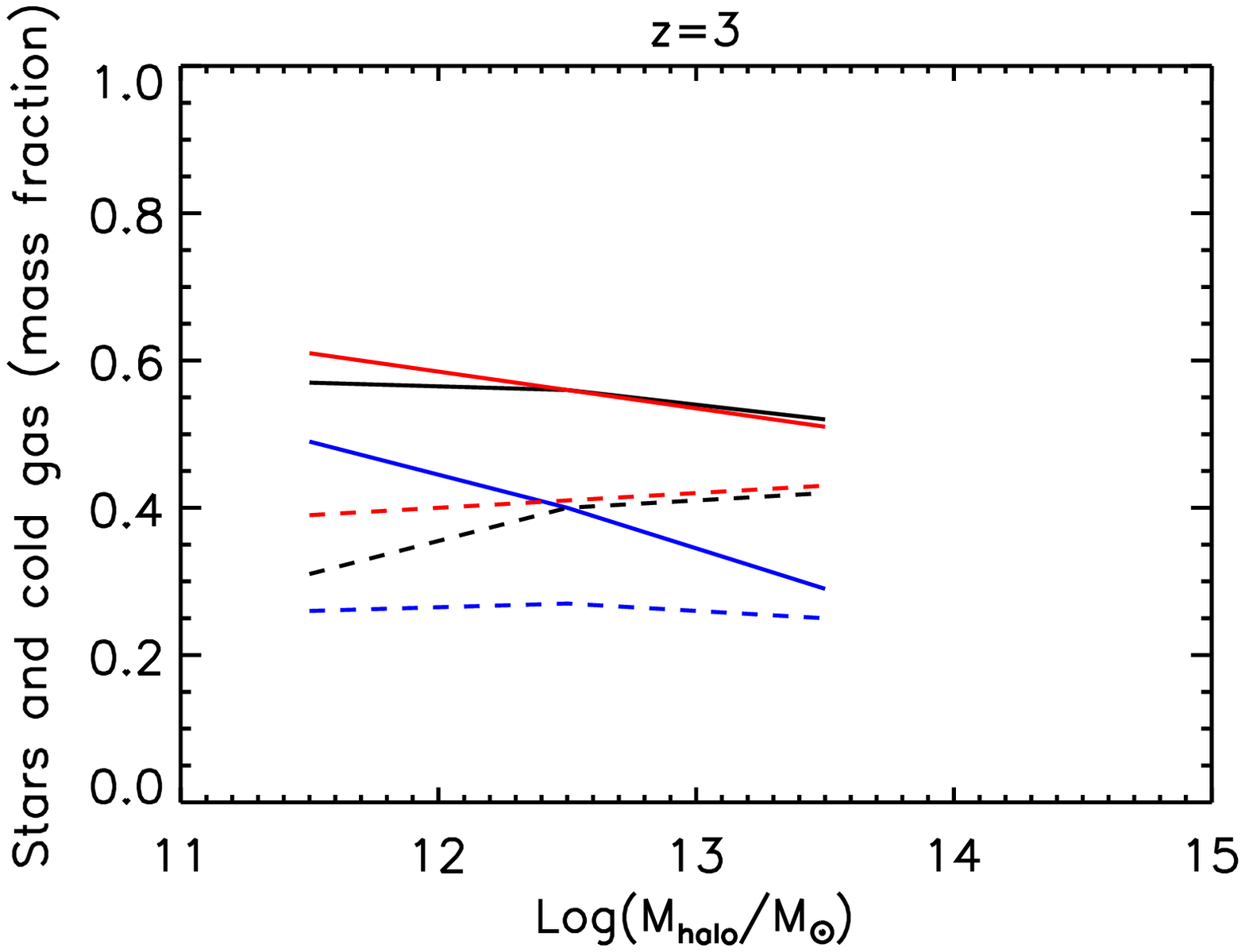,height=5.73cm,angle=0}
  }}
\end{minipage}\    \
%\hskip
\begin{minipage}{7.8cm}
  \centerline{\hbox{
      \psfig{figure=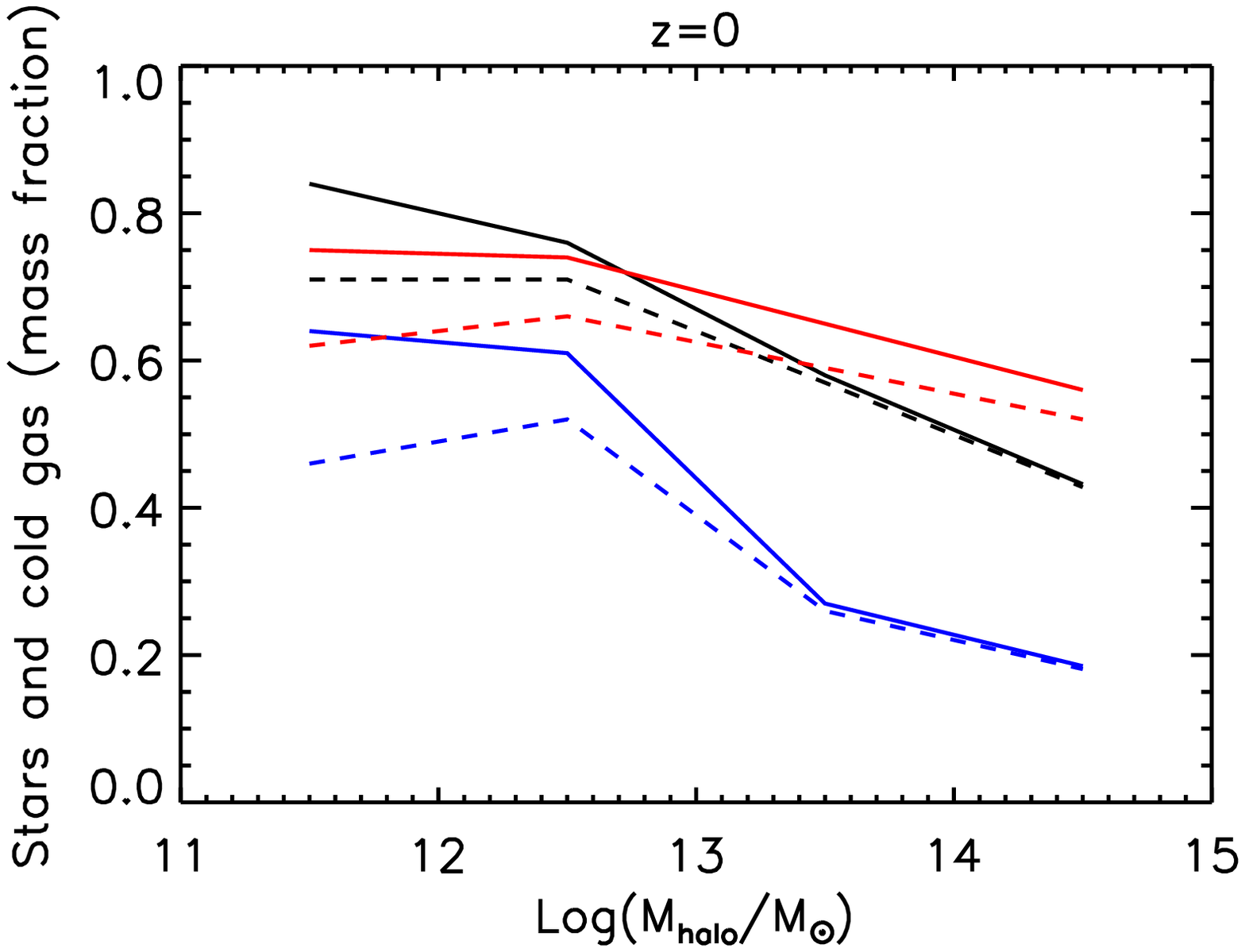,height=5.73cm,angle=0}
  }}
\end{minipage}\    \
\caption{The fraction of the total baryonic content of a halo in stars
      (dashed lines) and in cold gas plus stars (solid lines) as a
      function of halo mass.
      The results of the SPH simulations are printed in black.
      The results of the GalICS model without feedback are printed in red.
      The results of the standard GalICS model are printed in blue.
}
\end{figure*}
\begin{figure*}
\noindent
  \centerline{\hbox{
      \psfig{figure=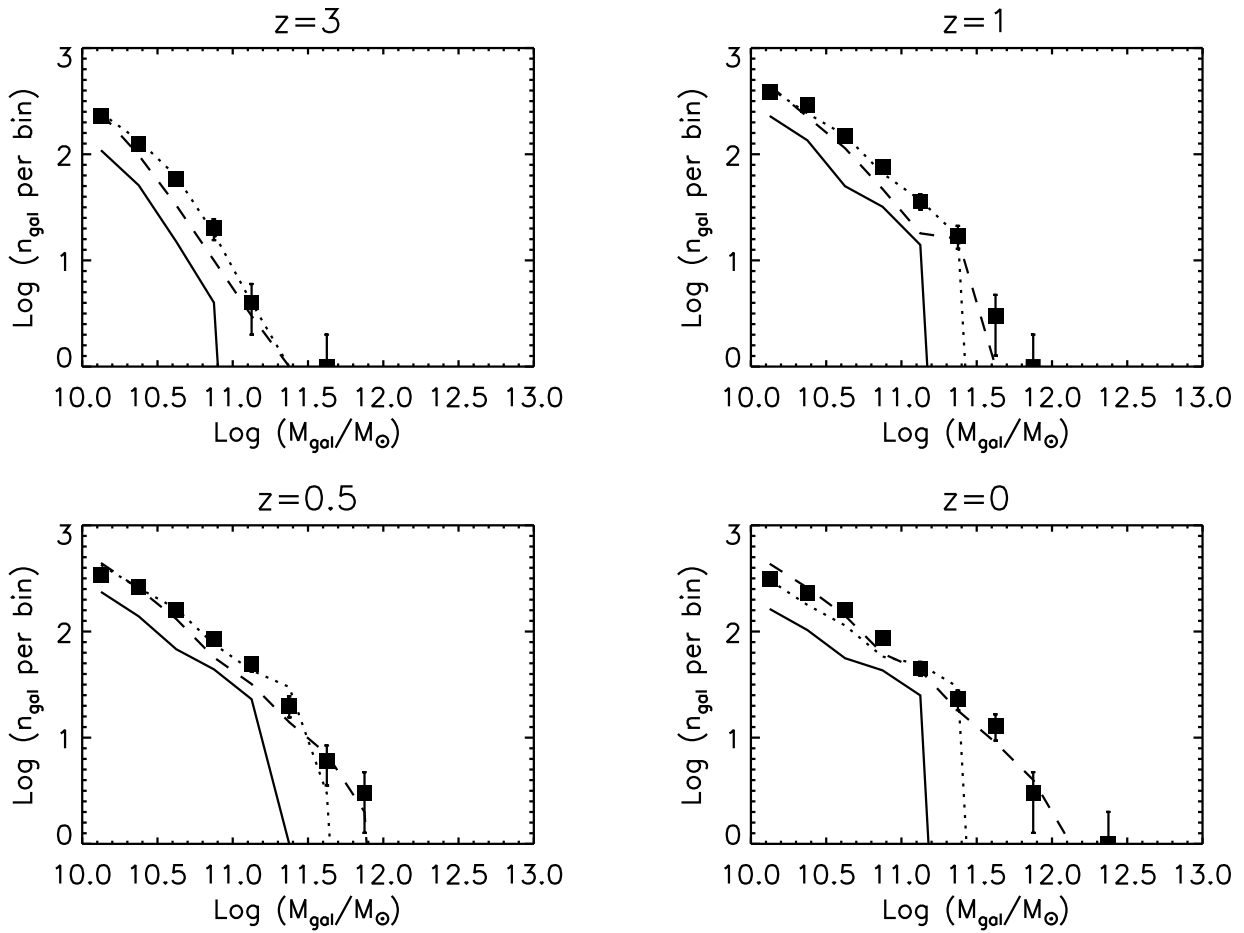,angle=0}
  }}
\caption{The mass function of galaxies in the SPH simulation (points)
      and in the GalICS model (lines) at four different redshifts.
      In each panel, the dashed line is GalICS without supernova or
      AGN feedback, the solid line is the standard GalICS model, and
      the dotted line is the standard GalICS model when all masses
      are multiplied by a factor of 2.1.
      The mass function is given as the number of galaxies
      per logarithmic bin of 0.25 dex. In contrast to
      Fig.~1, we have not divided the number of galaxies
      by the volume of the computational box, so that the numbers
      convey an idea of the statistics in
      each bin. At $M_{\rm gal}>3\times 10^{11}M_\odot$, the number of
      galaxies in the box is $\lsim 10$, and statistical effects
      (quantified by the Poisson error bars) are no longer negligible.}
\end{figure*}
\begin{figure*}
\noindent
\centerline{\hbox{
\psfig{figure=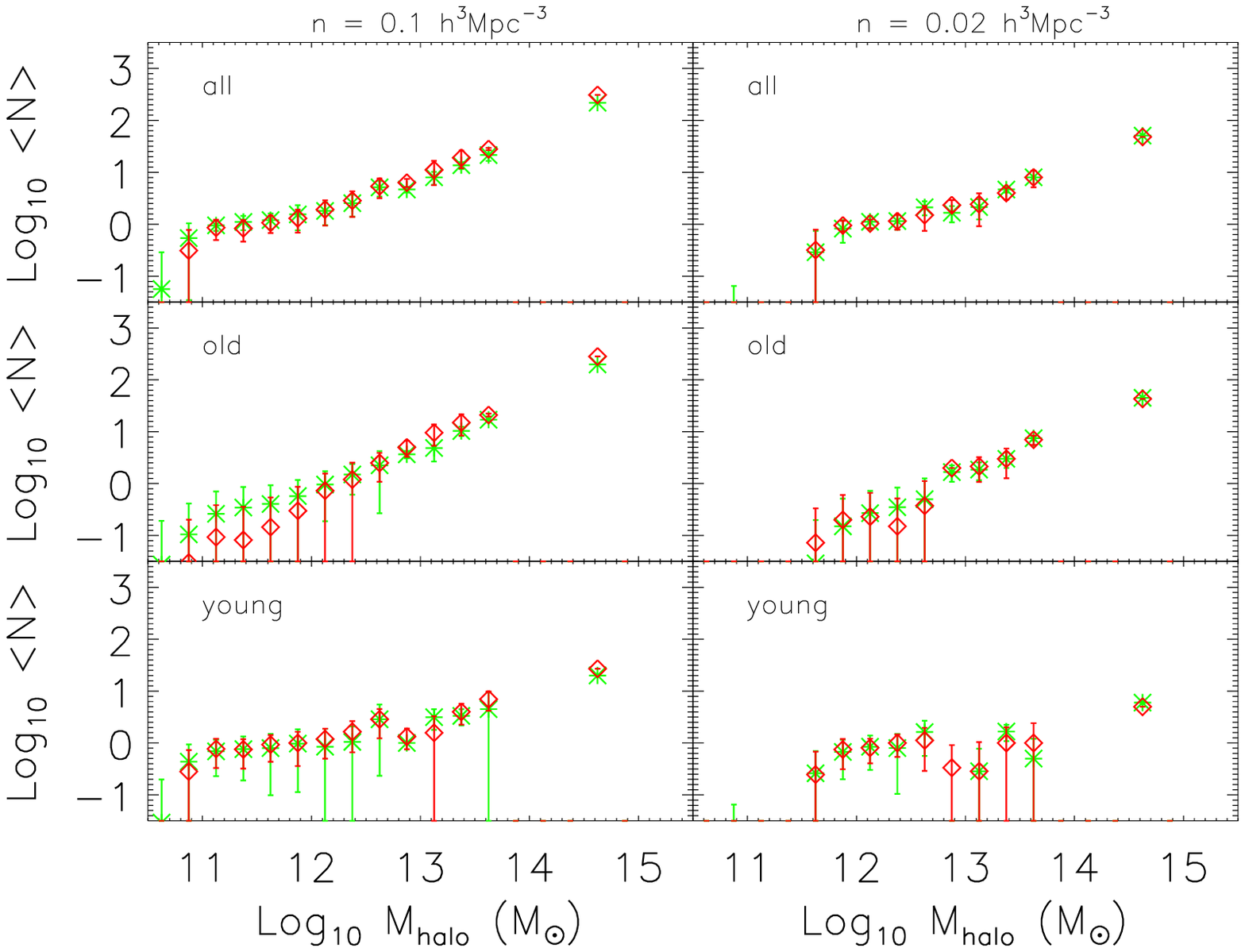,angle=0}
}}
\caption{
%dw Halo occupation distribution measured for the SPH galaxies (asteriks) 
Mean halo occupations measured for the SPH galaxies (asterisks)
and the full
\galics\ model galaxies (diamonds), at $z=0$. In the left
(resp. right) panels, galaxies where selected in baryonic mass so as
to yield a number density of $0.1 h^3\, {\rm Mpc}^{-3}$ (resp. $0.02
h^3\,{\rm Mpc}^{-3}$). From top to bottom, we show the 
%dw halo occupation distributions 
mean occupation functions
for the full, old, and young populations.
\label{fig:hod}}
\end{figure*}

Figure~2 portrays the galaxy populations of all haloes in each of
the three models, at $z=3$ and $z=0$.  The ``central'' galaxy
of each halo is marked by a blue point, while satellites in the same
halo are marked by red points directly underneath.  In \galics,
as in other semi-analytic models, central galaxies are treated
specially --- any cooling from the halo joins the central
galaxy, not the satellites.  The SPH simulation does not impose
this physics {\it a priori}, but central galaxies nonetheless emerge as
a distinct class.  The galaxy closest to the halo centre-of-mass
is almost always the most massive galaxy in the halo and it
usually has the oldest stellar population \citep{berlind_etal03}.
Since the centre-of-mass is affected by asymmetries in the halo
structure, the closest galaxy is occasionally a lower mass ``satellite,''
with the most massive galaxy nearby.
For convenience, we have simply {\it defined} the most massive
galaxy of each SPH halo to be the central galaxy. On closer inspection,
this identification appears reasonable in virtually
every case.  While every SPH halo thus has a central galaxy by
definition, a \galics\ halo may not, if the halo has experienced
a recent merger and there has not been sufficient time for
dynamical friction to drag the most massive galaxy to the bottom
of the halo potential well.

In all three models,
the central galaxy baryonic mass is tightly correlated with
halo mass, while satellite galaxy masses are broadly scattered with
little correlation.  Since most gas cooling in the SPH simulation
and all gas cooling in the \galics\ model occurs on the central galaxy,
its growth is tightly coupled to the growth of the host halo.
Most satellite galaxies, on the other hand, grew in smaller haloes,
and they stopped growing after merging into a larger halo
and becoming satellites.

At $M_{\rm halo}<3\times 10^{12}M_\odot$, both in the SPH model and in the
GalICS model without feedback, most of the blue points are aligned very
close to the $M_{\rm gal}=(\Omega_{\rm b}/\Omega_{\rm m})M_{\rm halo}$ 
black diagonal line, at least at $z=0$.
That means that most of the halo baryons are in the central galaxy.
At $M_{\rm halo}>3\times 10^{12}M_\odot$, this is less and less
true for two reasons:
a) in dense environments the cumulative baryonic content of
satellite galaxies is more important, and
b) as the cooling time
of the hot gas becomes longer, a larger fraction of the baryons
remain in the hot phase (Fig.~3).
For these reasons, we take $M_{\rm halo}=3\times 10^{12}M_\odot$
as the border that separates field galaxies from group galaxies.

The addition of supernova and black hole feedback has two important
effects.  At low halo masses, ejection of gas by supernova feedback
shifts the locus of central galaxy points down by about a factor
of two, with concomitant reductions in satellite galaxy masses.
The factor of two emerges because the ejected mass is, by construction,
about equal to the mass that forms stars.  This suppression
is evident both at $z=3$ and at $z=0$.  At $z=0$, black hole
feedback has a much more drastic effect on massive galaxies, introducing
a sharp cut-off at a total baryonic mass $\sim 3\times 10^{11}M_\odot$.
The most massive galaxies in the no-feedback model are up to a factor of
ten larger.

Figure~3 shows the mean mass of stars (dashed lines) and stars plus
cold gas (solid lines) in bins of halo mass, now including central
and satellite galaxy contributions.  The SPH and no-feedback \galics\
results agree fairly well at both $z=3$ and $z=0$, with \galics\
predicting somewhat higher masses in the most massive $z=0$ haloes.
The lower SPH stellar masses in low mass haloes are probably an
artefact of limited numerical resolution, which leads to underestimated
star formation rates and stellar masses in galaxies near the resolution
threshold.  We again see the factor of two reduction in baryonic masses
due to feedback in the full \galics\ model, which grows to a larger factor
at high halo masses.  Stellar masses are suppressed by about the
same factor.  Since the initial total gas content of each \galics\ halo
is $(\Omega_b/\Omega_m)M_{\rm halo}$ by assumption, the baryons
that are not in the cold gas plus stars phase are necessarily
in the hot phase.

Figure~4 compares the galaxy baryonic mass functions at $z=3$, 1, 0.5, and 0.
The generally good agreement between SPH and no-feedback \galics\
seen in Figure~1 extends to higher redshifts.  However, galaxy growth
in no-feedback \galics\ lags slightly (an effect also seen in Fig.~2),
so its mass function begins somewhat lower at $z=3$ and catches up
at low redshift.  This difference may reflect the high efficiency of
filamentary accretion (see \citealp{keres_etal05}) relative to the spherical
accretion assumed in \galics.  The dotted lines in Figure~4 show the full
\galics\ mass function (solid line) after all galaxy masses are multiplied
by a factor of 2.1.  This shift brings it into good agreement with the
SPH and no-feedback \galics\ mass functions up to
$M\sim 3\times 10^{11}M_\odot$, beyond which one sees the sharper
cut-off induced by AGN feedback.

%{\bf Here is what I think we should say about the halo occupation results.
%I think that one figure (along the lines described below) and this amount
%of text are appropriate; it's enough to get across the most important
%points of the HOD study without distracting too much from the narrative
%of this paper.  Note that this would be Figure 5, so other figures
%would have to be renumbered.}

Figure~\ref{fig:hod} compares the halo occupation statistics of galaxies in the SPH
simulation (asterisks) and the full \galics\ model (diamonds),
at $z=0$.  In each model, we select galaxies above a baryonic mass
threshold that yields a mean space density of $0.1 h^3\, {\rm Mpc}^{-3}$
(left panels) or $0.02 h^3\,{\rm Mpc}^{-3}$ (right panels).
We have divided the galaxies in each panel into an old and a young population
based on the median age of their stellar population.
The old/young threshold was chosen in such way that the two
subsets contain the same number of galaxies.
Top, middle, and bottom panels show the mean occupation function
(the average number of galaxies as a function of halo mass) for the
full, old, and young populations, respectively.
\begin{figure}
\noindent
\begin{minipage}{8.4cm}
  \centerline{\hbox{
      \psfig{figure=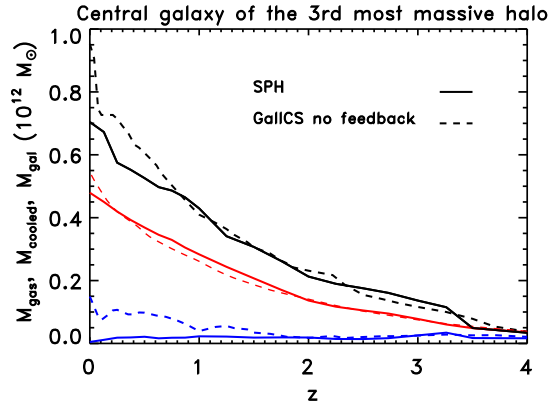,height=5.73cm,angle=0}
  }}
\end{minipage}\    \
\caption{Growth history of the central galaxy of the third most massive halo
($M_{\rm halo} \sim 3\times 10^{13}M_\odot$) in the SPH simulation
(solid lines) and in the GalICS model without feedback (dashed lines).
The black lines show the total galaxy mass (stars plus cold galactic gas),
while the red lines show mass of the gas that has cooled onto the galaxy since
the beginning of the simulation.
Galaxy mergers are the reason why the black lines are above the
red lines. The blue lines show the cold galactic gas content
in the SPH simulation and in the GalICS model without feedback.
%{\bf The $y-$axis is incorrectly labeled.  Is the initial offset
%between red and black
%just the mass the galaxy had when it was first identified?
%Wouldn't it be clearer to show merger growth rather than gas
%growth separately?  That would solve the initial offset problem,
%and the problem that a constant red-black offset on a
%logarithmic plot does not mean that no mergers are occuring.}
%AC: following your comment, I have switched from a logarithmic to a linear scale
%merger growth would be calculated indirectly by total growth - growth due to gas accretion
%therefore I think it is better to show cooled mass
%Your comment on the offset is correct.
%Hence I have renormalised the red lines so that they coincide with the black ones when the galaxy
%is firstly identified at z=4.5
}
\end{figure}
The SPH and \galics\ predictions agree impressively well, including the
separate halo occupation statistics of old and young populations,
which are quite different from each other. \citet{berlind_etal03}
found a similar level of agreement between a larger volume, lower
resolution SPH simulation and the semi-analytic galaxy formation
model of \citet{cole_etal00}.  Although we do not show it here,
the SPH and \galics\ models exhibit comparably good agreement for
the mean pair counts and triple counts
($\langle N(N-1) \rangle$ and $\langle N(N-1)(N-2) \rangle$)
as a function of halo mass.  The agreement of halo occupation
statistics essentially guarantees that the SPH method and \galics\
will make similar predictions for large scale clustering, even
in larger volumes that are more statistically representative,
and that this agreement will extend to the dependence of
clustering on galaxy age and to the 3-point correlation statistics.
On scales comparable to or smaller than the virial diameters
of the largest haloes, the predicted spatial distribution of
galaxies within haloes matters as well as the number of galaxies
per halo.  In investigating this issue, we found that
the \citet{hatton_etal03} prescription for computing satellite
galaxy positions predicts spatial distributions in large haloes
that are too diffuse, artificially suppressing small scale
clustering \citep[see][]{BlaizotEtal06}. However, a good agreement with the SPH simulation is achieved
by simply associating satellite galaxies with randomly selected
subsets of dark matter halo particles or dark matter sub-structures.

\section{Galaxy growth}

Galaxies grow by accreting gas and by merging with other galaxies.
In this Section, we compare semi-analytic and SPH results at three different levels. First, we discuss the mass assembly of an individual galaxy (the central galaxy of the third most massive halo in the simulation at $z=0$). We do this by following
 the ``main branch" of the galaxy tree, both in the SPH and in the GalICS outputs. That is, we recursively link the galaxy to its most massive progenitor. Second, we compare semi-analytic and SPH predictions for the total cooling rate integrated over this halo's history. In the SPH simulation, we measure this quantity as the cumulative gas accretion rate of all the {\it galaxy}'s resolved progenitors at a given time. In GalICS, we simply add up the mass of the gas that cools in all the progenitors of our {\it halo} at each time-step. Notice that these two estimates are differently affected by resolution effects. In the semi-analytic case, we only consider cooling in haloes that contain more than 20 dark matter particles. In the SPH case, the limit is more vague, because cooling is not affected by resolution in a straightforward way. Hence the comparison 
%dw remains semi-qualitative. 
is not precise, but it suffices for qualitative assessment.
 Third, we compare the total cooling rates in the entire simulation volume, for the SPH and the GalICS case. Here again, resolution affects  the results slightly differently in the two methods.

 \begin{figure}
\noindent
\begin{minipage}{8.4cm}
  \centerline{\hbox{
      \psfig{figure=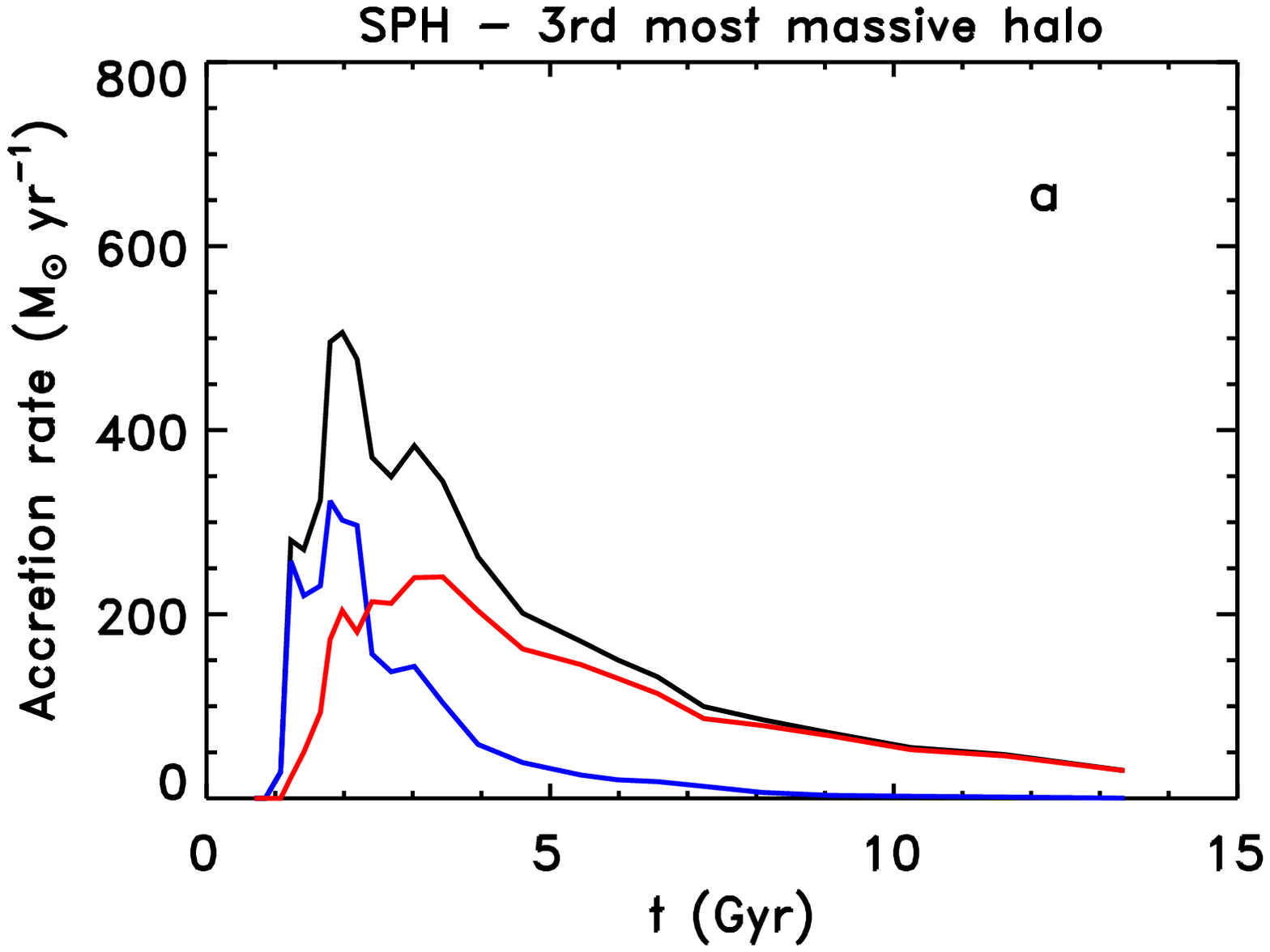,height=5.73cm,angle=0}
  }}
\end{minipage}\    \
%\hskip
\begin{minipage}{8.4cm}
  \centerline{\hbox{
      \psfig{figure=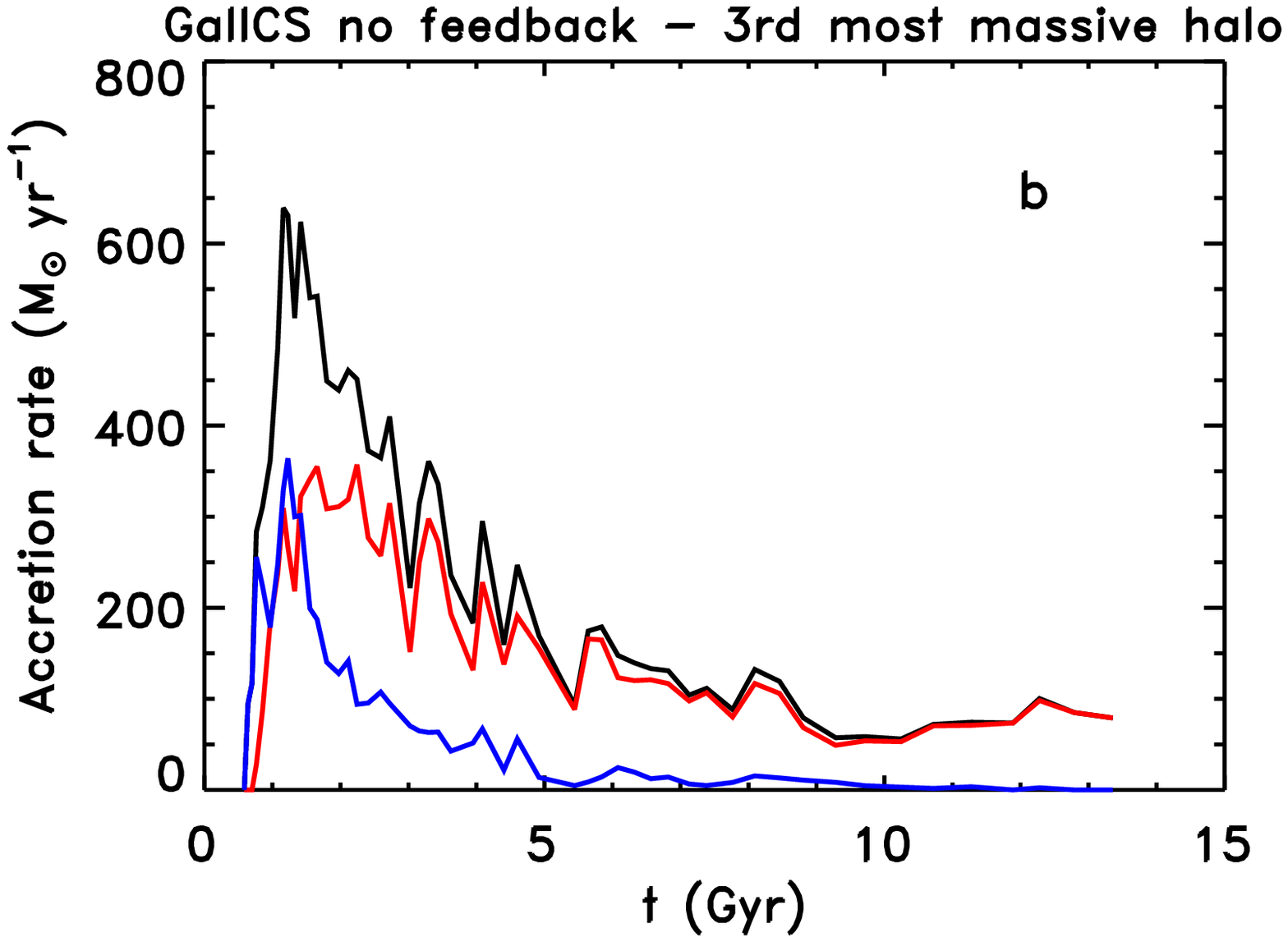,height=5.73cm,angle=0}
  }}
\end{minipage}\    \
\caption{Gas accretion history of the galaxies in the third most massive halo contained in
 the computational box.
The total amount of gas accreted by the galaxies in the halo at each cosmic time (black lines) has been
decomposed into the contributions of the cold (blue) and the hot (red) mode. See
Section~3.2 and Section~5 for an explanation of how we distinguish between
``cold"
%(infall-dominated)
and ``hot"
%(cooling-dominated)
accretion.}
\end{figure}
\begin{figure}
\noindent
\begin{minipage}{8.4cm}
  \centerline{\hbox{
      \psfig{figure=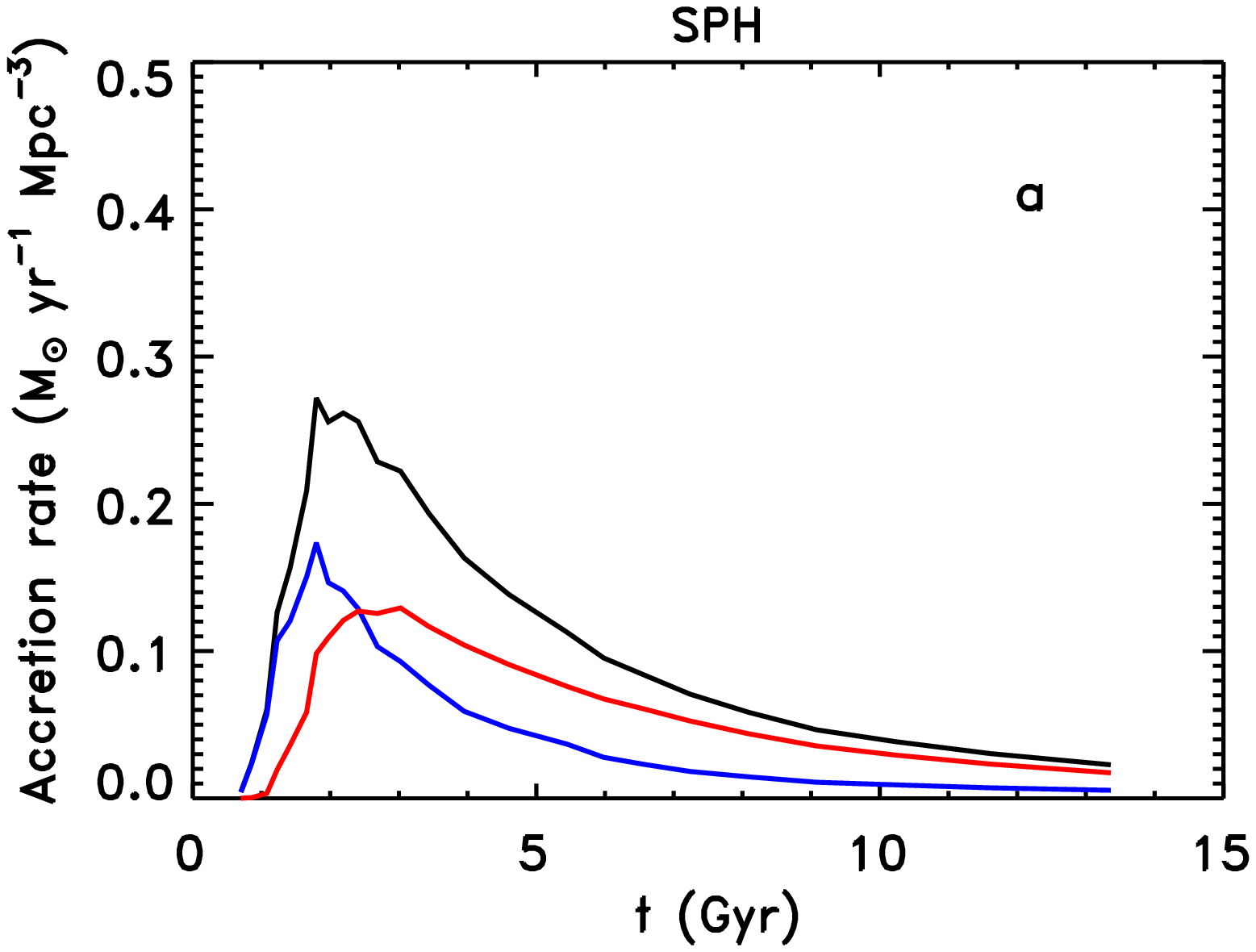,height=5.73cm,angle=0}
  }}
\end{minipage}\    \
%\hskip
\begin{minipage}{8.4cm}
  \centerline{\hbox{
      \psfig{figure=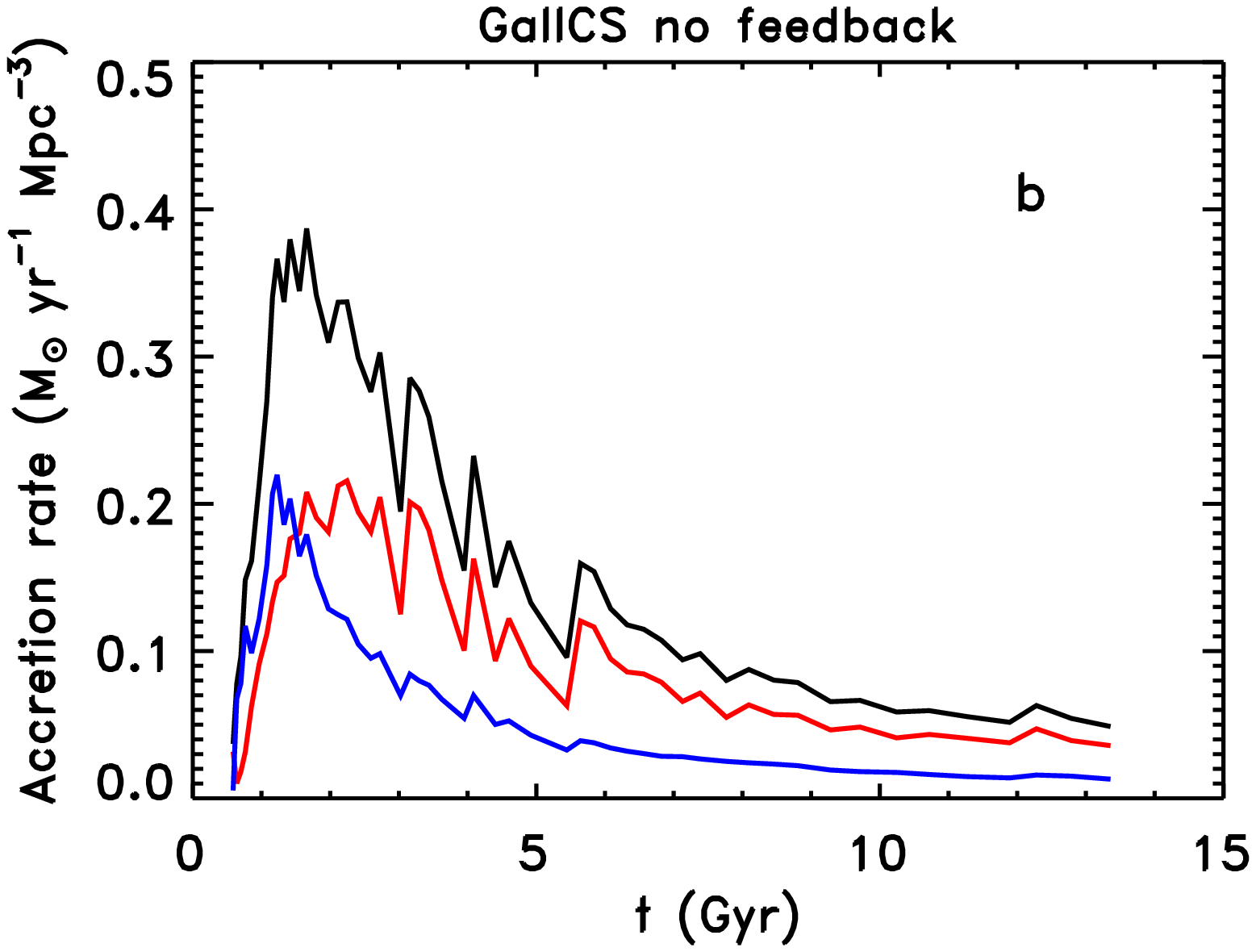,height=5.73cm,angle=0}
  }}
\end{minipage}\    \
%\hskip
\begin{minipage}{8.4cm}
  \centerline{\hbox{
      \psfig{figure=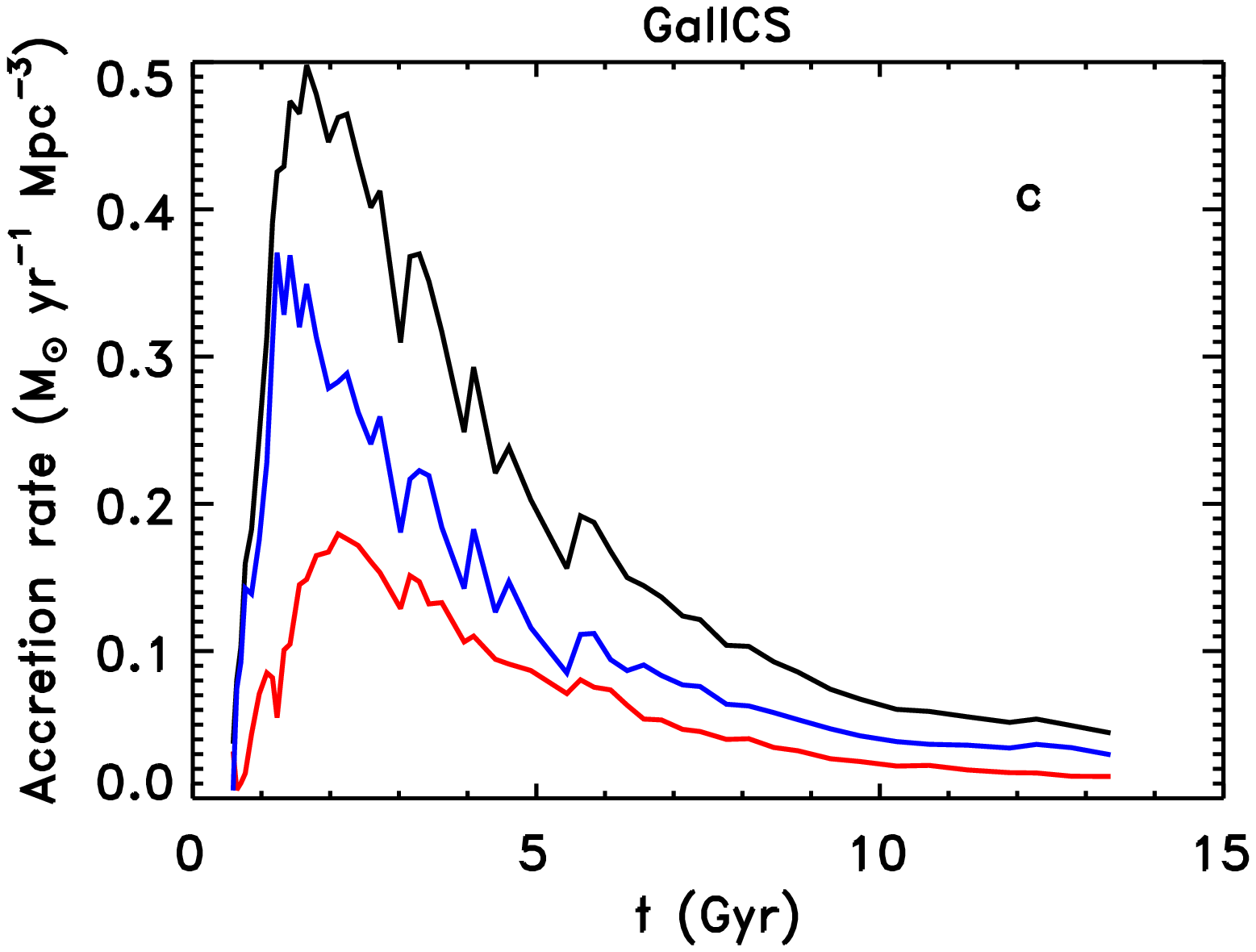,height=5.73cm,angle=0}
  }}
\end{minipage}\    \
\caption{The rate at which galaxies accrete gas in a cosmic comoving volume.
The total value (black lines) is decomposed into
the contributions of the cold mode (blue lines) and the hot mode (red lines).
%See Section~3.2 and Section~5 for an explanation of how we distinguish between
%cold and hot accretion.
}
\end{figure}

Figure~6 illustrates galaxy growth for the
central object of the second most
massive halo ($M_{\rm halo}\sim 3\times 10^{13}M_\odot$).
We start from the galaxy at $z=0$ and trace it back in time by
selecting the most massive progenitor at each step.
The black solid line shows the
growth of its total galaxy baryonic mass in the SPH simulation.
The no-feedback GalICS model, shown by the black dashed line, is in very
good agreement with the SPH model.
The only notable differences  are due to merging.
In the SPH model there is a sudden increase of the galaxy mass from
$5\times 10^{10}M_\odot$ to $1.3\times 10^{11}M_\odot$ at $z\sim 3.3-3.5$,
produced by a major merger between two outputs.
%{\bf Do we know this to be true?}
%AC:  for the SPH model, it is the only reasonable explanation
%and it fits in well with all the pieces of evidence that I have
% In GalICS I know for a fact that this galaxy has had 12 mergers at 0<z<3.7
The semi-analytic model contains a similar merging event but it occurs slightly earlier
(between the snapshot at $z\sim 3.7$ and the snapshot at $z\sim 3.5$)
and the growth is somewhat smaller
(from $4\times 10^{10}M_\odot$ to $8\times 10^{10}M_\odot$).
The final difference in mass between the two galaxies is almost entirely due to a merger
in the GalICS model at $z\sim 0.039-0.076$. In this merger, the twelfth since the first at
$z\sim 3.5-3.7$, the galaxy mass has risen from $\sim 7.4\times 10^{11}M_\odot$ to
$\sim 9.5\times 10^{11}M_\odot$.
However, for the most part, the two growth curves nearly overlap.

The red solid and dashed lines show the total amount of {\it gas} accreted
by the galaxy since the start of the simulation in the two calculations.
In the absence of mergers, the red and the black lines of each type
would coincide.  From the agreement between the red solid and dashed
lines, we see that the two methods agree on the relative amounts of
accretion growth and merger growth.  Finally, blue lines show the
amount of cold gas in the galaxy as a function of redshift.  The
two methods agree well down to $z=1.5$.  Below this redshift, the SPH
simulation predicts a lower gas fraction, but the gas fraction is
small in each case.  The residual gas fraction in this gas-poor regime
depends on the details of the star formation
prescription, so some disagreement is expected.

\citet{katz_etal03} and \citet{keres_etal05} have investigated gas accretion in
the same SPH simulation that we are considering here.
By tracing the temperature history of accreted particles, they have
identified two distinct modes of gas accretion.
About half of the gas follows the expected track in the conventional picture
of galaxy formation: it shock-heats to the virial temperature of the
galaxy's potential well ($T\sim 10^6\,$K for a Milky Way type galaxy)
before cooling, condensing, and forming stars.
However, the other half radiates its acquired gravitational energy at much
lower temperatures, typically $T<10^5\,$K.
Cold accretion is often directed along filaments, allowing
galaxies to draw gas efficiently from large distances, and it is
dominant for low mass galaxies ($M_{\rm halo}<3\times 10^{11}M_\odot$).
Hot accretion is quasi-spherical and dominates the growth of high mass
systems.

\citet{birnboim_dekel03} and
\citet{dekel_birnboim06} argue that the transition between
the cold and hot accretion regimes is determined by the stability
criterion for a shock at the halo's viral radius, which in turn
depends on the ratio of the post-shock cooling timescale to the
dynamical time.  Motivated by this argument, \citet{croton_etal05}
suggest that the cold and hot accretion regimes in hydrodynamic
simulations should be identified with the rapid cooling and slow cooling
regimes in semi-analytic models.
We investigate this identification
in Figures~7 and~8.

The black curve in Figure~7a shows the total
accretion rate onto all galaxies in the third most massive halo
of the SPH simulation as a function of time.  The blue and the red curves
show the cold and hot accretion rates, respectively.
Here cold accretion consists of gas particles whose maximum
temperature at any phase of their evolution is $T_{\rm max}<2.5\times 10^5\,$K.
Because the $T_{\rm max}$ histogram is bimodal \citep{keres_etal05},
the division of hot and cold modes is insensitive to the exact
choice of threshold.
Figure~7b presents analogous results for the same halo in the
no-feedback \galics\ model.  Here we identify cold accretion
by the $r_{\rm cool} > r_{\rm infall}$ criterion discussed
in Section~3.2.  With this identification, the qualitative agreement
between the SPH and \galics\ calculations is quite good.
In both cases there is a brief initial phase of cold accretion,
when the universe is less than $\sim 2$ Gyr old, but hot accretion dominates
after the halo becomes massive enough to support a virial shock.
In the SPH simulation, cold and hot accretion can co-exist for
the same galaxy (see \citealp{keres_etal05}), and in both
calculations cold accretion continues onto lower mass galaxies
that will subsequently join the main halo as satellites.
However, the central galaxy of this halo is massive ($\sim 10^{12}M_\odot$)
and contains a large fraction of the cooled baryons, so it
dominates the accretion statistics.  The cold accretion rate on
galaxies in this halo is negligible at $t>5$ Gyr, in both
calculations.

Figures~8a and~8b present a similar comparison for the accretion rates
averaged over the entire simulation volume.  The results are
analogous, but cold accretion remains important for a longer time
because most haloes are less massive than the one shown in
Figure~7.  The agreement between the SPH and the \galics\ calculations is
now even better.

Figure~8c shows the predictions of the full \galics\ model with feedback.
As discussed in Section~3.4, supernova feedback is modelled by assuming
that supernovae drive a wind with an outflow rate approximately equal to the
star formation rate, independent of the potential well's depth.
%{\bf Is my change of ``proportional'' to ``approximately equal'' correct?
%AC : correct
%Are there any other criteria not specified in 3.4, like a minimum
%SFR or surface density?
%AC: there is a threshold that I had forgot to mention (see addition to star formation section)
% but it is so low that its effect is negligible
% If not, I don't understand the statement
%that supernova feedback is ``more important'' for lower mass
%galaxies; it seems like the effect would be exactly equal everywhere.}
%AC: the point is that the mass loss from galaxies is v_vir independent,
%not so for the mass loss from haloes, which depends
%(although only logarithmically; Hatton et al. 2003) on the halo scale.
%In low mass haloes, baryons are not only lost from the galaxy, they are also lost from
%the surrounding hot haloes, hence they are not immediately available for cooling
Since ejected gas may cool and be accreted again, perhaps multiple times,
the total accretion rate in the full \galics\ model is higher
than that in Figures~8a or~8b, even though the final galaxy
masses are lower.  Gas entrained in a galactic wind is heated to
the halo virial temperature, but if its subsequent cooling time is
shorter than the infall time, we count it as cold accretion, just as
we do for infalling gas.  As most supernova feedback occurs in low
mass galaxies with short cooling times, it is the cold accretion
rate that rises significantly.
Black hole feedback is modelled with a cooling cut-off in groups
with $\Sigma M_{\rm bulge}>10^{11}M_\odot$, so it
acts to suppress hot accretion in massive haloes.
As a result, feedback reduces hot accretion in \galics, especially
at late times.  The new \galics\ model of
\citet{cattaneo_etal06} makes the assumption that {\it all} hot accretion
is suppressed by AGN feedback, though it incorporates a more
sophisticated calculation of where the cold-to-hot transition
occurs.

\section{The emergence of the galaxy bimodality}

We now turn to our principal subject, the origin of bimodality in
the galaxy population.  We must first understand the connection
between galaxy gas content and star formation rate (which we denote
either by SFR or by $\dot{M_*}$).
Figure~9 plots SFR against galaxy gas mass at $z=3$ and $z=0$,
for the SPH, no-feedback \galics, and full \galics\ models.
In each panel, red points represent satellite galaxies.  Solid
and dashed lines are the same in every panel and are provided
for visual reference.

\begin{figure*}
\noindent
\begin{minipage}{8.4cm}
  \centerline{\hbox{
      \psfig{figure=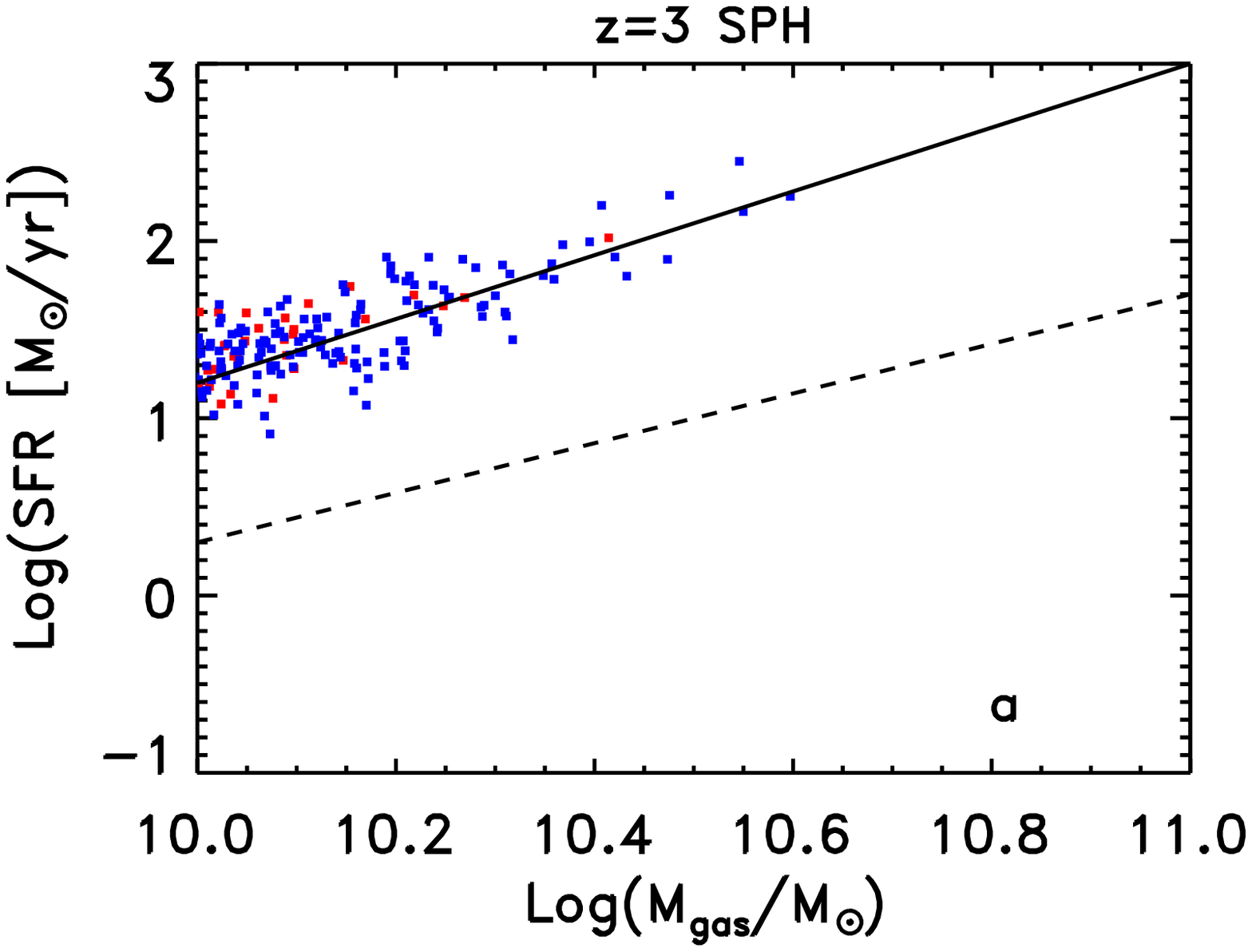,height=5.73cm,angle=0}
  }}
\end{minipage}\    \
%\hskip
\begin{minipage}{8.4cm}
  \centerline{\hbox{
      \psfig{figure=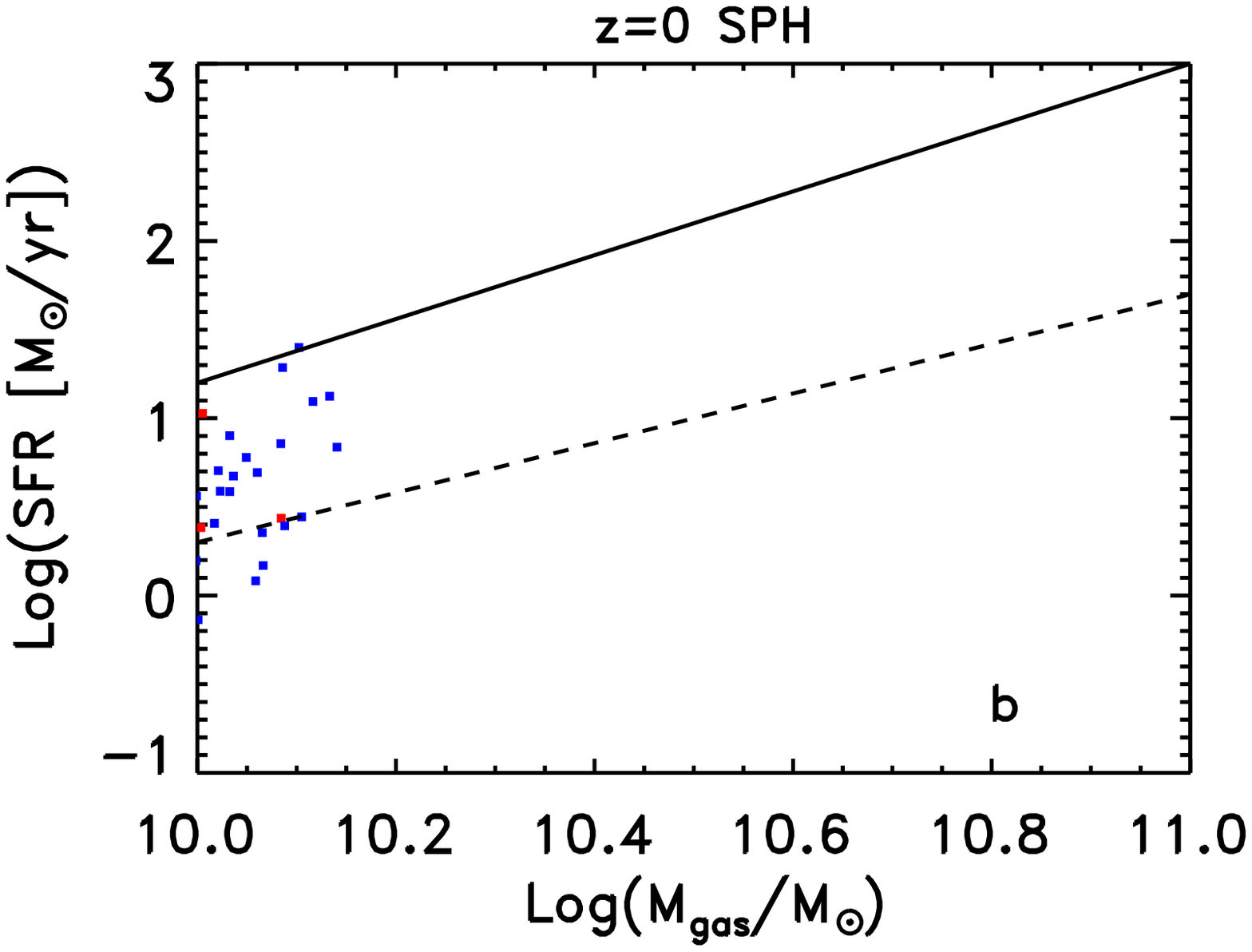,height=5.73cm,angle=0}
  }}
\end{minipage}\    \
%\hskip
\begin{minipage}{8.4cm}
  \centerline{\hbox{
      \psfig{figure=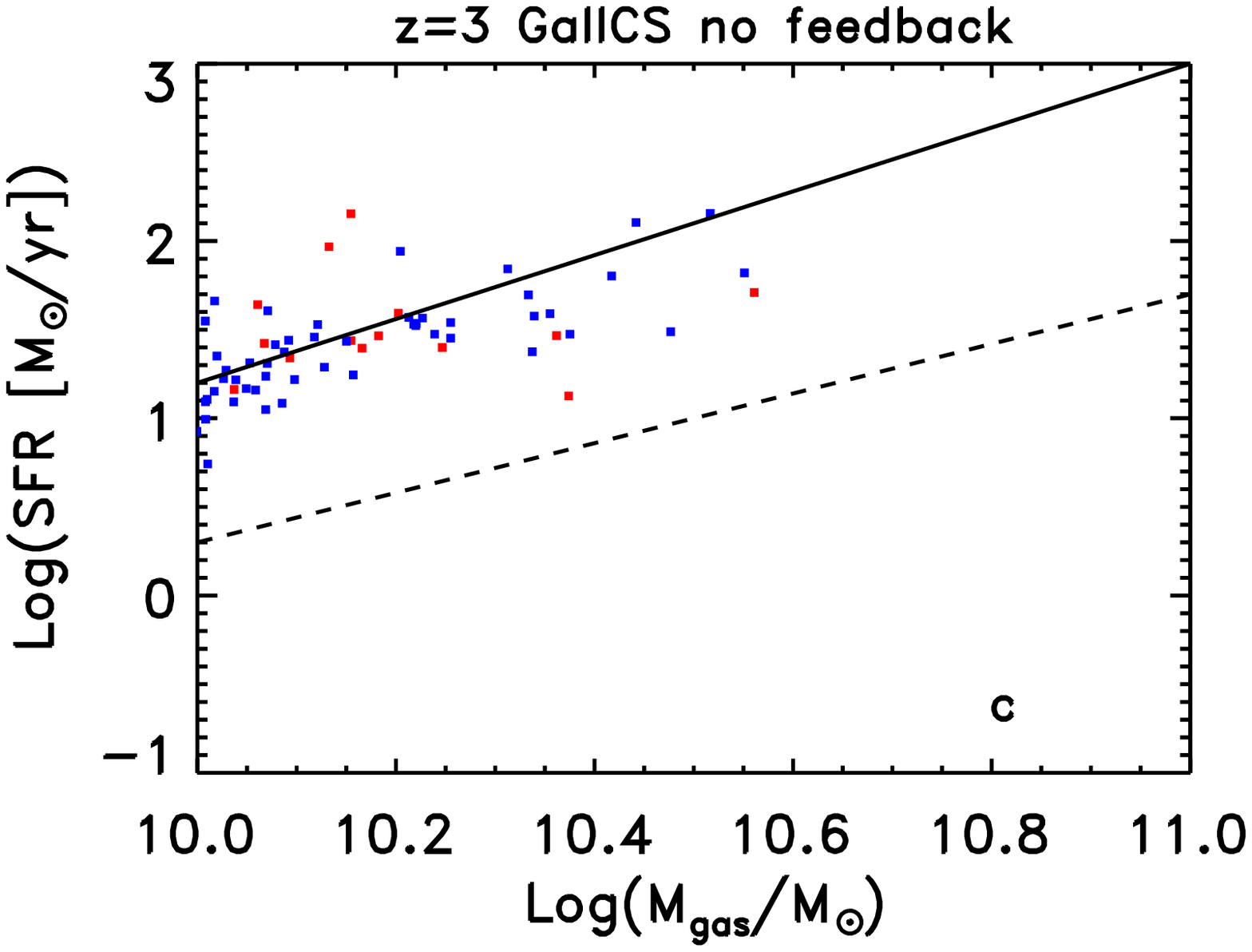,height=5.73cm,angle=0}
  }}
\end{minipage}\    \
%\hskip
\begin{minipage}{8.4cm}
  \centerline{\hbox{
      \psfig{figure=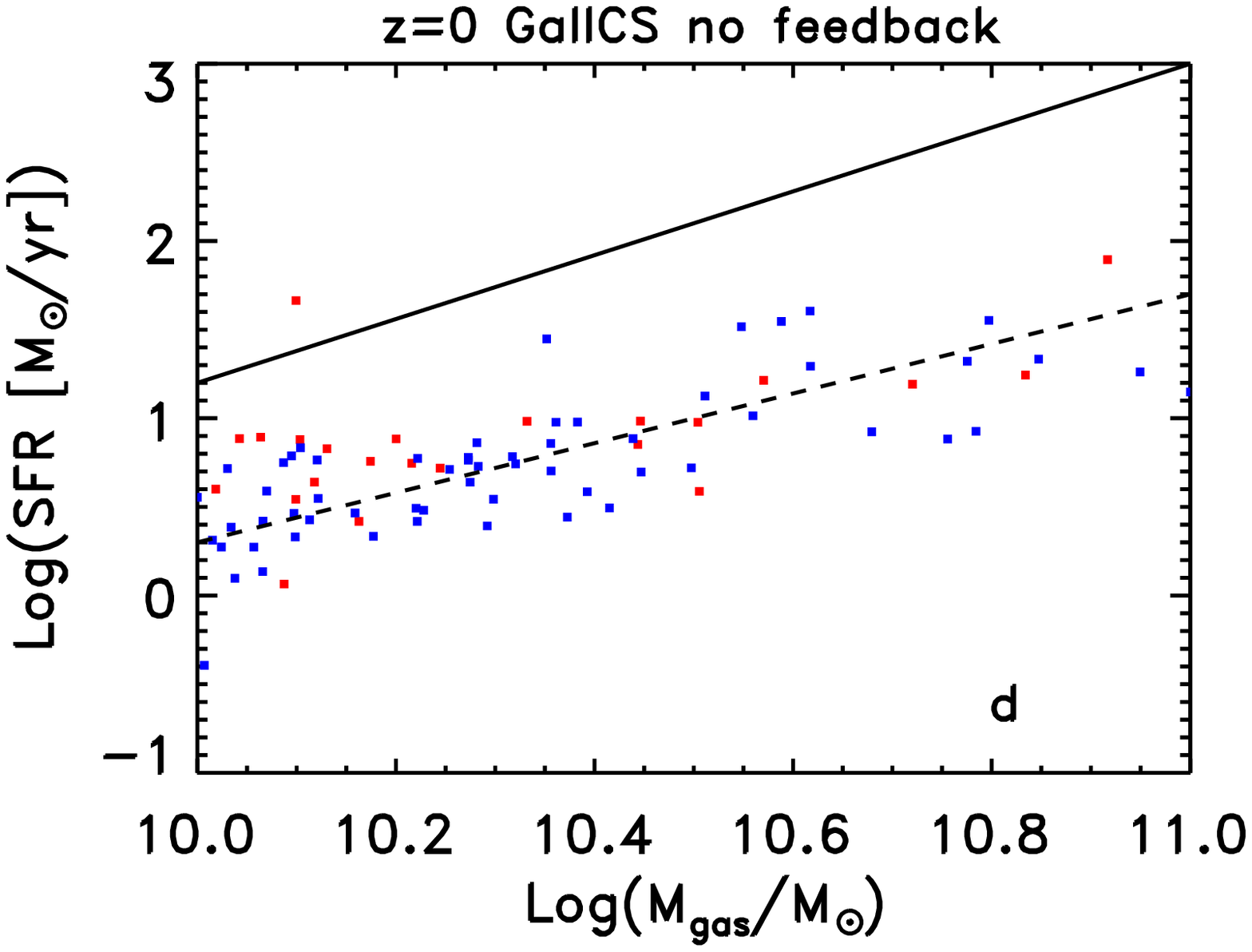,height=5.73cm,angle=0}
  }}
\end{minipage}\    \
%\hskip
\begin{minipage}{8.4cm}
  \centerline{\hbox{
      \psfig{figure=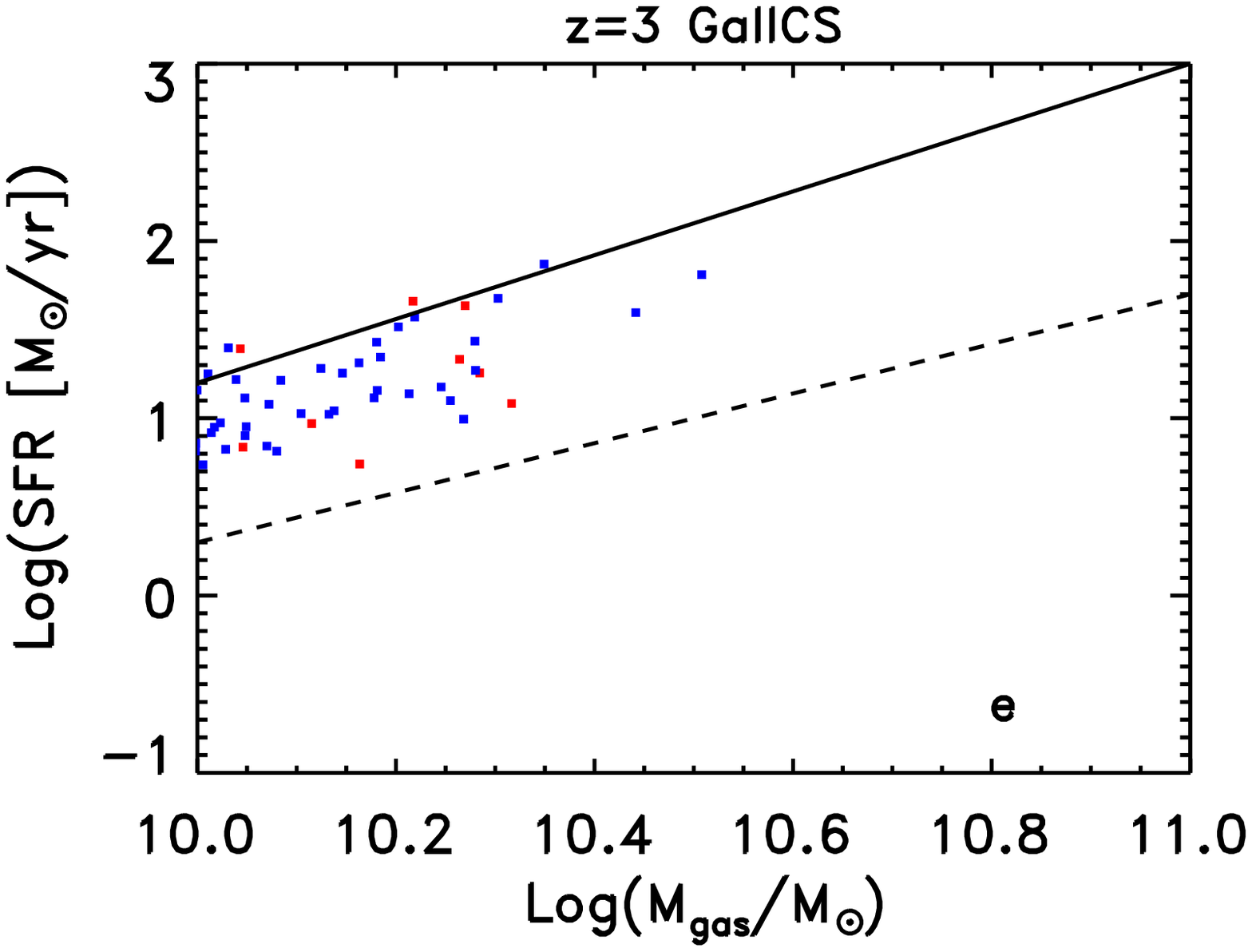,height=5.73cm,angle=0}
  }}
\end{minipage}\    \
%\hskip
\begin{minipage}{8.4cm}
  \centerline{\hbox{
      \psfig{figure=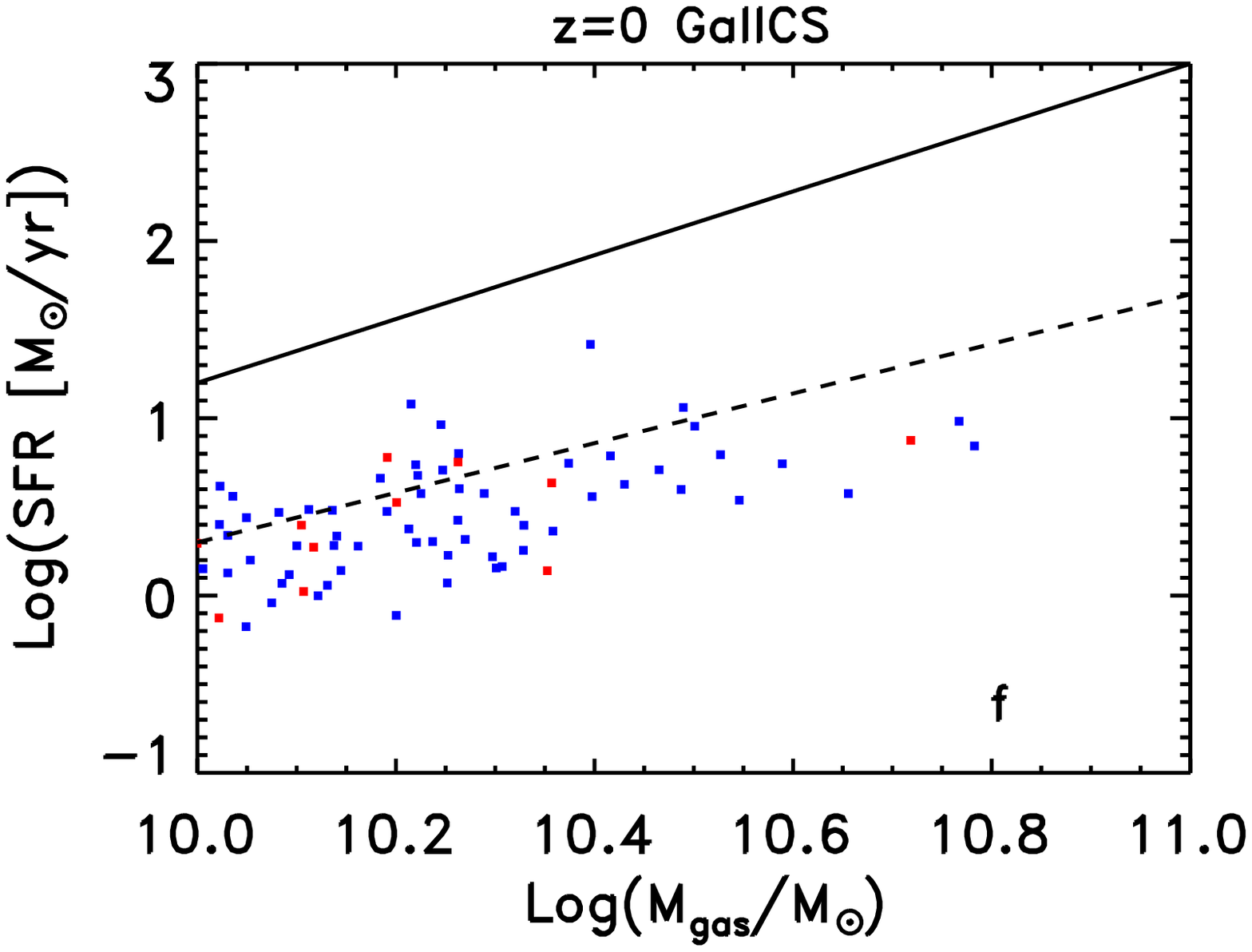,height=5.73cm,angle=0}
  }}
\end{minipage}\    \
\caption{The relation between gas content and star formation rate in the SPH
      simulation (top), in the GalICS model without supernova or AGN feedback
      (centre) and in the standard GalICS model (bottom). The blue points correspond to galaxies that are at the centre of their dark
      matter halo (isolated field galaxies, brightest
group or cluster galaxies), while the red points are
satellite galaxies in groups or clusters. The solid line and the dashed line
      correspond to $1.8{\rm\,Log}M_9-0.6$ and $1.4{\rm\,Log}M_9-1.1$ ,
      respectively ($M_9\equiv M_{\rm gas}/10^9M_\odot$).
They are not fits and
      are just shown to guide the eye.}
\end{figure*}

\begin{figure*}
\noindent
\begin{minipage}{8.4cm}
  \centerline{\hbox{
      \psfig{figure=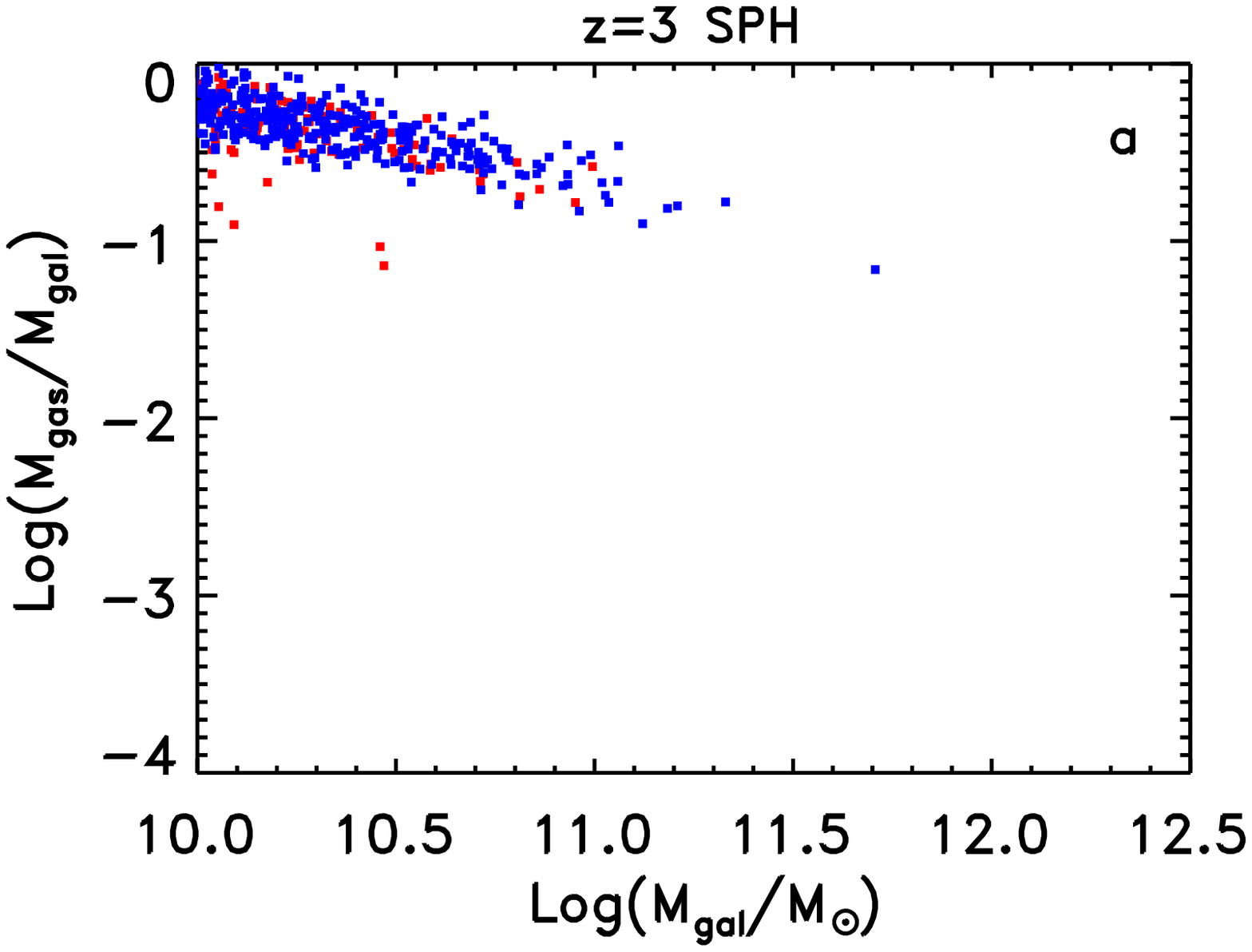,height=5.73cm,angle=0}
  }}
\end{minipage}\    \
%\hskip
\begin{minipage}{8.4cm}
  \centerline{\hbox{
      \psfig{figure=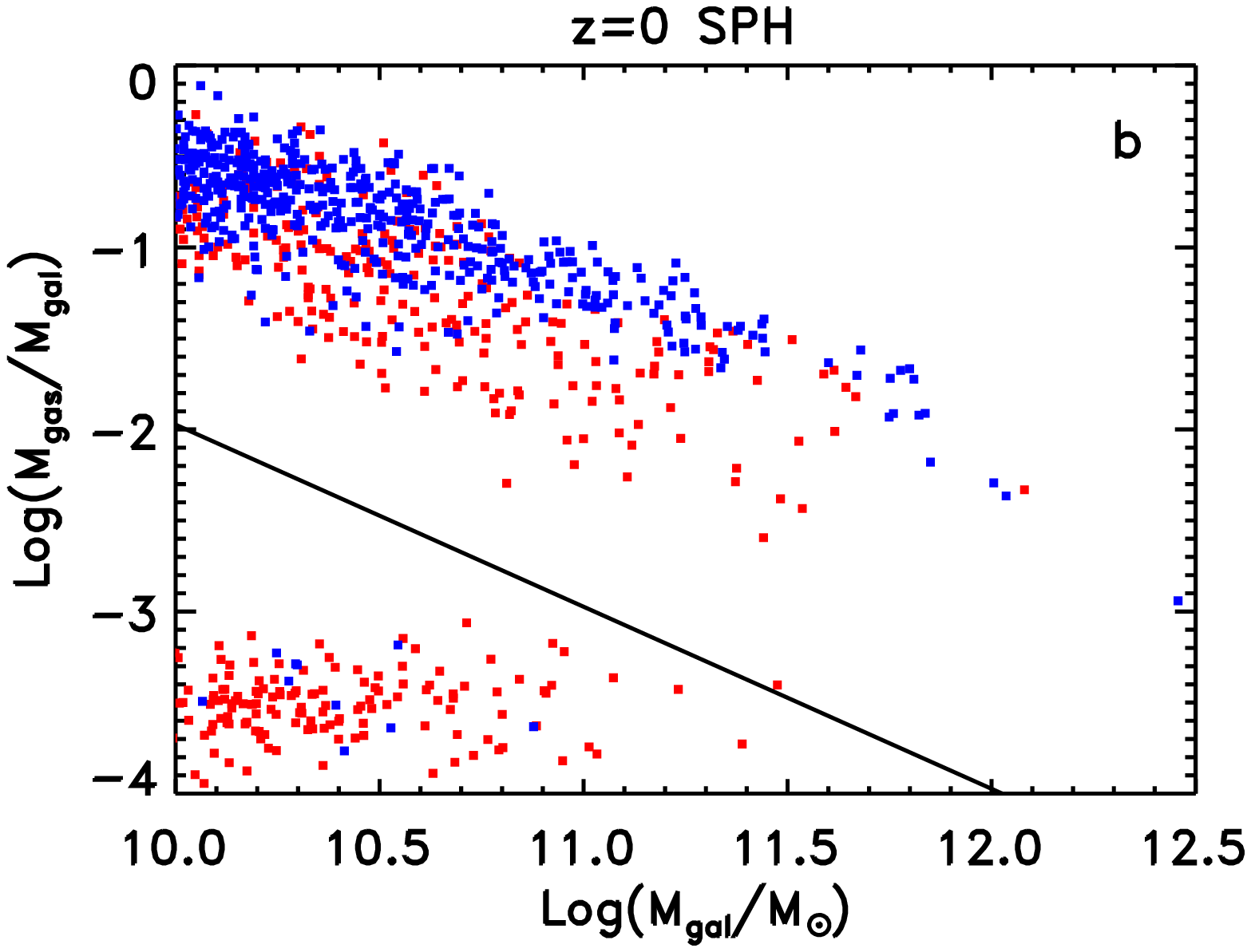,height=5.73cm,angle=0}
  }}
\end{minipage}\    \
%\hskip
\begin{minipage}{8.4cm}
  \centerline{\hbox{
      \psfig{figure=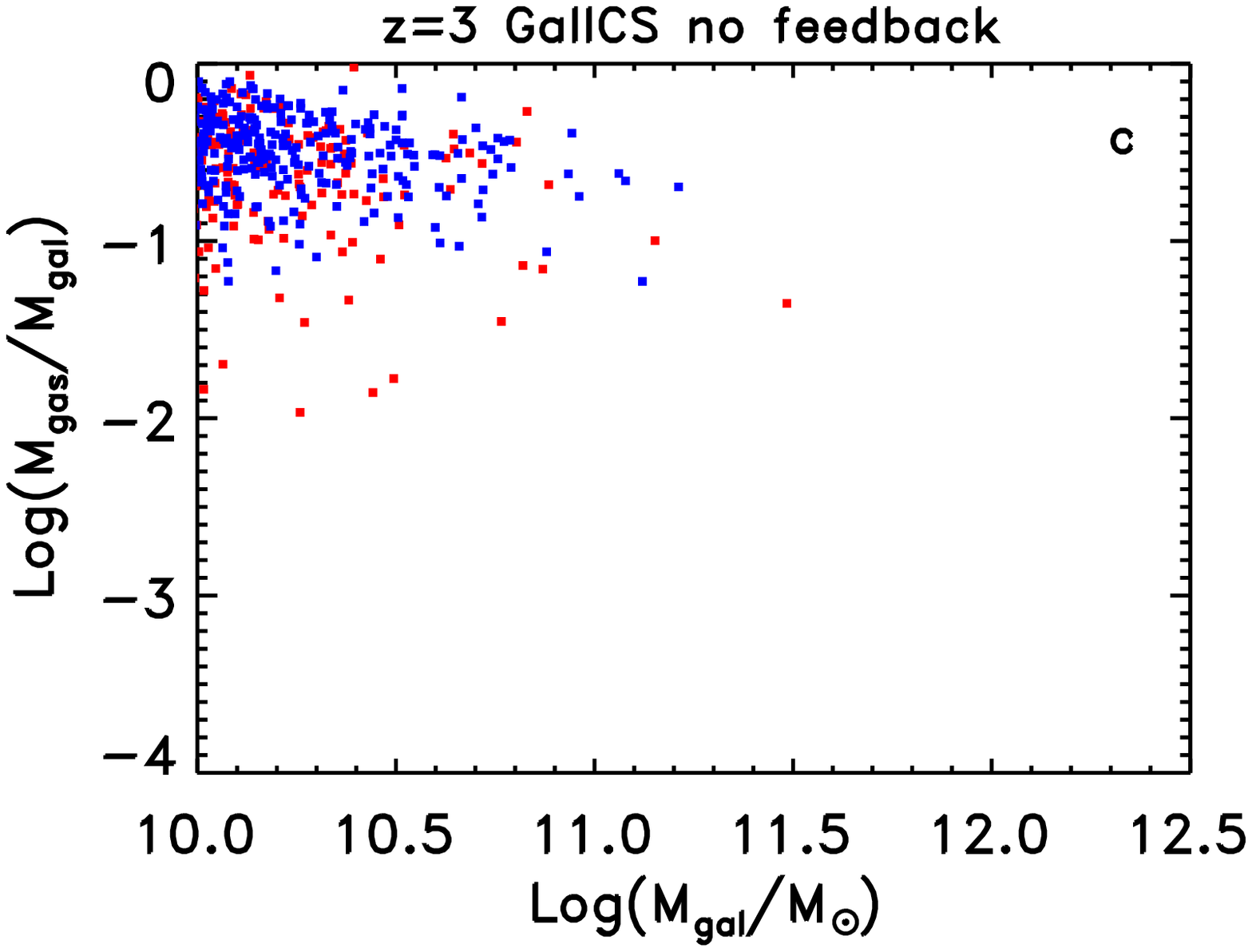,height=5.73cm,angle=0}
  }}
\end{minipage}\    \
%\hskip
\begin{minipage}{8.4cm}
  \centerline{\hbox{
      \psfig{figure=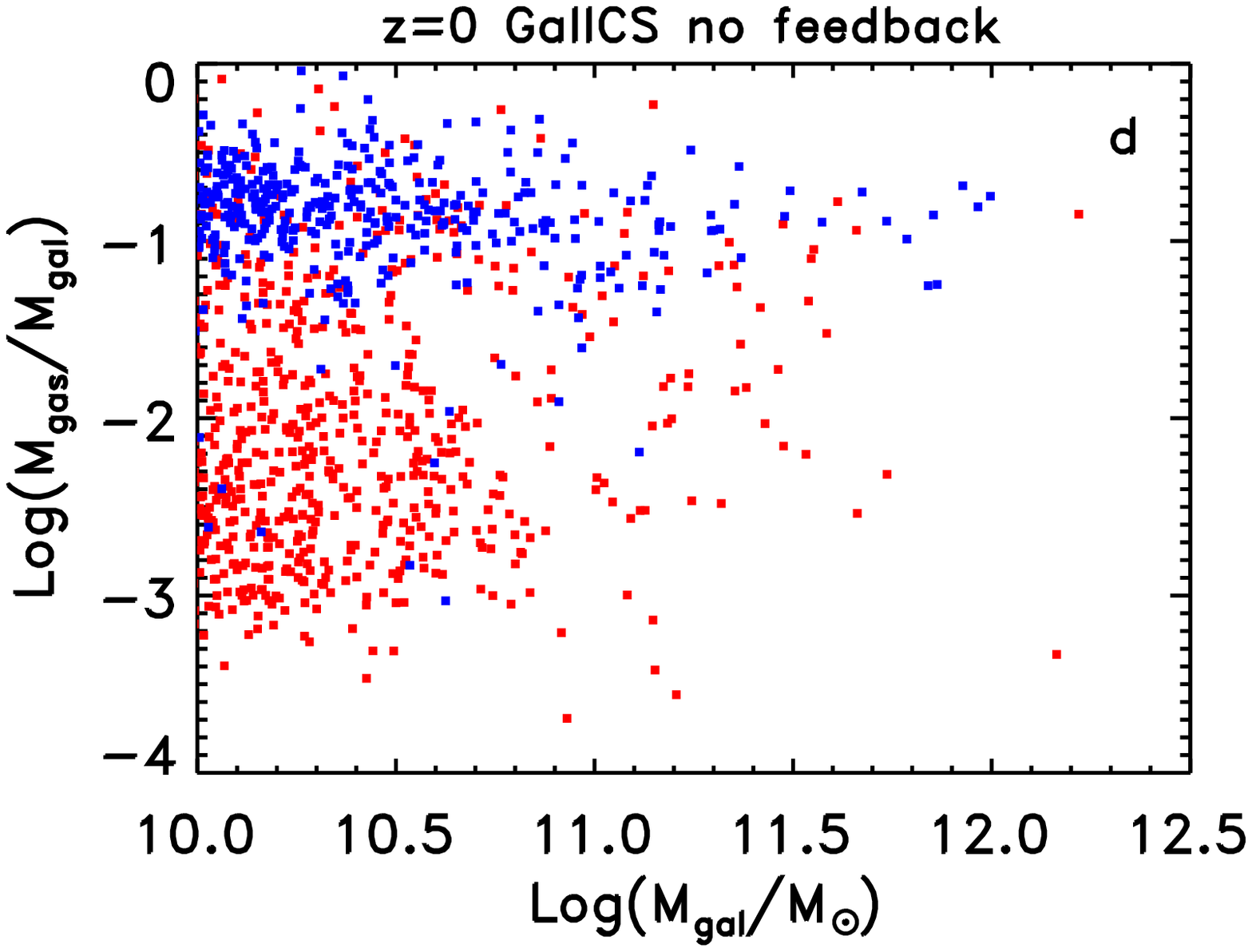,height=5.73cm,angle=0}
  }}
\end{minipage}\    \
%\hskip
\begin{minipage}{8.4cm}
  \centerline{\hbox{
      \psfig{figure=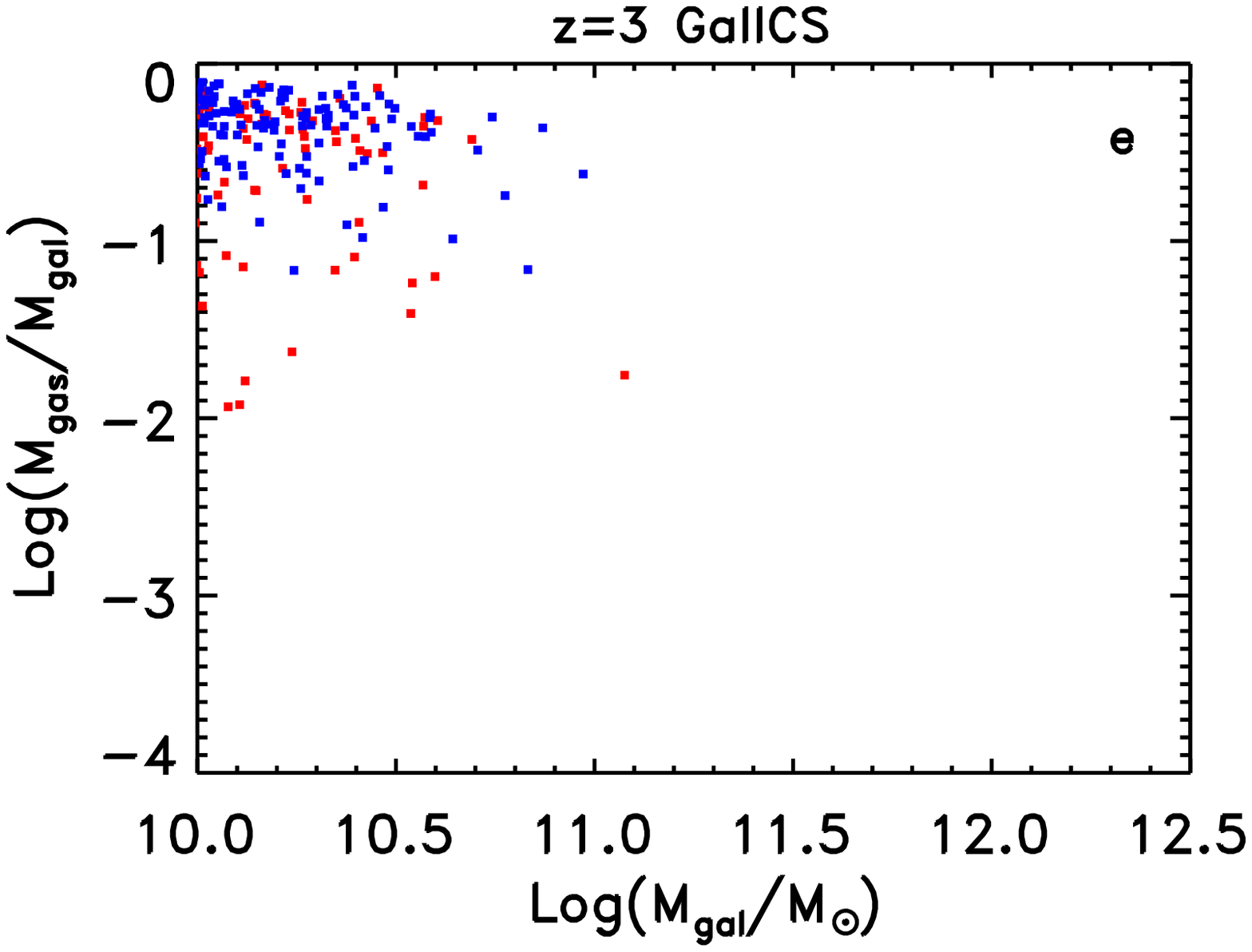,height=5.73cm,angle=0}
  }}
\end{minipage}\    \
%\hskip
\begin{minipage}{8.4cm}
  \centerline{\hbox{
      \psfig{figure=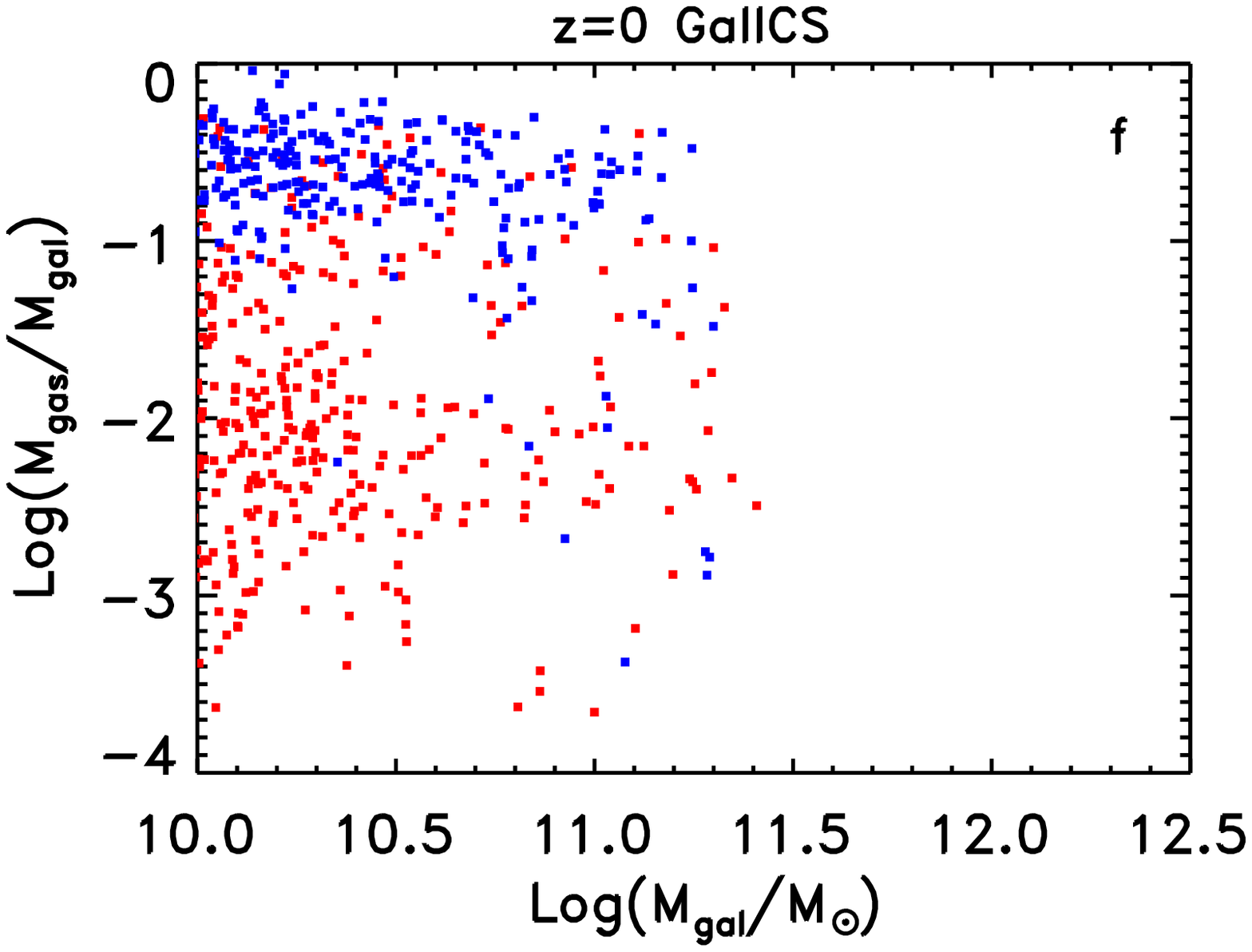,height=5.73cm,angle=0}
  }}
\end{minipage}\    \
\caption{The gas fraction as a function of galaxy mass and redshift. The blue points correspond to galaxies that are at the centre of their dark
      matter halo, while the red points are
satellite galaxies.
The diagonal line in (b) shows the position of galaxy with a given mass and one gas SPH particle only.
The galaxies below this line do not contain any measurable mass of gas.
They have been attributed a gas mass drawn from a $10^{-3.5\pm 0.2}$
log-normal distribution for the gas fraction to put them on the plot.
%{\bf Add a line showing the mass of 1 SPH particle to the top right panel.
%Note what has been done to SPH galaxies with no gas to put them
%on the plot.}
}
\end{figure*}

\begin{figure*}
\noindent
\begin{minipage}{5.7cm}
  \centerline{\hbox{
      \psfig{figure=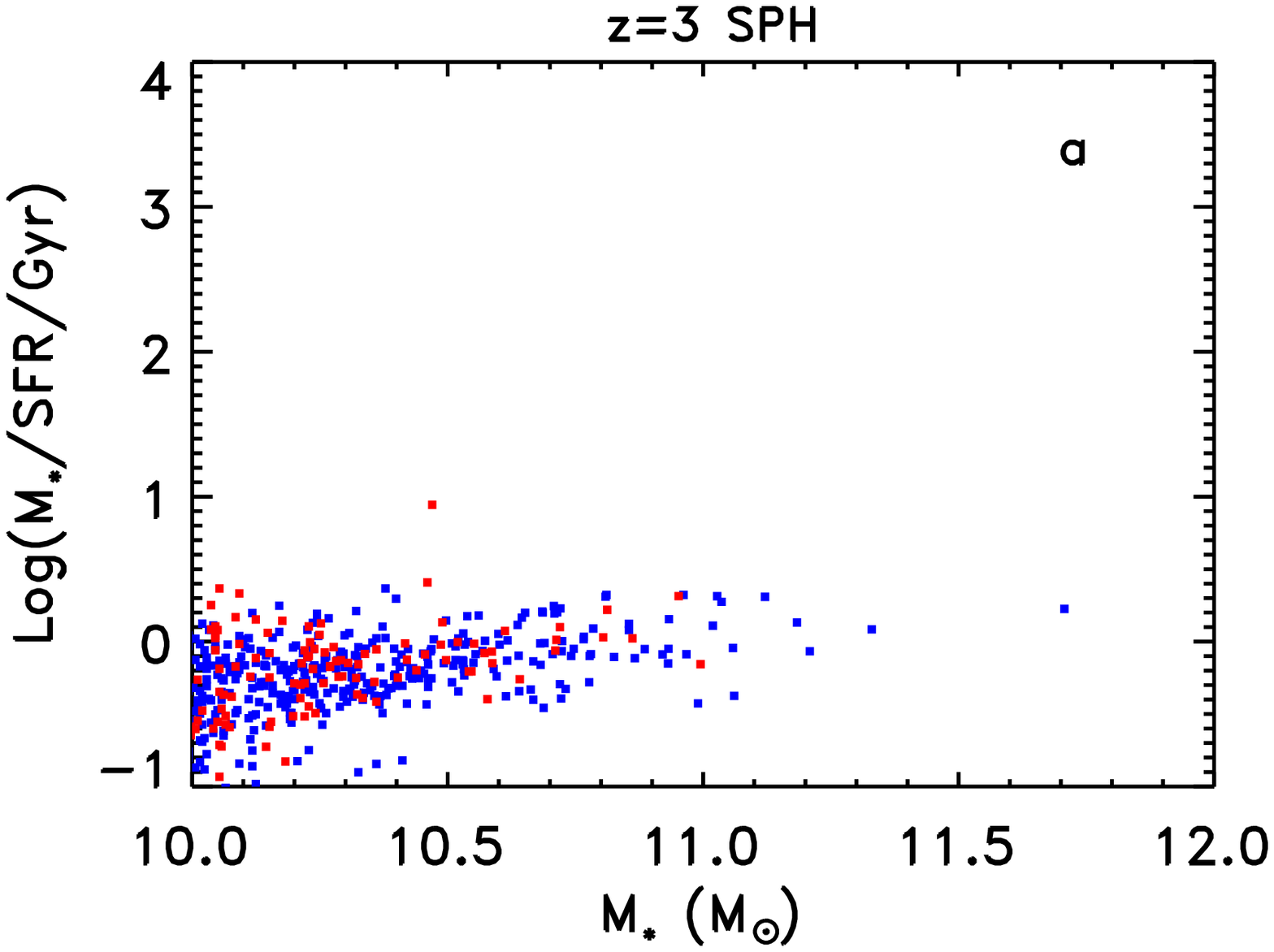,height=4.5cm,angle=0}
  }}
\end{minipage}\    \
%\hskip
\begin{minipage}{5.7cm}
  \centerline{\hbox{
      \psfig{figure=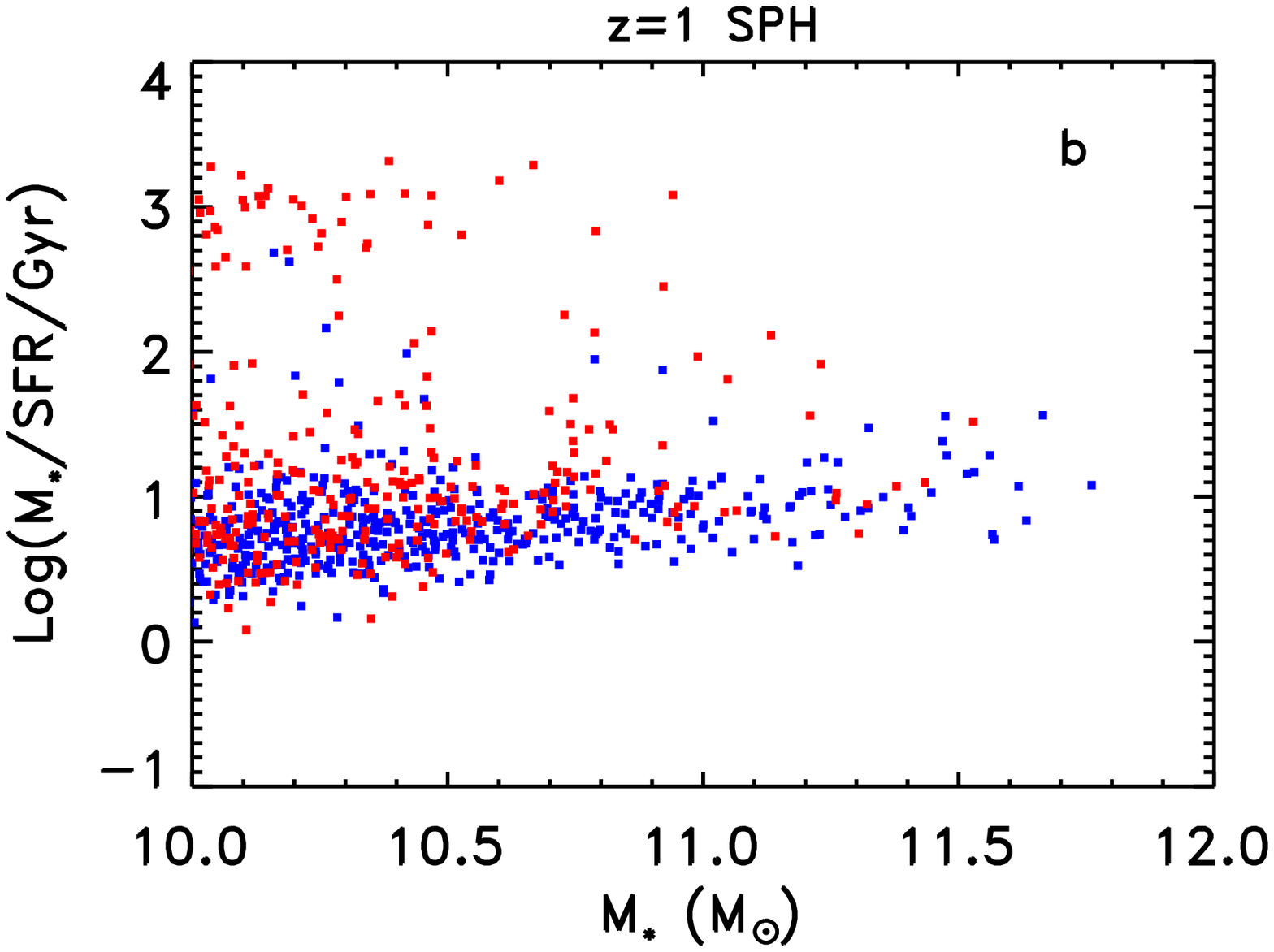,height=4.5cm,angle=0}
  }}
\end{minipage}\    \
%\hskip
\begin{minipage}{5.7cm}
  \centerline{\hbox{
      \psfig{figure=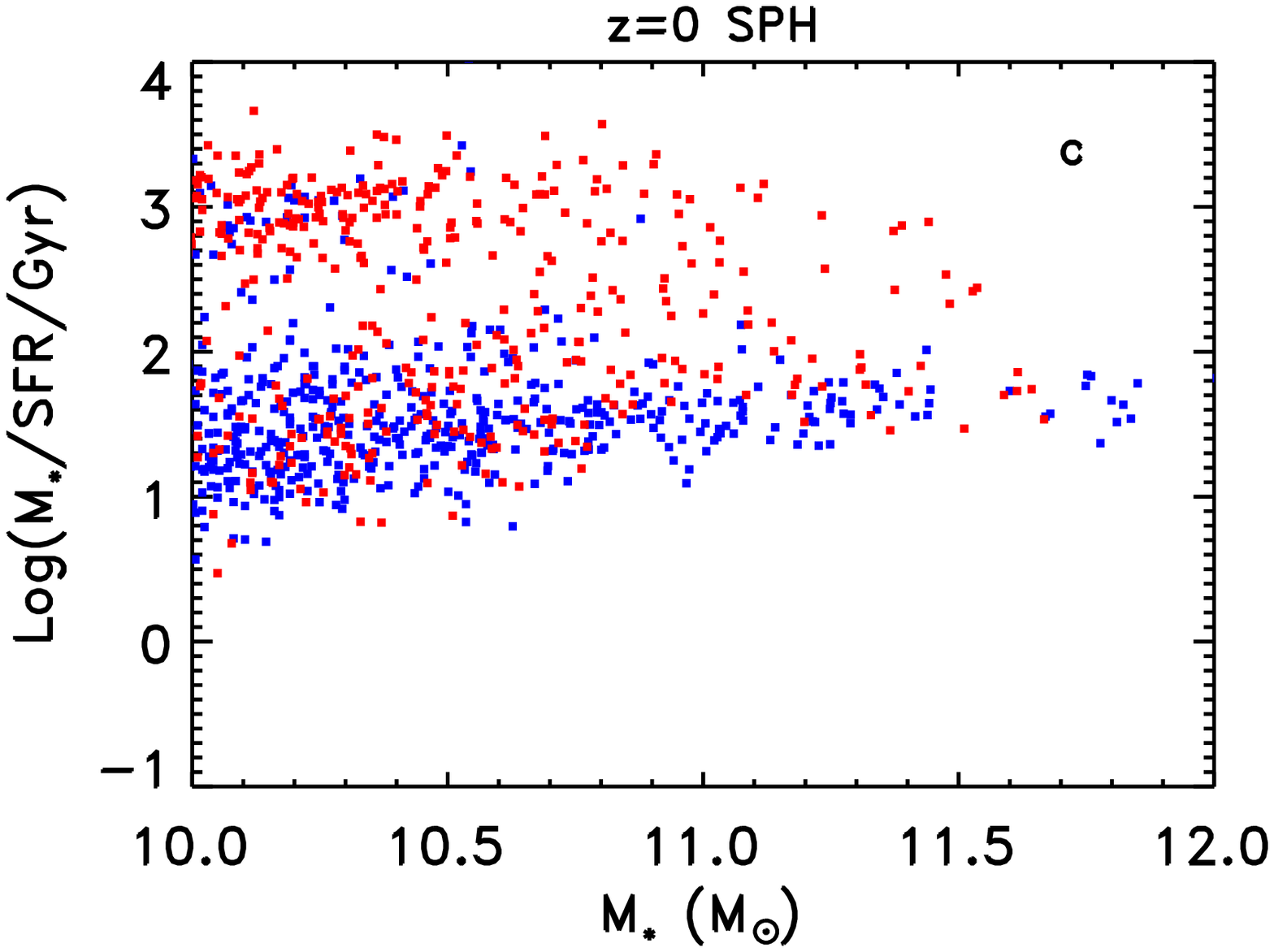,height=4.5cm,angle=0}
  }}
\end{minipage}\    \
%\hskip
\begin{minipage}{5.7cm}
  \centerline{\hbox{
      \psfig{figure=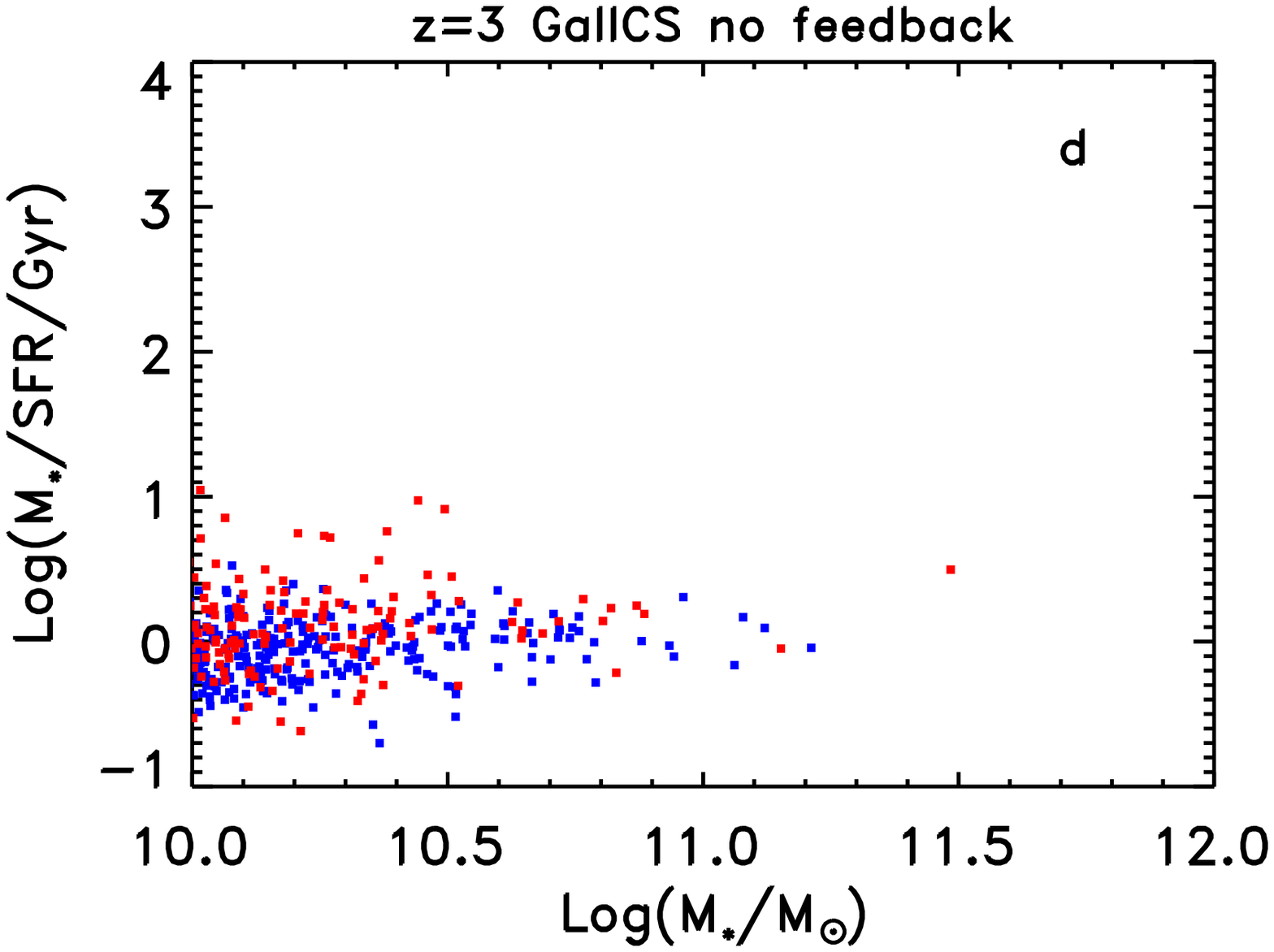,height=4.5cm,angle=0}
  }}
\end{minipage}\    \
%\hskip
\begin{minipage}{5.7cm}
  \centerline{\hbox{
      \psfig{figure=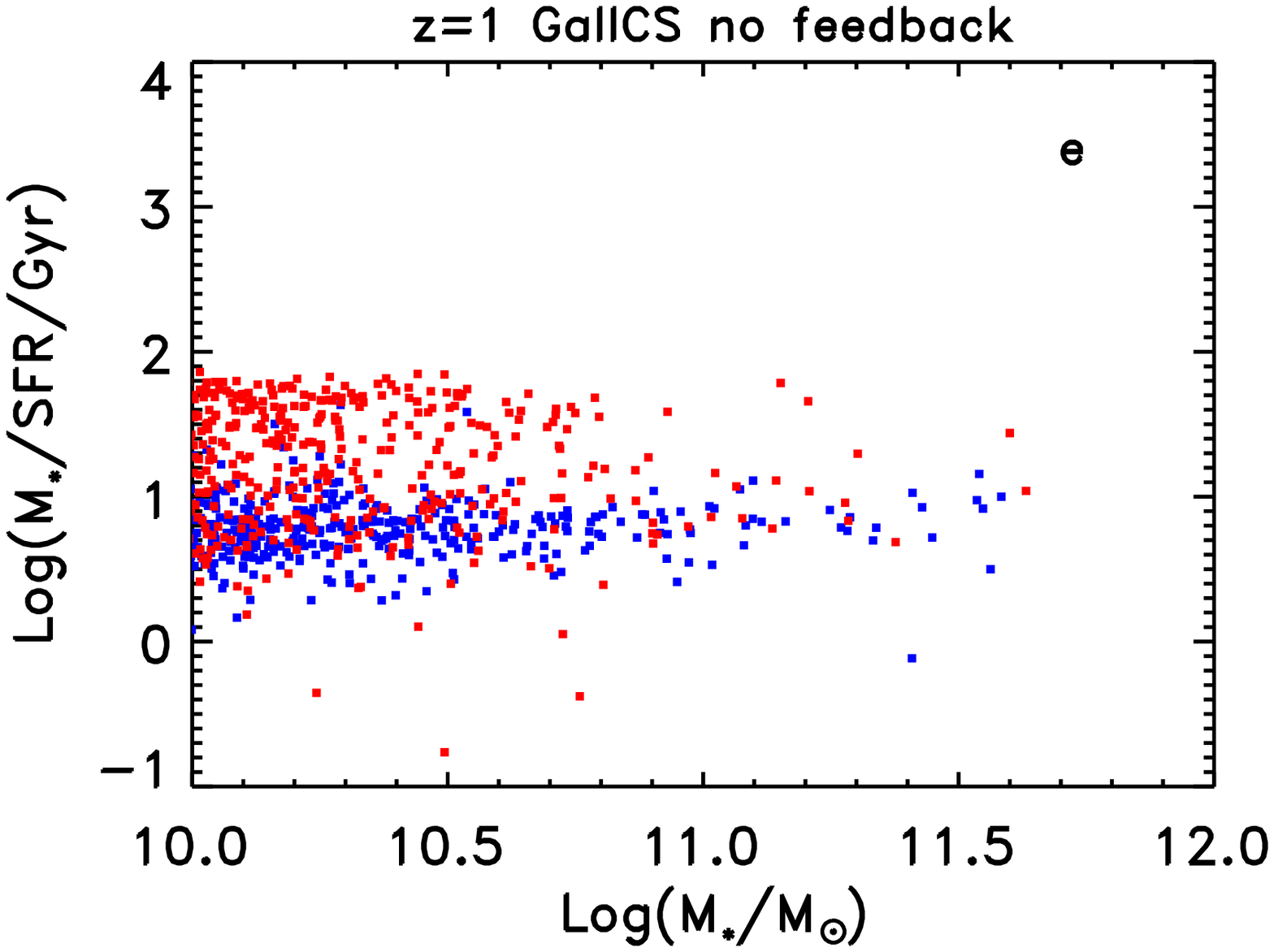,height=4.5cm,angle=0}
  }}
\end{minipage}\    \
%\hskip
\begin{minipage}{5.7cm}
  \centerline{\hbox{
      \psfig{figure=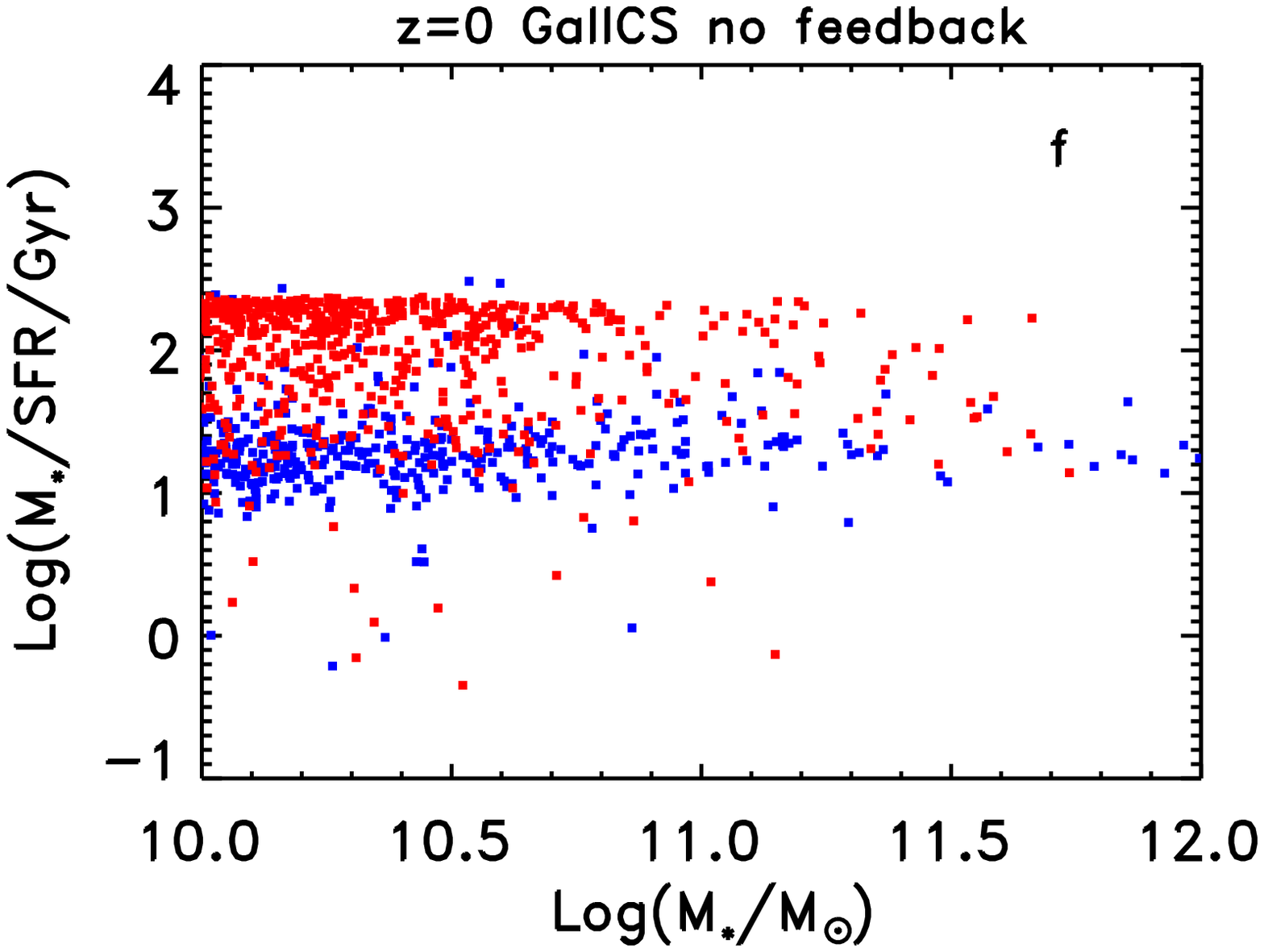,height=4.5cm,angle=0}
  }}
\end{minipage}\    \
\begin{minipage}{5.7cm}
  \centerline{\hbox{
      \psfig{figure=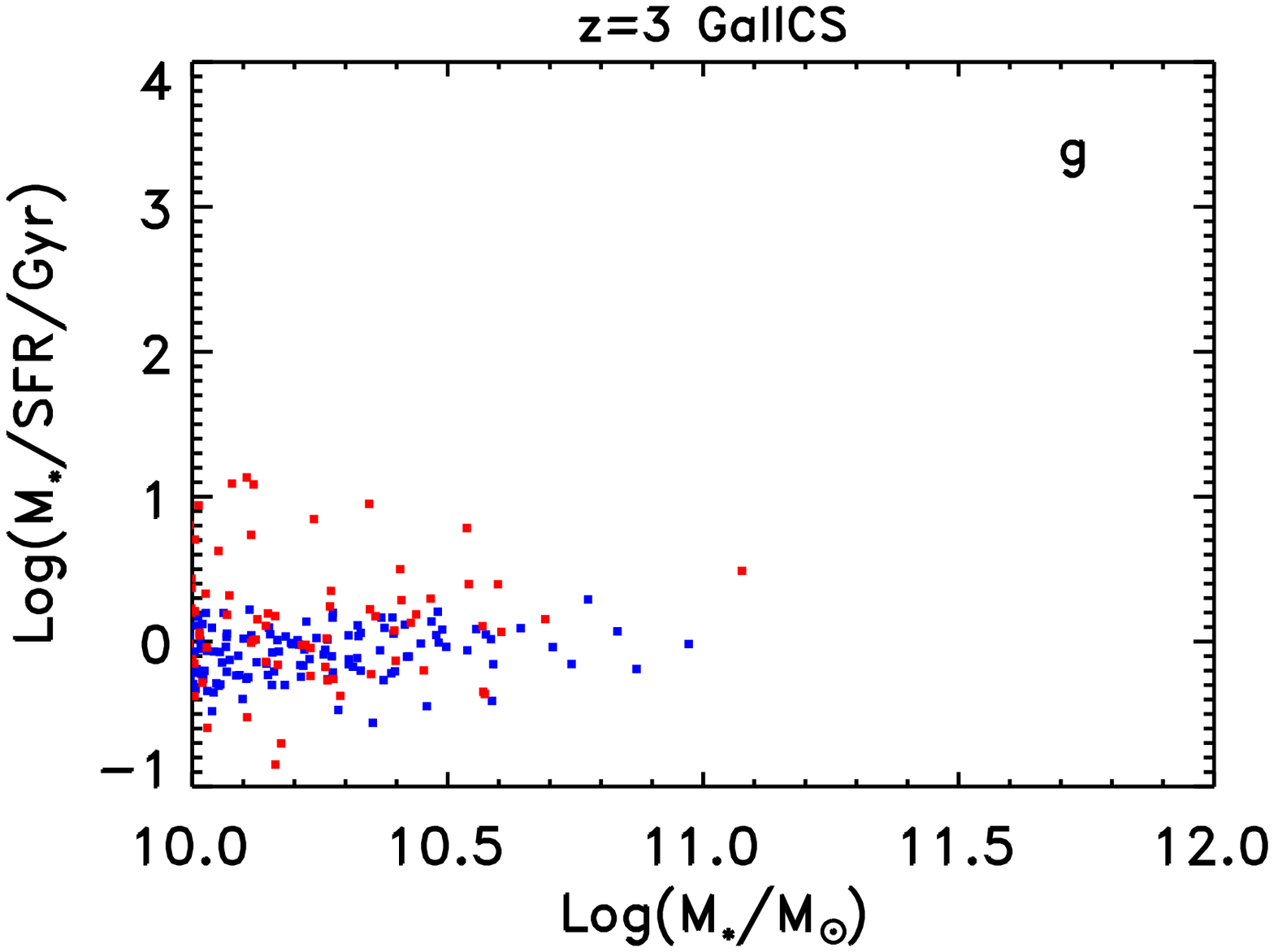,height=4.5cm,angle=0}
  }}
\end{minipage}\    \
%\hskip
\begin{minipage}{5.7cm}
  \centerline{\hbox{
      \psfig{figure=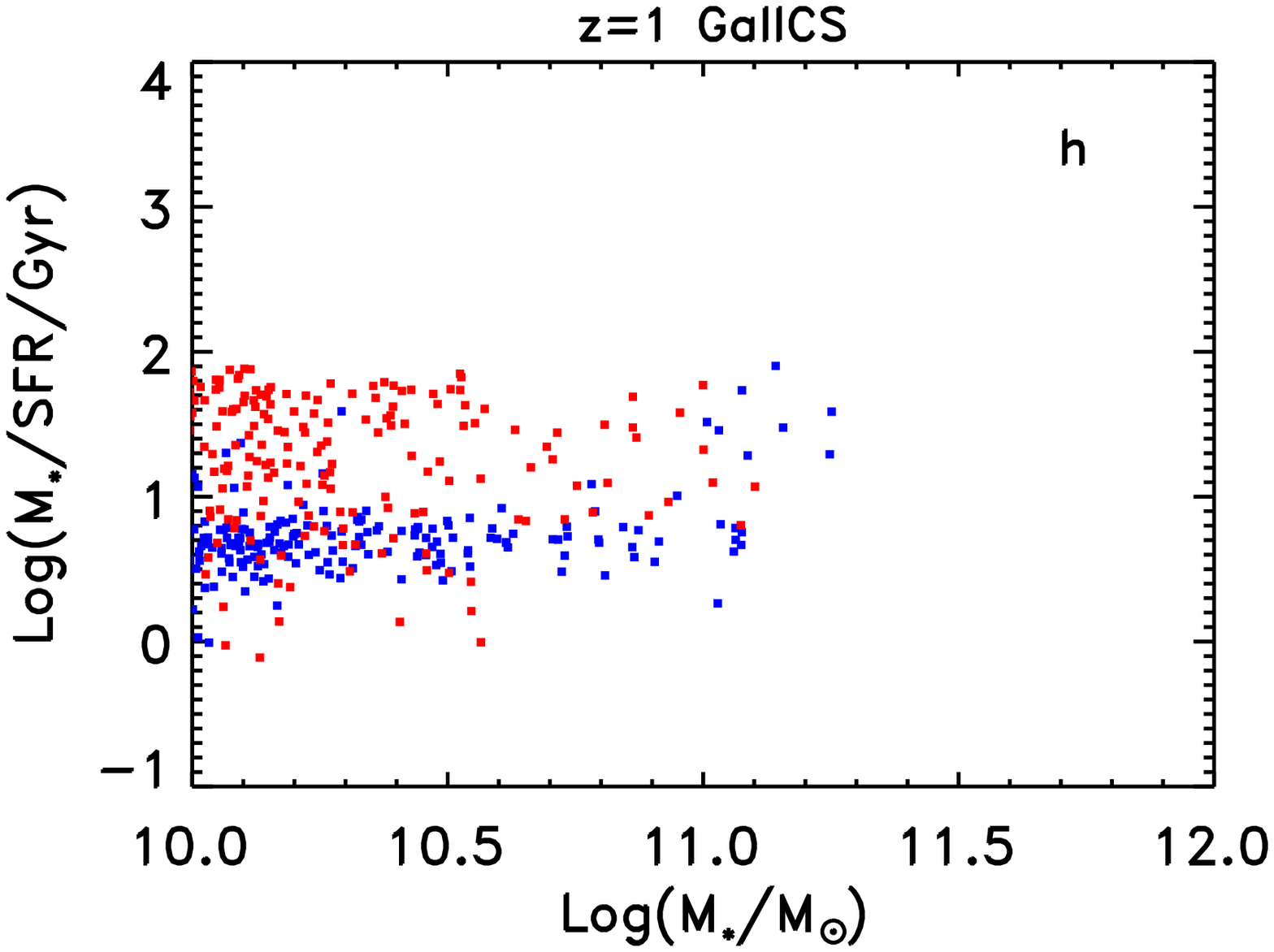,height=4.5cm,angle=0}
  }}
\end{minipage}\    \
%\hskip
\begin{minipage}{5.7cm}
  \centerline{\hbox{
      \psfig{figure=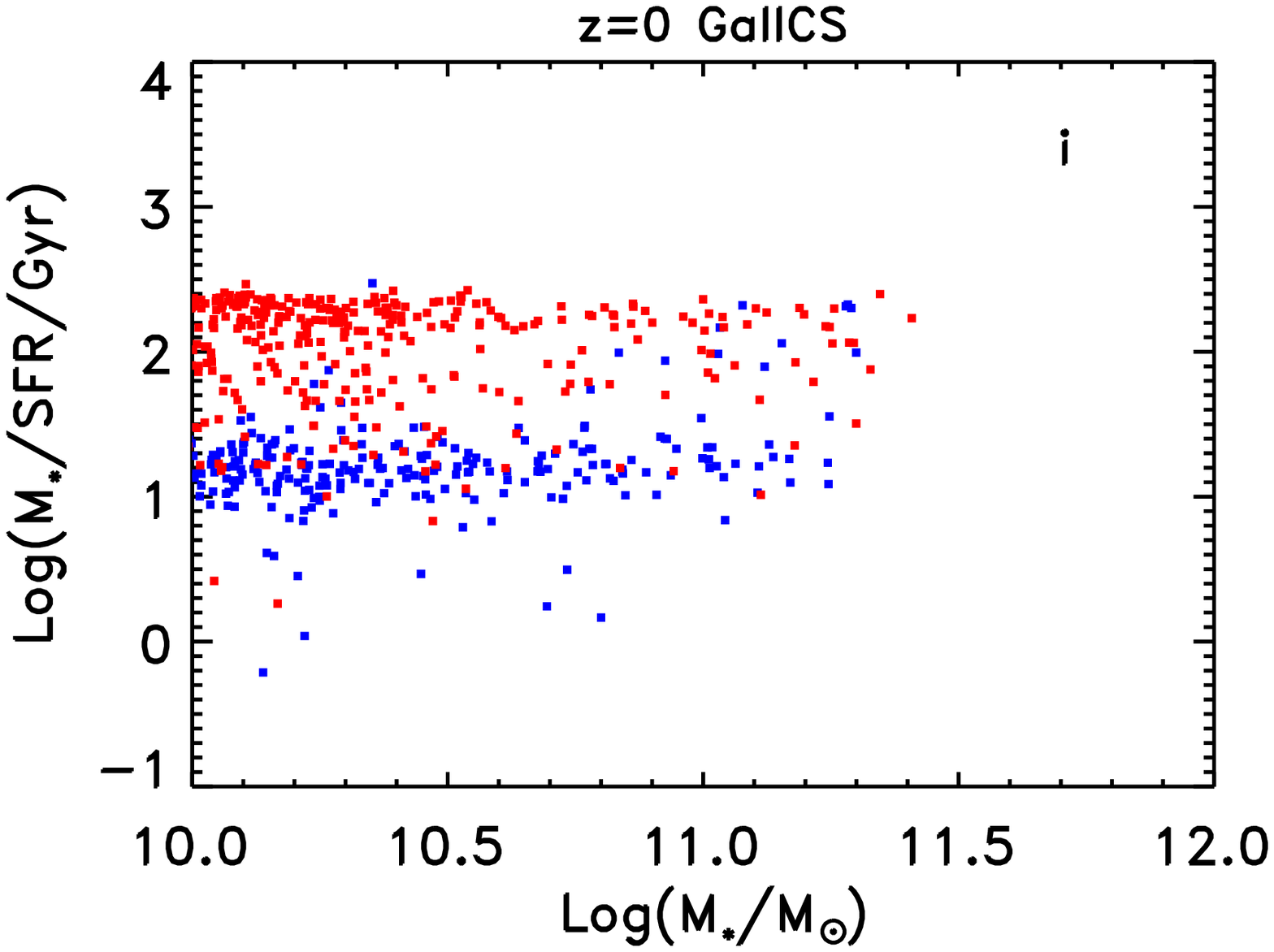,height=4.5cm,angle=0}
  }}
\end{minipage}\    \
\caption{The relation between galaxy mass and $M_*/\dot{M}_*$ at different redshifts in the SPH simulation (top),
in the GalICS model without supernova or AGN feedback (centre) and
in the standard GalICS model (bottom).
$M_*/\dot{M}_*$ is the time that a galaxy would have taken to form its stars
if it had formed them at its current star formation rate. A value of
$M_*/\dot{M}_*$ of the order of the Hubble time suggests a constant star
formation rate throughout the life of the Universe. A value
much longer than the Hubble time implies that the star formation rate has been
much higher in the past than it is now and that the current star formation is
negligible.
The blue points are galaxies at the centre of their halo and the red points are
satellite galaxies.
SPH galaxies with zero SFR have been displayed on the plot by attributing them a value of
${\rm Log}(M_*/\dot{M}_*/{\rm Gyr})$ drawn from a $3\pm0.2$ Gaussian distribution.}
%{\bf Correct the y-axis label of the top row.  Note what has been done
%for SPH galaxies with zero SFR to get them on the plot.}
\end{figure*}

At $z=3$, the SPH and no-feedback \galics\ populations show
similar trends between gas mass and SFR, roughly
SFR$=16 (M_{\rm gas}/10^{10}M_\odot)^{1.8} M_\odot\,{\rm yr}^{-1}$.
%{\bf Right?}
%AC: yes
There is substantial scatter about the mean relation but
no obvious difference between central and satellite galaxies
of the same gas mass.  The SFR at fixed gas mass in the full
\galics\ model is lower by a factor $\sim 2$, but the slope of
the trend is similar.  At $z=0$, the SFR for a given $M_{\rm gas}$
is lower than at $z=3$, as expected given the characteristically
larger sizes and lower surface densities of low redshift
galaxies.  The SFRs are slightly higher at fixed $M_{\rm gas}$
in the SPH simulation, but there are many fewer galaxies with
large gas masses because massive galaxies in the SPH simulation
convert nearly all of their gas into stars.
The trends in Figure~9 are determined largely by the
star formation law, which is qualitatively similar among
the three models (even though the feedback prescriptions
are different).

Figure~10 plots gas fraction against galaxy baryonic mass. Here
we see important 
%dw difference 
differences
among the three models and the
first indications of bimodal galaxy populations.
As in Figure~9, blue points represent central galaxies and red
points satellite galaxies.  At $z=0$, the no-feedback \galics\
model shows two distinct, albeit fuzzy sequences in this
$M_{\rm gas}/M_{\rm gal}$ vs. $M_{\rm gal}$ plane.
The gas rich sequence is centred on
$M_{\rm gas}/M_{\rm gal} \sim 0.2$, while the gas poor
sequence is centred on
$M_{\rm gas}/M_{\rm gal} \sim 0.005$,
with a large scatter.  Most central galaxies occupy the
gas rich sequence and most satellite galaxies occupy
the gas poor sequence, but there are exceptions in both
directions.  The most obvious effect of feedback in the full \galics\ model
is to reduce the baryonic masses of the most massive galaxies, to a
maximum of $\sim 3\times 10^{11} M_\odot$.  In contrast to the
no-feedback \galics\ case, many of the most massive central
galaxies are gas poor instead of gas rich.

In the SPH simulation, it is possible for a galaxy to have a
gas fraction of zero, if all of the particles that compose it
are purely stellar.  We have arbitrarily assigned these galaxies
a gas fraction of $10^{-3.5}$ with a 0.2 dex scatter,
so that one can see them in Figure~10b.  The dashed line in
this panel shows the gas fraction for a galaxy that contains
one gas SPH particle only.  There is a significant
gap between the bottom edge of the upper sequence of points and
this discreteness limit.  This gap suggests a genuinely bimodal
behaviour rather than an artificially divided continuum, with all
points above the line belonging to the gas rich sequence.
Within this sequence, satellite galaxies have lower
average gas fractions than central galaxies of the same mass.
There is also a trend of decreasing gas fraction with
increasing galaxy mass, but this is at least partly an artefact
of resolution: gas densities are underestimated in low mass 
%dw galaxies.  Gas consumption by star formation is thus underestimated, too.
galaxies, and gas consumption by star formation is thus underestimated as well.
Consequently, we do not regard this trend as a robust
prediction, at least for masses below $M_{\rm gal} \sim 2\times 10^{10}M_\odot$ (less than about 200 SPH particles).
Once we account for the resolution and discreteness effects,
the SPH results appear qualitatively similar to the no-feedback
\galics\ results, with the notable exception
that many SPH satellites occupy the
low end of the gas rich sequence rather than the separate gas
poor sequence.

Figure~11 presents the principal result of this paper, the distribution
of galaxies in the plane of stellar mass and star formation time-scale,
$M_*/\dot{M}_*$.  Dividing the Hubble time $t_0=13.4\,$Gyr by this
time-scale gives the ratio of the galaxy's current SFR to its time-averaged
SFR.  The light from galaxies with long star formation time-scales is
dominated by red giants from old stellar populations, while young
stars produce a significant amount of blue light in galaxies with
more current star formation.  Figure~11 is thus a theoretical version
of a galaxy ``colour-magnitude'' diagram.  At low redshift, many SPH
galaxies have zero star formation rate.  We have arbitrarily
assigned these galaxies time-scales of $10^3$ Gyr, with random scatter
of 0.2 dex, so that they appear on these plots.
The GalICS time-scale distributions have a sharp upper edge at
$M_*/\dot{M}_* \sim 10^{2.4}\,$Gyr.
To understand its origin, consider
the specific SFR,
$\dot{M}_*/M_*=(\dot{M}_*/M_{\rm gas})/(M_{\rm gas}/M_*)$.
The  first term is $\dot{M}_*/M_{\rm gas}\sim 2(M_{\rm gas}/10^{10}M_\odot)^{0.4} M_\odot{\rm yr}^{-1}$
(Figure~9). The second term is $M_{\rm gas}/M_*\gsim 10^{-3}$ in the GalICS model, but not in the SPH model, where many galaxies contain no gas at all (Figure~10).
The tendency of GalICS galaxies to retain a minimum gas fraction
explains the upper edge in the $M_*/\dot{M}_* $ distribution.
%is caused by {\bf ??}.

The main features of Figure~11 follow naturally from the trends in
Figures~9 and~10.  At $z=0$, the SPH simulation has a bimodal galaxy
distribution.  The old (red) sequence is populated mainly by satellite
galaxies, with a few lower mass central galaxies mixed in.  The young
(blue) sequence contains a mix of central and satellite galaxies, but
the consumption of gas in satellites is evidently fast enough to
leave few galaxies with intermediate star formation time-scales.
The most massive central galaxies all reside on the young
sequence because of their continuing gas accretion, as seen
already in Figure~10.  Results for the no-feedback \galics\ model
are similar, except that there are fewer satellites in the young
sequence.  In \galics, galaxies experience no gas accretion at all
once they fall into larger haloes and become satellites.  In the SPH
simulation, some satellites continue to accrete gas, e.g. because
they are the central objects of sub-haloes that have not yet dissolved
and lost their identity.  The complete shutoff of accretion in \galics\
satellites therefore appears overly drastic.

Results for the full \galics\ model are similar, with the crucial
distinction that many of the massive central galaxies are now
on the old sequence.  As we have already emphasised, this observational
success of the full \galics\ model is a consequence of AGN feedback, which
suppresses continuing gas accretion in the most massive haloes.
The SPH and no-feedback \galics\ populations do include some fairly
massive galaxies in the old sequence, but nearly all of these are
satellites in the most massive halo (see Figure~2).  This halo
is anomalously large for this simulation volume,
so a larger simulation would probably yield a smaller
fraction of massive old galaxies in these two models.

\citet{kauffmann_etal04} have investigated the distribution of
$\dot{M}_*/M_*$ in the SDSS, using spectral line diagnostics to
estimate star formation rates and stellar mass-to-light ratios.
They obtain a bimodal distribution with peaks corresponding to
$\log (M_*/\dot{M}_*/{\rm Gyr}) \sim 1.2$ and 2.6, in reasonable
agreement with the \galics\ predictions in Figure~11.
The typical star formation time-scales on the young sequence
in the SPH simulation appear slightly too long.

Moving to higher redshift, we see that all three models predict
a unimodal distribution of star formation time-scales at $z=3$,
with typical value $M_*/\dot{M}_* \sim 1\,$Gyr.  By $z=1$, the
time-scale on the young sequence has increased to about $4\,$Gyr,
and an old sequence has begun to emerge.  In the SPH and no-feedback
\galics\ populations, this sequence mostly contains relatively low
mass satellites.  However, in the full \galics\ model
%, however,
the central galaxies of the most massive $z=1$ haloes have already joined the
``red and dead'' population.

\section{Summary and Discussion}

%{\bf Note that much of the discussion is now incorporated into the
%results sections.  Rather than repeat it at the length that it
%was in the previous draft, I have gone for a more concise and punchy
%conclusion.  I think this will be more effective.}

We have compared the galaxy populations of an SPH simulation,
the hybrid \galics\ model of \citet{hatton_etal03}, and a stripped
down version of this model with no feedback from star formation
or AGN.  The \galics\ calculations are applied to the dark matter
halo population extracted from the SPH simulation, allowing
direct comparison of the results.

For the most part, the SPH and no-feedback \galics\ predictions
agree remarkably well.  In particular, the two methods predict
similar galaxy baryonic mass functions and similar dependence of these
mass functions on environment and redshift.  The SPH simulation predicts
somewhat more rapid growth at high redshift, probably a consequence
of efficient filamentary accretion and a star formation time-scale
that tracks the shorter dynamical time-scales of high redshift galaxies.
The global gas accretion histories are similar in the two calculations,
and this agreement extends to the individual galaxy level as shown
in Figure~6.  The breakdown between cold and hot accretion in the SPH
simulation is qualitatively similar to the breakdown between the
rapid cooling ($r_{\rm cool} > r_{\rm infall}$) and slow cooling
($r_{\rm cool} < r_{\rm infall}$) accretion in \galics.
This agreement supports the suggestion by \citet{croton_etal05}
that the cold/hot dichotomy seen in hydrodynamic simulations
\citep{katz92,katz_etal03,birnboim_dekel03,keres_etal05,dekel_birnboim06} can be
identified with the infall-dominated/cooling-dominated regimes of traditional
semi-analytic models \citep{white_frenk91,kauffmann_etal93,cole_etal94}.
\citet{keres_etal05} note some important caveats to this
identification, especially for models that incorporate a
photoionising background, which increases cooling time-scales in
low mass haloes.

Both the SPH and the no-feedback \galics\ models predict a bimodal
galaxy population with a ``blue'' sequence of gas-rich, star-forming
galaxies and a ``red'' sequence of gas-poor galaxies dominated by
old stellar populations.  The red sequence is populated mainly by
satellite galaxies, which stop accreting gas after they fall
into larger haloes.  A larger fraction of SPH satellites remain
on the gas rich, star-forming sequence, which suggests that the
truncation of gas accretion in \galics\ is unrealistically abrupt.
However, both calculations agree on a crucial point: the central
galaxies of massive haloes experience continuing gas accretion,
and they therefore remain on the star-forming sequence.
The masses of these galaxies are substantially larger than those
inferred for galaxies of similar space density in the real
universe --- i.e., both models drastically overpredict the high
end of the galaxy baryonic mass function.
There is also agreement that gas accretion by satellites in cluster-size haloes is negligible.

While the agreement between SPH and no-feedback \galics\ could in principle
reflect a conspiracy of errors, it seems more reasonable to conclude
that both methods do a fairly accurate job of modelling the growth
of galaxies from cosmological initial conditions.  This conclusion in
turn implies that the observational failings of these models are a
consequence of missing input physics.  The full \galics\ model
is much more successful in reproducing observations because of
its prescriptions for supernova and AGN feedback.
Supernova feedback reduces the masses of low and intermediate
mass galaxies by about a factor of two, as the gas outflow rate
is comparable to the star formation rate by construction.
AGN feedback produces a sharp cut-off in the baryonic mass
function by suppressing the cooling of gas in  high mass haloes.
Crucially, this mechanism also moves the central galaxies of
massive haloes to the old, gas-poor sequence.

Some mechanism that suppresses gas cooling in massive haloes appears
essential to explaining both the exponential cut-off in the observed
galaxy luminosity function \citep{kauffmann_etal93,benson_etal03}
and the observed form of bimodality in the galaxy population, where
nearly all of the most massive galaxies reside on the
red sequence \citep{blanton_etal03,kauffmann_etal03}.
As argued by
\citet{birnboim_dekel03},
\citet{katz_etal03},
\citet{binney04},
\citet{keres_etal05}, and \citet{dekel_birnboim06},
it seems natural to associate this mechanism with the
``hot mode'' of gas accretion that becomes dominant in haloes
more massive than $\sim 10^{12}M_\odot$, since the implied
mass scale is close to the transition mass inferred observationally
by \citet{kauffmann_etal03} and since the amounts of cool gas
in observed galaxy clusters are much lower than predicted by
the X-ray cooling rates in the absence of heating
(e.g., \citealp{kaastra_etal01}).
In this view, the factor that determines the effectiveness of
AGN feedback is not so much the mass of the black hole as the
state of the gas that it will affect.
Even a luminous AGN will have difficulty reversing the inflow
of dense, cold gas along filamentary streams, or ejecting the
interstellar medium of a galaxy disc.  However, an AGN can
couple more easily to the diffuse, quasi-spherical hot atmosphere
of a massive halo, and, with plausible accretion rates and
efficiencies, there is enough energy to suppress the gas
cooling that would otherwise be expected in these haloes
(see, e.g., \citealp{best_etal06}).

In this regard, 
%dw is it 
it is
interesting that the new \galics\ model of
\citet{cattaneo_etal06} is considerably more successful than the
\citet{hatton_etal03} version of \galics\ tested here.
The Hatton et al.\ implementation ties AGN feedback to black hole
mass and, thus, to bulge mass.
%(\citealp{marconi_hunt03} and \citealp{haering_rix04} present the
%observational evidence for the black hole-bulge mass relation, while
%\citealp{cattaneo_etal05} discuss this relation in the GalICS model).
This prescription suppresses cooling in massive
haloes, but not effectively enough.
Massive galaxies
do lie on a ``red'' sequence separated from the
``blue" sequence of star-forming galaxies. Nevertheless, a quantitative comparison with
the observations obtained using a stellar population synthesis model
shows that they are still not red enough.
Strong supernova feedback,
achieved by assuming that the mass loss rate through a supernova-driven wind
is approximately equal to the star formation
%dw rate independently of the depth of potential well,
%dw can compensates the incomplete shutdown of cooling in massive halo,
%dw but also ejects too many metals.
rate independently of the depth of the potential well,
can compensate the incomplete shutdown of cooling in massive haloes,
but it also ejects too many metals.
%, while they lie on the ``red'' sequence,
%are still not red enough, and the parameters of the supernova
%feedback prescription must be torqued to have large effect in
%high mass haloes so that supernovae can compensate for weak AGN
%feedback.
% {\bf Andrea, you should phrase this sentence more
%accurately.}
%AC: how do you find it now?
The \citet{cattaneo_etal06} model completely suppresses
gas accretion in haloes above the critical mass where virial shocks
arise, computed via the methods of \citet{dekel_birnboim06}.
This implementation yields excellent agreement with the
observed colour-luminosity distribution of galaxies at low and
high redshift.

\citet{croton_etal05} achieve a similar level of observational
success in a hybrid model that uses kinetic feedback from radio-loud
AGN to suppress gas cooling in massive haloes.  Because this
mechanism is assumed to affect only the quasi-hydrostatic atmospheres
that arise in haloes with long post-shock cooling times, the practical effect
of this scheme is quite similar to the sharp cut-off in the
\citet{cattaneo_etal06} model.  Very recently,
\citet{bower_etal06} have presented another model that is very similar to that
of 
%dw \citet{cattaneo_etal06} 
\citet{cattaneo_etal06},
both in the method and the key results.

There is still much to be understood
about the physics that suppresses gas cooling in massive haloes,
i.e., the detailed mechanism by which AGN energy couples to the
surrounding gas, or even whether AGN are really the source of
this suppression at all.  However, it is striking that the
theoretically derived mass scale separating the cold and hot
accretion regimes corresponds so well to the observed mass scale
that marks the transition between the blue and red galaxy populations,
and that the most successful models of the observed galaxy population
are those that incorporate suppressed hot accretion to identify
one transition with the other.

\section{Acknowledgements}
We acknowledge stimulating discussions with A.~Dekel.
A.C. has been supported by a Marie Curie Research Fellowship
at the IAP and by a Golda Meir Fellowship at the Hebrew University of Jerusalem.
Additional support for this research has come from NASA Grant
NAGS-13308.  D.W. thanks the IAP for hospitality during the initial
stages of this work.

%%%%%%%%%%%%%%%%%%%%%%%%%%%%%%%%%%%%%%%%%%%%%%%%%

\bibliographystyle{mn2e}

%\bibliography{dekel}
\bibliography{ref}
%%%%%%%%%%%%%%%%%%%%%%%%%%%%%%%%%%%%%%%%%%%%%%%%%

%\begin{thebibliography}{99}
%\end{thebibliography}

\end{document}